\documentclass[a4paper, 11pt]{article}
\usepackage{jheppub}
\usepackage{float}
\usepackage{amsmath,amssymb,amsfonts,bm}
\usepackage{hyperref}
\usepackage{physics}
\usepackage{mathrsfs}
\usepackage[dvipsnames]{xcolor}
\usepackage{tcolorbox}
\usepackage{graphicx}
\usepackage{multirow}
\usepackage{orcidlink}
\graphicspath{{./Figs/}}
\numberwithin{equation}{section}

\title{Gauss--Bonnet corrected string/black hole transition in large dimensions}

\author[a,b,c]{Bum-Hoon Lee\orcidlink{0009-0008-3322-2087},} 
\author[d]{Hocheol Lee\orcidlink{0000-0002-4420-5898},}
\author[a]{and Somyadip Thakur\orcidlink{0000-0003-2376-5906}}

\affiliation[a]{Center for Quantum Spacetime, Sogang University, Seoul 04107, Republic of Korea}
\affiliation[b]{Department of Physics, Sogang University, Seoul 04107, Republic of Korea}
\affiliation[c]{Department of Physics, Shanghai University, 99 Shangda Road, Baoshan District, Shanghai 200444, China}
\affiliation[d]{Department of Physics, Dongguk University, Seoul 04620, Republic of Korea}

\emailAdd{bhl@sogang.ac.kr, insaying@dongguk.edu, somyadip@sogang.ac.kr}

\abstract{We formulate a large-$D$ exterior description of the Horowitz--Polchinski thermal scalar in a heterotic-inspired Einstein--Gauss--Bonnet truncation. For a background satisfying the coupled equations at the retained order, and with boundary data specified at a stretched-horizon matching surface, near--far matching determines the normalised decaying winding profile, its decay exponent, and the relations among the near- and far-zone coefficients, including their Gauss--Bonnet dependence. The effective description applies in the weak-curvature, near-Hagedorn region and does not resolve the near the string-scale core region, which needs a careful non perturbative analysis. Finite-$D$ integrations illustrate local winding localisation within the truncated equations, without establishing global solutions with the required asymptotics.  We also organise a conditional exterior Landau functional; its complete physical coefficients require core matching data, the full metric--dilaton response, and the ensemble boundary terms.}


\begin{document}
\maketitle
\flushbottom

\section{Introduction} \label{sec:intro}
    Black holes sit at the crossroads of classical geometry, thermodynamics, and quantum theory. Their equilibrium properties are elegantly captured by Euclidean gravity, yet their microscopic origin remains subtle and theory dependent. String theory offers a particularly concrete arena where both viewpoints coexist: at weak coupling, highly excited \emph{long strings} provide a statistical description of entropy, while at larger masses gravitational backreaction produces black holes. A sharp formulation of this crossover is the Horowitz--Polchinski (HP) correspondence~\cite{Horowitz:1996nw,Horowitz:1997jc}, which asserts that a fundamental string transitions into a black hole when its self--gravity becomes comparable to the string scale. At the correspondence point, the entropies, free energies, and characteristic length scales of the two descriptions match parametrically, suggesting a continuous interpolation between string microstates and coarse-grained geometry.

    The HP mechanism is most naturally framed in \emph{Euclidean signature}, where thermal physics is encoded by compactifying time on a circle of period $\beta=1/T$. In the Euclidean worldsheet description, strings can wind the thermal circle; the lightest winding excitation---\emph{the thermal scalar}---governs the approach to the Hagedorn transition~\cite{Atick:1988si,Kruczenski:2005pj,Metsaev:1987zx}. As the temperature is raised, the thermal scalar becomes light and can turn tachyonic, signaling the breakdown of the perturbative string gas. On the gravitational side, a Euclidean black hole is a smooth saddle of the thermal path integral, whose regularity at the horizon fixes the periodicity of the Euclidean time circle and reproduces the Hawking temperature~\cite{Gibbons:1976ue,Gibbons:2004ai}. In this language, the HP crossover becomes a competition between Euclidean saddles in a common ensemble: a thermal long string phase and a Euclidean black hole phase, with the winding sector providing the order--parameter-like degree of freedom that distinguishes them.

    The target space Horowitz--Polchinski effective theory is controlled in the near-Hagedorn, weak curvature regime, where $\alpha'|m_{\rm eff}^2|\ll1$ and $\alpha'{\cal R}\ll1$; it is not a pointwise description of a string-scale Euclidean tip.  The dilaton and higher-derivative operators are retained because they enter the target space equations at the relevant orders in the joint expansion in $\beta-\beta_{\rm H}$, $\alpha'$, and $1/D$.  In the heterotic string, the complete $\mathcal O(\alpha')$ action also contains the antisymmetric tensor, gauge fields, and the supersymmetric Chern--Simons completion~\cite{Gross:1986iv,Metsaev:1987zx,Bergshoeff:1989de,Chemissany:2007he}. Here we isolate the curvature-squared Lovelock sector and retain the zeroth-order thermal scalar operator specified in Sec.~\ref{sec:Action}.

    Two additional developments sharpen the motivation for such a framework. First, the large-$D$ expansion connects higher-dimensional Euclidean Schwarzschild black holes to an \emph{exact} two-dimensional worldsheet background: in an appropriate near-horizon scaling limit, the geometry reduces to the $SL(2)_k/U(1)$ cigar, with the coset level $k$ set by the black hole temperature/horizon scale~\cite{Chen:2021emg}. In particular, one has $\beta/2\pi = 2r_0/(D-3)$ and $k=(2r_0/(D-3))^2$ (in $\alpha'=1$ units), so that $\beta/2\pi=\sqrt{k}$ in this limit~\cite{Chen:2021emg}. This provides a clean physical interpretation of ``Large-$k$'' versus ``small--$k$'' as a dial on the stringiness of the Euclidean cap: for large $k$ the winding condensate is sharply localized near the tip, while decreasing $k$ causes the stringy atmosphere to spread outward and, at a critical value, become non--normalizable and make a formally divergent contribution to the entropy \cite{Chen:2021emg}. Second, motivated by FZZ duality, it has been observed that the HP effective string equations admit a nontrivial \emph{first-order} reformulation for the metric--dilaton--winding sector, and that in higher dimensions an analogous first-order structure persists in the cap/low temperature regime; remarkably, at a critical winding amplitude the Euclidean cigar develops a puncture and the entropy carried by the winding condensate reproduces the Schwarzschild Bekenstein--Hawking entropy~\cite{Krishnan:2024zax}. Together, these insights suggest that a universal, effectively two-dimensional cap dynamics controls the onset of the HP transition, while the exterior region primarily supplies boundary data and organizes corrections.

    Black holes corrected by higher-derivative interactions often display a much richer thermodynamic landscape than their two-derivative counterparts. In particular, such corrections can generate novel critical behavior and thermodynamic phase structure—including first-order/second-order transitions and re-entrant phenomena—and may also trigger topological transitions, where the qualitative structure of the spacetime/Euclidean saddle (and the associated thermodynamic branch data) changes discontinuously. For recent discussions of these exotic features in concrete higher-derivative setups, see Refs.~\cite{Ahn:2014fwa, Khimphun:2016gsn, Lee:2018zym, Lee:2021uis, Jeon:2024yey}.

    In this paper we study a conditional exterior problem for the Horowitz--Polchinski thermal scalar within the stated EGB truncation. The large-$D$ expansion separates the radial regions and organises the local equations at fixed stretched horizon matching data. A physical interpretation requires the background to satisfy the coupled metric and dilaton equations at the retained order, together with the EFT validity conditions throughout the region being used. The exterior calculation does not determine the string-scale core data, the physical condensate normalisation, or a complete comparison of the competing thermodynamic saddles.

\subsection*{Broader Physical Connections: Superconductivity, QCD, and Fuzzballs}
    The thermal scalar perspective also illuminates broader connections that extend beyond the HP correspondence. Its onset near the Hagedorn point behaves as an emergent order parameter undergoing a symmetry--breaking transition, evocative of condensate formation in superconductivity. In holographic settings, Euclidean scalar condensation near black holes generates an effective Ginzburg--Landau potential for a boundary order parameter~\cite{Hartnoll:2008vx,Hartnoll:2008kx}; here, the thermal scalar provides a natural worldsheet origin for such Euclidean instabilities and suggests a stringy analogue of superconducting order.

    The connection with confinement/deconfinement is especially sharp in holographic AdS models.  There the Hawking--Page transition between thermal AdS and an AdS black hole is the large-$N$ confinement/deconfinement transition of the dual gauge theory~\cite{Witten:1998zw}; weak coupling and AdS/QCD realizations likewise connect Hagedorn growth to deconfinement~\cite{Rama:1998jm,Aharony:2003sx,Herzog:2006ra,Cai:2007zw}.  Recent string-star analyses make the relation to Horowitz--Polchinski saddles and black hole/string transitions in confining holographic backgrounds more direct~\cite{Urbach:2023wkw}.  We use this literature as motivation and do not claim to derive the transition of a specific QCD-like boundary theory.

    A third connection arises with proposals for black hole microstructure, particularly the fuzzball paradigm. In fuzzball models, individual microstate geometries replace the classical horizon~\cite{Mathur:2005zp,Skenderis:2008qn,Bena:2022rna}. The HP framework instead provides a thermal interpolation in which a long string, once sufficiently self--gravitating, evolves into a Euclidean black hole. The large-$D$ viewpoint sharpens this bridge by isolating the universal cap region where winding condensation and tip dynamics are controlled by an effectively two-dimensional system, thereby linking microscopic string structure to coarse-grained geometry in a particularly transparent way.

\subsection*{Large-\texorpdfstring{$D$}{D} and Large-\texorpdfstring{$k$}{k} Motivation}
    Technically, the large-$D$ expansion is useful because radial gradients localize in a layer of thickness $r_h/D$.  This produces separate near- and far-zone expansions with a parametrically large overlap, so the exterior winding profile and its backreaction can be organized order by order in $1/D$~\cite{Emparan:2013moa,Emparan:2014aba,Emparan:2015hwa,Chen:2018nbh,Dandekar:2019hyc}.  The exact finite-$D$ Boulware--Deser geometry supplies the vacuum far-zone background; matching is a consistency check for that known geometry and becomes nontrivial only for the dilaton, thermal scalar, and their backreaction.

    The relation to the $SL(2)_k/U(1)$ cigar provides complementary worldsheet motivation~\cite{Chen:2021emg,Krishnan:2024zax}.  Subleading $1/D$ terms capture geometric finite-$k$ effects within the metric--dilaton--thermal scalar truncation, but they do not automatically reproduce the full finite-$k$ string spectrum or all higher-$\alpha'$ corrections.  We therefore use the cigar correspondence as a guide to the exterior cap dynamics, not as a claim that the large-$D$ effective theory resolves the string-scale tip.

    In this work, we analyse the exterior HP configuration in an EGB truncation at fixed stretched horizon matching data. We derive the near-zone equations for the metric, dilaton, and thermal scalar, adopt the known Boulware--Deser geometry in the far zone, and relate the exterior profiles in the overlap region. Within the stated approximations, the calculation determines the thermal scalar decay exponent, relative matching coefficients, and Gauss--Bonnet dependence of the plateau marginality condition. The amplitude at the matching surface and the core contribution to the action remain undetermined. Numerical integrations over finite radial intervals provide local diagnostics of the truncated equations. The paper is organised as follows:
\begin{itemize}
    \item  In section ~\ref{sec:hphd}, we introduce the Horowitz--Polchinski model and the higher-derivative correction.  We display the exact string frame metric, dilaton, and thermal scalar equations and then identify the terms that enter beyond the retained large-$D$ order.

    \item In section~\ref{sec:solld}, we construct the analytic exterior solution in a large-$D$ expansion.  The near and far descriptions are matched in the common region $1\ll\rho\ll D$, while the string-scale core is replaced by data at the stretched horizon.  At leading order the background has the universal large-$D$ cigar form, the dilaton is constant, and the thermal scalar obeys a simplified radial equation.  At next-to-leading order the scalar sources perturbative metric and dilaton corrections.  We also organize a cutoff-regulated Landau functional into core and exterior pieces, without claiming that the complete coefficients are fixed by the exterior calculation.

    \item In section~\ref{sec:numsol}, we present finite interval numerical integrations of the full radial equations in unit radial gauge. We discuss the local field profiles and distinguish the displayed trajectories from solutions satisfying prescribed asymptotic boundary conditions. Section~\ref{sec:disc} summarises the exterior results and the unresolved requirements for a global solution.

    \item We have also supplemented with Apps.~\ref{app:oldresults},~\ref{app:subnum},~and~\ref{app:NLO-details} explaining some of the solutions methodology and some quick reviews. 
\end{itemize}

\section{Higher-Derivative Horowitz--Polchinski Model} \label{sec:hphd}
    The HP correspondence~\cite{Horowitz:1996nw,Horowitz:1997jc} provides a controlled framework for analysing the transition between highly excited long strings and black holes. In the Euclidean thermal ensemble, obtained by compactifying time on a circle of circumference $\beta=1/T$, the dominant degrees of freedom near the Hagedorn temperature arise from singly wound strings. Integrating out all non-winding oscillators yields an effective spacetime description in terms of the \emph{thermal scalar}~\cite{Atick:1988si,Kruczenski:2005pj,Zakharov:2016fba}. Its effective mass,
\begin{equation}
	m_{\rm th}^2
	= \frac{\beta^2 g_{\tau\tau}}{4\pi^2\alpha'^2}
	- \frac{2}{\alpha'} + \cdots ,
\end{equation}
    becomes tachyonic as $T\to T_H$, making the thermal scalar the natural order parameter for the HP transition. Because the Euclidean time circle shrinks near a black hole horizon, the thermal scalar localises in the near-tip region and is sensitive to the detailed structure of the Euclidean geometry.

\subsection*{Dilaton Contribution}
    The dilaton $\phi$ controls the local string coupling and enters the target space equations at the same order as the metric and winding mode in the near-Hagedorn expansion.  We retain it for this order-by-order consistency, while requiring $\alpha'(\nabla\phi)^2\ll1$ in the region where the effective theory is applied.

\subsection*{Higher-Derivative Corrections and the Gauss--Bonnet Term}
    We use the Gauss--Bonnet (GB) density as a controlled representative of the leading curvature-squared gravitational sector,
\begin{equation}
	R_{\rm GB}^2
	= R_{\mu\nu\rho\sigma}R^{\mu\nu\rho\sigma}
	- 4 R_{\mu\nu}R^{\mu\nu}
	+ R^2.
\end{equation}
    This Lovelock combination has second-order, ghost-free gravitational field equations in $D>4$~\cite{Zwiebach:1985uq,Boulware:1985wk}.  In the heterotic string the purely gravitational curvature-squared sector can be brought to GB form modulo field redefinitions and suitable truncations, but the complete $\mathcal O(\alpha')$ supersymmetric action contains additional fields and operators~\cite{Duff:1986pq,Gross:1986iv,Metsaev:1987zx,Bergshoeff:1989de,Chemissany:2007he}.  Type-II theories instead have their leading purely gravitational correction at order $\alpha'^3R^4$~\cite{Gross:1986mw}.

\subsection*{Gauss--Bonnet Coupling versus \texorpdfstring{$\alpha'$}{alpha'}}
    Accordingly, the model is best described as a large-$D$, heterotic-inspired EGB (Lovelock) truncation coupled to the thermal scalar.  The coefficient $\alpha$ is treated as an independent EFT Wilson coefficient, and we do not claim that the isolated GB term is the complete leading correction of any critical string theory.  This restriction isolates the response of the exterior thermal scalar configuration to a tractable curvature-squared deformation while avoiding higher-derivative gravitational equations and associated field redefinition ambiguities~\cite{Zwiebach:1985uq,Callan:1988hs,Cai:2001dz}.

\subsection*{Large-\texorpdfstring{$D$}{D} Motivation}
    The coupled Euclidean system involving the metric, dilaton, thermal scalar, and Gauss--Bonnet corrections is analytically intractable at fixed dimension. The large-$D$ expansion~\cite{Emparan:2013moa,Emparan:2014aba,Emparan:2015hwa} provides a powerful simplification: gravitational fields localise in a thin near-zone region of width $r_h/D$, the thermal scalar becomes effectively one-dimensional, and higher-derivative terms acquire controlled $D$-scaling. The far-zone is governed by Einstein--Gauss--Bonnet equations, while the near-zone reduces to a universal Schr\"odinger-type problem that determines the thermal scalar decay exponent. Although the dynamical membrane paradigm~\cite{Bhattacharyya:2015dva,Bhattacharyya:2015fdk} is extremely useful for real-time perturbations, the HP transition is \emph{static} in Euclidean signature. Thus only the static near-/far-zone decomposition is required. 
    
    At this point, let us make a short statement clear: \textit{Even though critical string theory is defined in a fixed target space dimension ($D=10$ for type~II and heterotic strings and $D=26$ for the bosonic string), the large-$D$ expansion can be used as a controlled analytic continuation in the spirit of Ref.~\cite{Chen:2021dsw}.  One treats $D$ as a parameter of the effective equations, computes observables in a systematic $1/D$ expansion, and assesses any continuation to a physical value of $D$ from the size of subleading terms.  A Euclidean string-theoretic embedding can be obtained by supplementing the Schwarzschild factor with a timelike linear dilaton whose central charge contribution is tunable.  This makes large $D$ a useful analytic laboratory for near-Hagedorn Euclidean thermodynamics, but does not extend the target space EFT into the excised string-scale tip or supply a standard Lorentzian continuation.}

\subsection{Euclidean EGB--Dilaton--Thermal Scalar Action} \label{sec:Action}
    To study these ingredients within the stated truncation, we adopt the Euclidean action
\begin{equation}
	I = -\int d^Dx\,\sqrt{g}\,e^{-2\phi}\Big[
	\frac{1}{2\kappa}\big(R - 2\Lambda + \alpha R_{\rm GB}^2 + 4(\partial\phi)^2\big)
	- (\partial\chi)^2
	- \frac{1}{4\pi^2\alpha'^2}
	\big(\beta^2 g_{\tau\tau} - \beta_{\rm H}^2\big)\chi^2
	\Big],
	\label{eq:action}
\end{equation}
    For later convenience we define inverse length variables
\begin{equation}
    \tilde{\beta}\equiv\frac{\beta}{2\pi\alpha'},
    \qquad
    \tilde{\beta}_{\rm H}\equiv\frac{\beta_{\rm H}}{2\pi\alpha'},
    \qquad
    m_{\rm th}^2(r)=\tilde{\beta}^2g_{\tau\tau}(r)-\tilde{\beta}_{\rm H}^2.
    \label{eq:beta-bar-units}
\end{equation}
    This restores the required factor $1/\alpha'^2$ in every mass term.  The displayed action is a low-amplitude, long-wavelength effective theory.  Its simultaneous control conditions are
\begin{equation}
    \alpha'|m_{\rm eff}^2|\ll1,\qquad
    \alpha'{\cal R}\ll1,\qquad
    \alpha'(\nabla\phi)^2\ll1,\qquad
    \kappa|\chi|^2\ll1,
    \label{eq:EFT-validity}
\end{equation}
    together with slowly varying fields and weak string coupling.  The first condition makes the winding mode light, the next two control the target space derivative expansion, and the last justifies the quadratic treatment of the condensate.

    In a general curved background the thermal scalar operator can contain both mass-type and derivative curvature corrections~\cite{Mertens:2013zya,Mertens:2014kba,Mertens:2014nca,Harmark:2024qrf}.  Representative structures are
\begin{align}
    m_{\rm th}^2
    &\ \longrightarrow\
    m_{\rm th}^2+c_RR
    +\frac{c_\tau}{\alpha'}R_{\mu\nu}\tau^\mu\tau^\nu+\cdots,
    \nonumber\\[-2pt]
    -g^{\mu\nu}\nabla_\mu\nabla_\nu
    &\ \longrightarrow\
    -\bigl(g^{\mu\nu}+c_K\alpha'R^{\mu\nu}+\cdots\bigr)
     \nabla_\mu\nabla_\nu,
    \label{eq:thermal-curvature-coupling}
\end{align}
    where $\tau^\mu$ is the thermal winding vector, of length ${\cal O}(\beta)$, and the coefficients depend on the string theory, background, and field-redefinition scheme.  The explicit calculation of Ref.~\cite{Harmark:2024qrf} fixes Ricci tensor structures for the backgrounds considered there; it does not establish a universal, isolated $\xi_RR$ term.  Near the Hagedorn point $\tau^2\sim\beta_H^2\sim\alpha'$, so the displayed corrections are of relative order $\alpha'{\cal R}$.  At the accuracy claimed here we retain the zeroth order of this thermal scalar operator expansion.  This is an explicit additional truncation, not a universal large-$D$ suppression.  These operators must be included when computing the first curvature correction to the thermal scalar spectrum or Hagedorn temperature.

    The thermal scalar EFT is applied only outside the string scale core. At the classical Euclidean horizon, $g_{\tau\tau}(r_h)=0$, the quantity $\alpha'm_{\rm eff}^2$ is generically of order unity. We therefore introduce a stretched horizon matching surface $r=r_{\rm c}$, with
\begin{equation}
    s(r_{\rm c}) \gtrsim \ell_s,
    \qquad
    \ell_s = \sqrt{\alpha'},
    \qquad
    \chi_{\rm c} \equiv \chi(r_{\rm c}),
    \label{eq:stretched-horizon}
\end{equation}
    where $s$ is the proper distance from the classical tip. The matching surface must be chosen so that the EFT conditions~\eqref{eq:EFT-validity} hold in the region described by the exterior calculation. The proper distance estimate alone does not ensure EFT validity. The amplitude $\chi_{\rm c}$ and the core contribution to the action are external matching data supplied by the unresolved region. The homogeneous quadratic thermal scalar equation does not determine $\chi_{\rm c}$, and its exterior solution does not provide a prediction for $\chi(r_h)$. Resolving the core requires additional microscopic input.

    In the following sections, we derive the large-$D$ exterior equations from~\eqref{eq:action} and analyse the near- and far-zone profiles at fixed matching data. The resulting matching relations do not determine the complete free energy or the physical string/black hole correspondence point.

\subsection{Equations of Motion} \label{sec:EOM}
    We first display the exact covariant equations obtained by varying the Euclidean EGB--dilaton--thermal scalar action~\eqref{eq:action}.  The large-$D$ truncation used in the analytic calculation is imposed only afterwards.  This distinction is important because the string frame prefactor $e^{-2\phi}$ also varies through the Gauss--Bonnet term.

\subsubsection{Metric Equation of Motion} 
    Varying~\eqref{eq:action} with respect to the inverse metric gives
\begin{equation}
	G_{\mu\nu}+\Lambda g_{\mu\nu}
	+2\nabla_\mu\nabla_\nu\phi
	-2g_{\mu\nu}\nabla^2\phi
	+2g_{\mu\nu}(\nabla\phi)^2
	+\alpha\bigl(H_{\mu\nu}+\Delta^{(\phi {\rm GB})}_{\mu\nu}\bigr)
	=2\kappa\,\Theta^{(\chi)}_{\mu\nu},
	\label{eq:EinsteinEq}
\end{equation}
    where the common factor $e^{-2\phi}$ has been divided out and
\begin{equation}
	\Theta^{(\chi)}_{\mu\nu}
	=
	\partial_\mu\chi\,\partial_\nu\chi
	-\frac12 g_{\mu\nu}\bigl[(\partial\chi)^2+m_{\rm th}^2\chi^2\bigr]
	-\tilde{\beta}^2\chi^2 g_{\tau\mu}g_{\tau\nu},
	\label{eq:thermal-stress-exact}
\end{equation}
    with
\begin{equation}
	m_{\rm th}^2 
	=\tilde{\beta}^2g_{\tau\tau}-\tilde{\beta}_{\rm H}^2.
\end{equation}
	The last term in~\eqref{eq:thermal-stress-exact} follows from varying the explicit $g_{\tau\tau}$ in the winding mass; writing it with $g_{\tau\mu}g_{\tau\nu}$ keeps the expression tensorial.

    The Gauss--Bonnet contribution $H_{\mu\nu}$ is
\begin{equation}
	H_{\mu\nu}
	=
	2 R R_{\mu\nu}
	- 4 R_{\mu\rho}R^\rho{}_\nu
	- 4 R_{\mu\rho\nu\sigma} R^{\rho\sigma}
	+ 2 R_{\mu}{}^{\rho\sigma\lambda} R_{\nu\rho\sigma\lambda}
	-\frac12 g_{\mu\nu}R_{\rm GB}^2 .
	\label{eq:Hmunu}
\end{equation}

    The dilaton-dependent Gauss--Bonnet prefactor contributes the additional tensor~\mbox{\cite{Guo:2008OhtaTorii,Maeda:2010Sasagawa}}
\begin{align}
	\Delta^{(\phi {\rm GB})}_{\mu\nu}
	&\equiv
	4e^{2\phi}P_{\mu\rho\nu\sigma}
	\nabla^\rho\nabla^\sigma e^{-2\phi}
	\nonumber\\
	&=
	16P_{\mu\rho\nu\sigma}\nabla^\rho\phi\nabla^\sigma\phi
	-8P_{\mu\rho\nu\sigma}\nabla^\rho\nabla^\sigma\phi,
	\label{eq:phiGB-extra}
\end{align}
    where
\begin{equation}
	P_{\mu\nu\rho\sigma}
	=R_{\mu\nu\rho\sigma}
	+2g_{\mu[\sigma}R_{\rho]\nu}
	+2g_{\nu[\rho}R_{\sigma]\mu}
	+R g_{\mu[\rho}g_{\sigma]\nu},
	\qquad
	\nabla^\mu P_{\mu\nu\rho\sigma}=0.
	\label{eq:P-div-free}
\end{equation}

	Equation~\eqref{eq:EinsteinEq}, rather than the nondilatonic EGB equation by itself, is the exact metric Euler--Lagrange equation of~\eqref{eq:action}.  Nevertheless, the extra tensor~\eqref{eq:phiGB-extra} lies beyond the order retained in our analytic large-$D$ metric calculation.  Indeed, in the exterior near zone $\partial_r\sim D/r_h$, while Eqs.~\eqref{eq:largeD-fixed-data} and~\eqref{eq:phi-expansion-LO} keep $\widetilde\alpha\sim D^2\alpha$ fixed and assign $\phi-\phi_0=D^{-2}\phi_1$.  Consequently,
\begin{equation}
	P_{\mu\rho\nu\sigma}=\mathcal O(D^2/r_h^2),
	\qquad
	H_{\mu\nu}=\mathcal O(D^4/r_h^4),
	\qquad
	\nabla\nabla\phi=\mathcal O(r_h^{-2}),
	\qquad
	(\nabla\phi)^2=\mathcal O(D^{-2}r_h^{-2}),
\end{equation}
    and hence
\begin{equation}
	\frac{\alpha\Delta^{(\phi {\rm GB})}_{\mu\nu}}
	     {\alpha H_{\mu\nu}}
	=\mathcal O(D^{-2}).
	\label{eq:phiGB-suppression}
\end{equation}
	It vanishes identically on the LO constant-dilaton background and first contributes at relative order $D^{-2}$ (NNLO) to the metric equation.  Thus it may be omitted from the LO and relative $1/D$ analytic metric equations, but it must be restored in an NNLO calculation or in an exact finite-$D$ treatment.

\subsubsection{Dilaton Equation of Motion}
    Variation with respect to $\phi$ gives
\begin{equation}
	0
	=
	\Big[
	4 \nabla^2\phi 
	- 4 (\partial\phi)^2
	+ R - 2\Lambda + \alpha R_{\rm GB}^2
	\Big]
	-2\kappa\bigl[(\partial\chi)^2+m_{\rm th}^2\chi^2\bigr].
	\label{eq:DilatonEOM}
\end{equation}
    This is the string frame generalisation of the usual dilaton beta function equation.  Every term couples universally because of the global prefactor $e^{-2\phi}$ in the action.

\subsubsection{Thermal Scalar Equation of Motion}
    Varying with respect to $\chi$ yields
\begin{equation}
	\nabla_\mu\!\left(e^{-2\phi}\nabla^\mu\chi\right)
	- e^{-2\phi}\,m_{\rm th}^2\,\chi
	= 0.
	\label{eq:ChiEOM}
\end{equation}

    Expanding the prefactor gives the explicitly dilaton modified Schr\"odinger equation,
\begin{equation}
	\nabla^2\chi - 2\,\partial_\mu\phi\,\nabla^\mu\chi
	- m_{\rm th}^2\,\chi = 0.
	\label{eq:ChiSchrodinger}
\end{equation}
    This is the equation that reduces, in the large-$D$ near-zone, to the universal Schr\"odinger problem governing the HP instability.

    The full set of equations is:
\begin{equation}
\boxed{
\begin{aligned}
	&G_{\mu\nu} + \Lambda g_{\mu\nu}
	+ 2\nabla_\mu\nabla_\nu\phi - 2 g_{\mu\nu}\nabla^2\phi
	+2g_{\mu\nu}(\nabla\phi)^2
	+\alpha\bigl(H_{\mu\nu}+\Delta^{(\phi {\rm GB})}_{\mu\nu}\bigr)
	= 2\kappa\, \Theta^{(\chi)}_{\mu\nu}, \\[4pt]
	&4 \nabla^2\phi - 4 (\partial\phi)^2
	+ R - 2\Lambda + \alpha R_{\rm GB}^2
	= 2\kappa\,\big[(\partial\chi)^2 + m_{\rm th}^2\chi^2\big], \\[4pt]
	&\nabla^2\chi - 2\,\partial_\mu\phi\,\nabla^\mu\chi
	- m_{\rm th}^2\,\chi = 0.
\end{aligned}
}
    \label{eq:mseqn}
\end{equation}
    These exact equations are the starting point.  In the analytic metric sector below we then apply the explicit power counting~\eqref{eq:phiGB-suppression}; no claim is made that the omitted NNLO tensor vanishes at finite $D$.

\section{Solutions in large number of dimensions} \label{sec:solld}
    We solve the equations of Sec.~\ref{sec:EOM} in a large-$D$ expansion.  To remove any ambiguity in the notation, throughout the analytic calculation we use the areal gauge ansatz
\begin{equation}
    ds^2=F(r)d\tau^2+\frac{dr^2}{F(r)}+r^2d\Omega_{D-2,k}^2,
    \qquad \phi=\phi(r),\qquad\chi=\chi(r),
    \label{eq:section3-metric}
\end{equation}
    where $F(r)=g_{\tau\tau}(r)$ is the blackening function, $F(r_h)=0$, and $k=1,0,-1$ labels spherical, planar, and hyperbolic horizons.  EGB gravity is dynamical for $D\ge5$; the additional assumption $n\equiv D-3\gg1$ is used only for the large-$D$ expansion.

\subsection{Large-\texorpdfstring{$D$}{D} regions, scaling, and matching}
    We use one dimensionless near-zone coordinate everywhere in this section,
\begin{equation}
    \rho\equiv n\log\!\left(\frac r{r_h}\right),
    \qquad r=r_he^{\rho/n},
    \qquad \partial_r=\frac nr\partial_\rho .
    \label{eq:regions-rho}
\end{equation}
    The radial domains are
\begin{align}
    &\text{near zone:} && r-r_h\ll r_h,\quad \rho\ll n,
    \qquad
    r-r_h=\mathcal O(r_h/n),\ \rho=\mathcal O(1)
    \ \text{in the scaling layer},\nonumber\\
    &\text{far zone:} && r-r_h\gg r_h/n,\quad \rho\gg1,\nonumber\\
    &\text{overlap:} && \frac{r_h}{n}\ll r-r_h\ll r_h,
    \quad 1\ll\rho\ll n.
    \label{eq:largeD-regions}
\end{align}
    Matching means expanding the near-zone solution at large $\rho$ and the far-zone solution toward the horizon inside the last window.  For the vacuum geometry this procedure is not needed to construct the solution: the exact finite-$D$ Boulware--Deser metric is already known and is used as the far-zone background.  Vacuum matching is therefore a consistency check that relates the near-zone integration constants to the parameters of that exact metric.  It becomes genuinely nontrivial for the dilaton, thermal scalar, and their $1/D$ backreaction.  Because the quadratic thermal scalar equation is homogeneous, matching selects the normalisable branch and relates near- and far-zone coefficients, but it does not fix the overall condensate normalization.

    The large-$D$ limit is taken at fixed
\begin{equation}
    \widetilde\alpha=(D-3)(D-4)\alpha,
    \qquad \ell,
    \qquad \frac{\widetilde\alpha}{\ell^2},
    \label{eq:largeD-fixed-data}
\end{equation}
    so that the bare parameters scale as
\begin{equation}
    \alpha=\frac{\widetilde\alpha}{(D-3)(D-4)},
    \qquad
    \Lambda=-\frac{(D-1)(D-2)}{2\ell^2}.
    \label{eq:largeD-bare-scaling}
\end{equation}
    Only the variables used in the calculation are retained: at fixed $\rho$ the near-zone fields are expanded directly as functions of $\rho$, while the corresponding far-zone fields are expanded at fixed $r$.

\subsection{Leading-order fields in the large-\texorpdfstring{$D$}{D} expansion} \label{subsec:LO-cigar-all}
	In this subsection we collect the leading-order large-$D$ data for the Euclidean Einstein--Gauss--Bonnet (EGB) black hole, the dilaton, and the complex thermal scalar that propagates on the resulting cigar geometry. Throughout we work in the static ansatz~\eqref{eq:section3-metric}, take $D = n+3$ with $n \gg 1$, and employ the standard near- and far-zone coordinates adapted to the large-$D$ limit.

\subsubsection{Geometry: Boulware--Deser background and large-\texorpdfstring{$D$}{D} cigar} \label{sec:LO-geometry}
\paragraph{Boulware--Deser background at large-\texorpdfstring{$D$}{D}}
	The vacuum EGB$+\Lambda$ equation with the static ansatz
\begin{equation}
	ds^2
	=
	F(r)\,d\tau^2
	+ \frac{dr^2}{F(r)}
	+ r^2 d\Omega^2_{D-2,k},
	\qquad
	k = 1,0,-1,
	\label{eq:metric-ansatz-LO}
\end{equation}
    admits the Boulware--Deser solution
\begin{equation}
	F_0(r)
	=
	k + \frac{r^2}{2\tilde{\alpha}}
	\left[
	1 - \sqrt{1+4\tilde{\alpha}\left(\frac{\mu}{r^{D-1}}
	{-\frac{1}{\ell^2}}\right)}
	\right],
	\qquad
	\tilde{\alpha}
	\equiv (D-4)(D-3)\alpha,
	\label{eq:BD-LO}
\end{equation}
	where $\mu$ is an integration constant proportional to the mass and 	$\ell$ is the AdS radius. The horizon radius $r_h$ is defined by
\begin{equation}
	F(r_h) = 0.
	\label{eq:horizon-def}
\end{equation}
	At leading order in $1/D$ the geometry of interest is precisely~\eqref{eq:BD-LO}; all matter fields enter only at subleading orders.

\paragraph{Near-zone coordinate and cigar profile}
	We now zoom into the near-horizon region by introducing the standard large-$D$ near-zone coordinate
\begin{equation}
	D = n+3,\qquad n\gg 1,
	\qquad
	\rho \equiv n\ln\frac{r}{r_h},
	\qquad
	r = r_h e^{\rho/n},
	\qquad
	\partial_r = \frac{n}{r}\,\partial_\rho.
	\label{eq:rho-def-geom}
\end{equation}
	The horizon is at $\rho=0$, and the near-zone corresponds to 	$r-r_h = \mathcal{O}(r_h/n)$ at fixed $\rho=\mathcal{O}(1)$. Near the horizon we expand the BD blackening function as
\begin{equation}
	F(r)
	=
	F(r_h) + F'(r_h)(r-r_h) + \mathcal{O}\bigl((r-r_h)^2\bigr)
	=
	F'(r_h)(r-r_h) + \cdots,
	\label{eq:F-near-horizon}
\end{equation}
	using $F(r_h)=0$ from~\eqref{eq:horizon-def}. Inserting 	$r-r_h \simeq (r_h/n)\,\rho$ at fixed $\rho$ gives
\begin{equation}
	F(r) \simeq \frac{r_h}{n}F'(r_h)\,\rho,
\end{equation}
	so the leading near-zone behaviour is linear in $\rho$. It is convenient to define the dimensionless \emph{cigar scale}
\begin{equation}
	K \equiv \frac{r_h}{n}\,F'(r_h)\Big|_{\text{BD}} > 0,
	\label{eq:K-def-geom}
\end{equation}
	For reference, evaluating the derivative of the exact solution gives
\begin{equation}
	K=\frac{1}{n}\left[-2k+
	    \frac{n+2}{1+2\widetilde\alpha k/r_h^2}
	    \left(k+\frac{\widetilde\alpha k^2}{r_h^2}+\frac{r_h^2}{\ell^2}\right)\right],
    \qquad n=D-3.
    \label{eq:K-explicit}
\end{equation}
	This encodes all dependence on $(\tilde{\alpha},\ell,k,\mu)$ through
	the BD derivative $F'(r_h)$. Solving the LO radial Einstein--Gauss--Bonnet
	equation in the scaling limit~\eqref{eq:rho-def-geom} with this
	near-horizon behaviour yields a universal first-order ODE for $F_0(\rho)$,
	whose solution is
\begin{equation}
	F(r)
	=
	F_0(\rho) + \mathcal{O}\!\left(\frac{1}{n}\right),
	\qquad
	F_0(\rho) = K\bigl(1-e^{-\rho}\bigr).
	\label{eq:F0-NZ-final}
\end{equation}
	By construction,
\begin{equation}
	F_0(0) = 0,
	\qquad
	F_0(\rho)\sim K\rho \quad (\rho \to 0),
	\qquad
	F_0(\rho)\to K \quad (\rho \to \infty),
\end{equation}
	so the Euclidean time circle closes smoothly at the tip ($\rho=0$) and opens into a plateau of constant $F_0=K$ deeper in the near-zone. The LO geometry in this region is therefore a standard ``large-$D$ cigar'' with scale set by $K$.

\paragraph{Far-zone expansion and overlap check for the known geometry}
    The far-zone metric is not a second, unspecified solution: it is the exact finite-$D$ Boulware--Deser metric~\eqref{eq:BD-LO}.  For comparison with the near-zone expansion, it is convenient to use
\begin{equation}
	R \equiv \left(\frac{r}{r_h}\right)^{D-2}
	= \left(\frac{r}{r_h}\right)^{n+1},
	\qquad
	r = r_h R^{1/(D-2)}.
	\label{eq:R-def-geom}
\end{equation}
    The exact metric approaches its asymptotic flat or AdS form for $r-r_h=\mathcal O(r_h)$.  To localize it to the overlap region, we instead expand toward the horizon with
\begin{equation}
    r=r_h e^{\rho/n}
    =r_h\left[1+\frac{\rho}{n}
    +\mathcal O\!\left(\frac{\rho^2}{n^2}\right)\right],
    \qquad 1\ll\rho\ll n.
    \label{eq:far-toward-overlap}
\end{equation}
    Equivalently, $1\ll R\ll e^n$ and $r_h/n\ll r-r_h\ll r_h$.  In this common domain, the exact far-zone metric and the large-$\rho$ near-zone metric both reduce at leading order to
\begin{equation}
	F(r) \simeq K,
	\label{eq:F-plateau}
\end{equation}
    up to subleading corrections.  The coordinate relation in the overlap is
\begin{equation}
	\rho = n\ln\frac{r}{r_h},
	\qquad
	\ln R = (D-2)\ln\frac{r}{r_h} = \frac{n+1}{n}\,\rho
	= \rho + \mathcal{O}\!\left(\frac{\rho}{n}\right),
\end{equation}
    Thus the comparison only verifies the relation between $K$ and the global Boulware--Deser parameters.  Its practical purpose is to prepare the coupled matter problem, for which the normalisable and non-normalisable coefficients must genuinely be matched.

\paragraph*{Boxed leading-order metric}
	For later reference we summarise the LO geometry in compact form:
\begin{equation}
    \boxed{
		\begin{aligned}
			& \text{\bf Background:} &&
			ds^2
			=
			F(r)\,d\tau^2
			+ \frac{dr^2}{F(r)}
			+ r^2 d\Omega^2_{D-2,k},
            \\[0.4em]
			&&&
			F(r)
			=
			k + \frac{r^2}{2\tilde{\alpha}}
			\left[
			1 - \sqrt{1+4\tilde{\alpha}\left(\frac{\mu}{r^{D-1}}
			{-\frac{1}{\ell^2}}\right)}
			\right],
            \\[0.4em]
			&&&
            \tilde{\alpha}
			= (D-4)(D-3)\alpha,
            \\[0.6em]
			& \text{\bf Near-zone:} &&
			\rho = n\ln\frac{r}{r_h},\qquad D=n+3,\qquad
			F(r) = F_0(\rho) + \mathcal{O}\!\left(\frac{1}{n}\right),
            \\[0.2em]
			&&&
            F_0(\rho) = K\bigl(1-e^{-\rho}\bigr),\qquad
			K \equiv \frac{r_h}{n}\,F'(r_h)\Big|_{\text{BD}} > 0,
            \\
			&&&
			K\ \text{is given explicitly in Eq.~\eqref{eq:K-explicit}}
            \\
            & \text{\bf {Matching far-zone (LO)}:} &&
            \\
			&&&R = \left(\dfrac{r}{r_h}\right)^{D-2},\qquad
			F_0(r) \simeq K
			\quad\text{for} 
			1\ll \rho \ll n,\;\; 1\ll R \ll e^n.
        \end{aligned}
    }	
	\label{eq:LO-metric-box}
\end{equation}
	This LO geometric background will be used below as the fixed large-$D$ cigar on which the dilaton and thermal scalar propagate.

\subsubsection{Dilaton: leading-order constant profile} \label{sec:LO-dilaton}
\paragraph{Dilaton equation and large-\texorpdfstring{$D$}{D} scaling}
	The dilaton equation following from the HP+GB action~\eqref{eq:action} can be written as
\begin{equation}
	R - 2\Lambda + \alpha R_{\rm GB}^2
	+ 4\Box\phi - 4(\partial\phi)^2
	- 2\kappa\Bigl(|\partial\chi|^2
	+ (\tilde{\beta}^2 g_{\tau\tau}-\tilde{\beta}_{\rm H}^2)|\chi|^2\Bigr)
	= 0.
	\label{eq:dilaton-eq-full}
\end{equation}
	We now separate three ingredients:
\begin{itemize}
	\item the purely geometric combination
\begin{equation}
		\mathcal{G}_{\rm HP}
		\equiv R - 2\Lambda + \alpha R_{\rm GB}^2,
\end{equation}
    which is $\mathcal{O}(D^2)$ on a black hole background;
	
    \item the dilaton kinetic piece $4\Box\phi - 4(\partial\phi)^2$;
	
    \item the matter source
\begin{equation}
	\mathcal{S}_\chi
	\equiv 2\kappa\Bigl(|\partial\chi|^2
	+ (\tilde{\beta}^2 g_{\tau\tau}-\tilde{\beta}_{\rm H}^2)|\chi|^2\Bigr).
\end{equation}
\end{itemize}
	We assume that the \emph{background} geometry is a vacuum solution with thermal scalar switched off, $\chi=0$. In that case the dilaton equation reduces to
\begin{equation}
	\mathcal{G}_{\rm HP}
	+ 4\Box\phi_0 - 4(\partial\phi_0)^2 = 0.
	\label{eq:dilaton-background}
\end{equation}
	We take the Boulware--Deser solution~\eqref{eq:BD-LO} on the branch	such that $\mathcal{G}_{\rm HP}=0$ with a \emph{constant} background dilaton $\phi_0=\phi_\infty$. This is the standard HP construction: the EGB black hole with a constant string coupling $g_s = e^{\phi_\infty}$ solves the tree-level string equations with the thermal scalar turned off.
	
	When the thermal scalar is turned on, we work in a probe approximation for $\chi$:
\begin{equation}
	\chi = \mathcal{O}(1),
	\qquad
	\mathcal{S}_\chi = \mathcal{O}(1),
	\qquad
	\mathcal{G}_{\rm HP} = \mathcal{O}(D^2).
\end{equation}
    In the large-$D$ counting, $\mathcal{G}_{\rm HP}$ is parametrically larger than $\mathcal{S}_\chi$ by a factor of $D^2$. Thus,
\begin{itemize}
	\item the Einstein equation is dominated by $\mathcal{G}_{\rm HP}$ and is solved by the BD metric at LO (as used in Sec.~\ref{sec:LO-geometry});
    
	\item the dilaton backreaction sourced by $\chi$ can be consistently postponed to subleading orders in $1/D$.
\end{itemize}
	Concretely, we expand
\begin{equation}
	\phi(x)
	=
	\phi_0 + \frac{1}{D^2}\phi_1(x) + \mathcal O(D^{-3}),
	\qquad
	\phi_0 = \phi_\infty,\qquad
	\phi_1 = \mathcal{O}(1),
	\label{eq:phi-expansion-LO}
\end{equation}
	This makes the $D^{-2}$ suppression explicit.  In the NLO equations below, $\phi_1$ denotes the coefficient after this overall factor has been extracted. We keep only $\phi_0$ at leading order. The non-trivial profile $\phi_1$ is then determined at next-to-leading order by~\eqref{eq:dilaton-eq-full} on the fixed background and constitutes a small backreaction effect which we will not need for the LO analysis of the thermal scalar. At strict LO we therefore set
\begin{equation}
	\phi(x) = \phi_0 = \phi_\infty = \text{const},
	\label{eq:phi0-constant-ansatz}
\end{equation}
	and verify that this is consistent with regularity and normalisability in both near- and far-zones.

\paragraph{Near-zone constant solution}
    We now work in the near-zone coordinate defined in~\eqref{eq:rho-def-geom},
\begin{equation}
	D = n+3,\qquad
	\rho = n\ln\frac{r}{r_h},
	\qquad
	r = r_h e^{\rho/n},
\end{equation}
	and assume an $s$-wave dilaton $\phi_0=\phi_0(\rho)$. For any scalar $X=X(\rho)$, the near-zone Laplacian on the LO cigar background is (see App.~\ref{app:NLO-details})
\begin{equation}
	\Box X(\rho)
	\simeq \frac{n^2}{r_h^2}\,e^{-\rho}
	\frac{d}{d\rho}\Bigl(e^{\rho}F_0(\rho)X'(\rho)\Bigr),
	\qquad
	X'(\rho)\equiv\frac{dX}{d\rho},
	\label{eq:Box-NZ-dilaton}
\end{equation}
	with
\begin{equation}
	F_0(\rho) = K\bigl(1-e^{-\rho}\bigr),
\end{equation}
	as in~\eqref{eq:F0-NZ-final}. Since $\phi_0$ is taken to be LO and the thermal scalar backreaction is subleading, the LO dilaton equation is simply
\begin{equation}
	\Box\phi_0 = 0,
	\label{eq:phi0-NZ-eq}
\end{equation}
	on the fixed cigar background. Using~\eqref{eq:Box-NZ-dilaton}, we obtain
\begin{equation}
	e^{-\rho}
	\frac{d}{d\rho}\Bigl(e^{\rho}F_0(\rho)\phi_0'(\rho)\Bigr) = 0,
\end{equation}
	which integrates to
\begin{equation}
	e^{\rho}F_0(\rho)\phi_0'(\rho) = C_\phi,
	\label{eq:phi0-NZ-int}
\end{equation}
	with $C_\phi$ a constant of integration.
	
	Near the tip, $\rho\to 0$, we have
\begin{equation}
	F_0(\rho) \sim K\rho,
	\qquad
	e^{\rho}\sim 1+\rho,
\end{equation}
	so
\begin{equation}
	e^{\rho}F_0(\rho)\phi_0'(\rho)
	\sim K\rho\,\phi_0'(\rho).
\end{equation}
	Regularity at the tip requires $\phi_0'(\rho)$ to remain finite as $\rho\to 0$, which implies
\begin{equation}
	C_\phi = 0
	\quad\Longrightarrow\quad
	\phi_0'(\rho) = 0
	\quad\Longrightarrow\quad
	\phi_0(\rho) = \phi_\infty.
\end{equation}
	Thus, the only regular near-zone LO dilaton profile is a constant, consistent with our starting assumption~\eqref{eq:phi0-constant-ansatz}.

\paragraph{Far-zone behaviour and matching}
	We now confirm that the same constant solution is the unique
	normalisable LO profile in the far-zone. Using the far-zone variable
	defined in~\eqref{eq:R-def-geom},
	\begin{equation}
		R \equiv \left(\frac{r}{r_h}\right)^{D-2},
		\qquad
		r = r_h R^{1/(D-2)},
	\end{equation}
	and assuming again an $s$-wave $\phi_0=\phi_0(R)$, the radial Laplacian
	takes the form (App.~\ref{app:NLO-details})
	\begin{equation}
		\Box X(R)
		\simeq F(r)\,\frac{(D-2)^2}{r_h^2}
		\frac{d}{dR}\bigl(R^2 X'(R)\bigr),
		\qquad
		X'(R)\equiv\frac{dX}{dR}.
	\end{equation}
	In the matching far-zone the blackening function is already on its
	plateau, $F(r)\simeq K$, so the LO dilaton equation
	$\Box\phi_0=0$ becomes
	\begin{equation}
		\frac{d}{dR}\bigl(R^2\phi_0'(R)\bigr) = 0.
		\label{eq:phi0-FZ-eq}
	\end{equation}
	Integrating once gives
	\begin{equation}
		R^2\phi_0'(R) = A_\phi,
	\end{equation}
	with $A_\phi$ a constant. Regularity at the horizon ($R\to 1$) allows
	for a finite derivative there, but normalisability at infinity requires
	\begin{equation}
		\phi_0'(R) \xrightarrow{R\to\infty} 0
		\quad\Rightarrow\quad
		A_\phi = 0.
	\end{equation}
	Hence
	\begin{equation}
		\phi_0'(R) = 0
		\quad\Rightarrow\quad
		\phi_0(R) = \phi_\infty,
	\end{equation}
	in agreement with the near-zone result.
	
	In the overlap region $1 \ll \rho \ll n$, $1 \ll R \ll e^n$ we have $\ln R \simeq \rho$ (\textit{cf.}, Sec.~\ref{sec:LO-geometry}), so the near-zone constant solution $\phi_0(\rho)=\phi_\infty$ and the far-zone constant solution $\phi_0(R)=\phi_\infty$ trivially match.

\paragraph*{Boxed leading-order dilaton}
    We summarise the LO dilaton result in a compact form:
\begin{equation}
	\boxed{
		\begin{aligned}
			&\text{\bf Background:} &&
			\text{EGB Boulware--Deser black hole~\eqref{eq:BD-LO} on the branch with}~~ \mathcal{G}_{\rm HP}=0,\\[0.4em]
			&\text{\bf Dilaton expansion:} &&
			\phi(x) = \phi_0 + D^{-2}\phi_1(x) + \mathcal O(D^{-3}),
			\qquad
			\phi_0 = \phi_\infty = \text{const},
			\qquad
			\phi_1 = \mathcal{O}(1),\\[0.4em]
			&\text{\bf Near-zone (LO):} &&
			\Box\phi_0 = 0
			\quad\Longrightarrow\quad
			\phi_0(\rho) = \phi_\infty
			\quad\text{(regular at the tip)},\\[0.4em]
			&\text{\bf Far-zone (LO):} &&
			\Box\phi_0 = 0
			\quad\Longrightarrow\quad
			\phi_0(R) = \phi_\infty
			\quad\text{(normalisable at infinity)},\\[0.4em]
			&\text{\bf Global LO solution:} &&
			\phi_0(x) = \phi_\infty
			\quad\text{throughout the cigar}.
		\end{aligned}
	}
	\label{eq:LO-dilaton-box}
\end{equation}
	
	At next-to-leading order in the combined $1/D$ and probe expansion, the
	thermal scalar sources a small but non-trivial dilaton profile
	$\phi_1(x)$ through~\eqref{eq:dilaton-eq-full}, which can be solved on
	the fixed LO cigar background. This backreaction, however, is genuinely
	subleading and does not modify the LO constant-dilaton result used in
	our analysis of the thermal scalar at leading order.
	
\subsubsection{Thermal scalar: leading-order mode at large-\texorpdfstring{$D$}{D}}  \label{sec:LO-thermal-scalar}
\paragraph{Equation of motion and scaling}
	We now analyse the leading-order behaviour of the complex thermal scalar $\chi$ on the large-$D$ EGB cigar, with the metric and dilaton fixed at LO by the BD solution with a constant dilaton,
\begin{equation}
	ds^2
	=
	F(r)\,d\tau^2
	+ \frac{dr^2}{F(r)}
	+ r^2 d\Omega_{D-2,k}^2,
	\qquad
	F(r) = F_{\rm BD}(r),
	\qquad
	\phi_0 = \phi_\infty.
\end{equation}
	We work in the same large-$D$ limit $D=n+3$ and treat $\chi$ in the	probe approximation: its backreaction on the geometry and on the dilaton is parametrically subleading in $1/D$. The thermal scalar equation from the HP+GB action can be written as
\begin{equation}
	0
	=
	\Box\chi - 2\,\partial^\mu\chi\,\partial_\mu\phi
	- \bigl(\tilde{\beta}^2 g_{\tau\tau}-\tilde{\beta}_{\rm H}^2\bigr)\chi,
	\label{eq:chi-eq-full}
\end{equation}
	Here $\tilde{\beta}$ and $\tilde{\beta}_{\rm H}$ are the inverse-length variables defined in~\eqref{eq:beta-bar-units}; hence every term in~\eqref{eq:chi-eq-full} has the required mass dimension. We use $g_{\tau\tau}=F(r)$.
	
	At LO we set
\begin{equation}
	\phi(x) = \phi_0 = \phi_\infty = \text{const},
\end{equation}
	so that $\partial_\mu\phi_0=0$ and the mixed term $-2\,\partial^\mu\chi\,\partial_\mu\phi$ drops out. The LO thermal scalar equation thus reduces to
\begin{equation}
	\Box\chi_0 - \bigl(\tilde{\beta}^2 F(r)-\tilde{\beta}_{\rm H}^2\bigr)\chi_0 = 0,
	\label{eq:chi-eq-LO}
\end{equation}
	on the fixed BD background. We restrict to $s$-wave configurations, $\chi_0=\chi_0(r)$, which dominate in the HP limit.
	
	The two competing scales in~\eqref{eq:chi-eq-LO} are:
\begin{itemize}
	\item the radial kinetic term, which scales as $\Box\chi_0\sim D^2$ in the near-horizon region;
    
	\item the thermal mass term $m_{\rm eff}^2(r)\chi_0$, with
\begin{equation}
	m_{\rm eff}^2(r)
	\equiv \tilde{\beta}^2 F(r)-\tilde{\beta}_{\rm H}^2.
\end{equation}
\end{itemize}
	To obtain a non-trivial LO interplay between these terms, we adopt the standard Horowitz--Polchinski scaling: we work near the Hagedorn point and keep the following dimensionless combination fixed as $D\to\infty$,
\begin{equation}
	\Xi^2
	\equiv \frac{r_h^2}{(D-2)^2 K}\bigl(\tilde{\beta}^2 K - \tilde{\beta}_{\rm H}^2\bigr),
	\qquad
	\Xi^2 = \mathcal{O}(1),
	\label{eq:Xi-def-TS}
\end{equation}
	where $K$ is the near-horizon plateau value of the blackening function (see below). This scaling ensures that the kinetic and mass terms in~\eqref{eq:chi-eq-LO} are of the same order in the large-$D$ expansion.

	The following solution is an exterior EFT solution.  Although the coordinate horizon is at $\rho=0$, the effective equations are used only for $\rho\ge\rho_{\rm c}>0$, where $r_{\rm c}$ obeys the stretched horizon condition~\eqref{eq:stretched-horizon}.  The interval $0\le\rho<\rho_{\rm c}$ is an unresolved string-scale core.

\paragraph{Near-zone solution}
	We first analyse the thermal scalar in the near-zone, defined by $r-r_h=\mathcal{O}(r_h/D)$ at fixed $D\gg 1$. As in the geometry and dilaton analysis, we introduce
\begin{equation}
	D = n+3,
	\qquad
	\rho \equiv n\ln\frac{r}{r_h},
	\qquad
	r = r_h e^{\rho/n},
	\qquad
	\rho=0\ \Leftrightarrow\ r=r_h.
	\label{eq:rho-def-TS}
\end{equation}
	In this regime the BD blackening function takes the universal LO form
\begin{equation}
	F(r) = F_0(\rho) + \mathcal{O}\!\left(\frac{1}{n}\right),
	\qquad
	F_0(\rho) = K\bigl(1-e^{-\rho}\bigr),
	\label{eq:F0-NZ-TS}
\end{equation}
	with $K$ an effective curvature scale (constant in the near-zone) determined by the Gauss--Bonnet coupling, the cosmological constant and the horizon topology. For an $s$-wave scalar $X=X(r)$, in the metric
\begin{equation}
	ds^2 = F(r)\,d\tau^2 + \frac{dr^2}{F(r)} + r^2 d\Omega_{D-2,k}^2,
\end{equation}
	the Laplacian is
\begin{equation}
	\Box X
	= \frac{1}{\sqrt{g}}\,\partial_r\bigl(\sqrt{g}\,g^{rr}\partial_r X\bigr)
	= \frac{1}{r^{D-2}}\partial_r\bigl(r^{D-2}F(r)\partial_r X\bigr),
\end{equation}
	with $\sqrt{g}\propto r^{D-2}$. In terms of the near-zone variable $\rho$, one finds (see also App.~\ref{app:NLO-details})
\begin{equation}
	\Box X(\rho)
	\simeq \frac{n^2}{r_h^2}\,e^{-\rho}
	\frac{d}{d\rho}\Bigl(e^{\rho}F_0(\rho)X'(\rho)\Bigr),
	\qquad
	X'(\rho)\equiv\frac{dX}{d\rho},
	\label{eq:Box-NZ-TS}
\end{equation}
	to leading order in $1/n$. Assuming $\chi_0=\chi_0(\rho)$ and using~\eqref{eq:F0-NZ-TS} and~\eqref{eq:Box-NZ-TS}, the LO thermal scalar equation~\eqref{eq:chi-eq-LO} becomes
\begin{equation}
	e^{-\rho}
	\frac{d}{d\rho}\Bigl(e^{\rho}F_0(\rho)\chi_0'(\rho)\Bigr)
	- \frac{r_h^2}{n^2}\bigl(\tilde{\beta}^2 F_0(\rho)-\tilde{\beta}_{\rm H}^2\bigr)
	\chi_0(\rho) = 0.
	\label{eq:chi-NZ-exact}
\end{equation}
	In the ``deep'' near-zone away from the very tip, $e^{-\rho}\ll 1$ and $F_0(\rho)$ has already reached its plateau,
\begin{equation}
	F_0(\rho) \simeq K,
	\qquad
	\tilde{\beta}^2 F_0(\rho)-\tilde{\beta}_{\rm H}^2
	\simeq \tilde{\beta}^2 K - \tilde{\beta}_{\rm H}^2.
\end{equation}
	Keeping the combination~\eqref{eq:Xi-def-TS} fixed, the equation~\eqref{eq:chi-NZ-exact} reduces, to LO in $1/n$, to a constant-coefficient ODE:
\begin{equation}
	\chi_0''(\rho) + \chi_0'(\rho) - \Xi^2\,\chi_0(\rho) = 0,
	\label{eq:chi-NZ-ODE}
\end{equation}
	with $\Xi^2$ given by~\eqref{eq:Xi-def-TS}. The characteristic equation
\begin{equation}
	\lambda^2 + \lambda - \Xi^2 = 0
\end{equation}
	has roots
\begin{equation}
	\lambda_{\pm}
	= \frac{-1 \pm \sqrt{1+4\Xi^2}}{2},
	\qquad
	\lambda_+>0,\quad
	\lambda_-<0
	\quad(\text{for } \Xi^2>0).
	\label{eq:lambda-pm-def}
\end{equation}
	The general near-zone solution in this regime is
\begin{equation}
	\chi_0^{\rm (NZ)}(\rho)
	= A_{\rm NZ} e^{\lambda_+\rho} + B_{\rm NZ} e^{\lambda_-\rho}.
\end{equation}
    Normalisability towards the outer part of the cigar (large $\rho$) forbids the growing mode $e^{\lambda_+\rho}$, so we set $A_{\rm NZ}=0$ and write
\begin{equation}
	\chi_0^{\rm (NZ)}(\rho)
	= \chi_h\,e^{\lambda_-\rho},
	\qquad
	\lambda_-<0,
	\label{eq:chi-NZ-sol}
\end{equation}
	Here $\chi_h$ is retained only as a label for the undetermined normalization of the exterior mode; it is not the literal value $\chi(r_h)$.  Its relation to the stretched horizon datum is $\chi_{\rm c}=\chi_h e^{\lambda_-\rho_{\rm c}}$.  The approximation~\eqref{eq:chi-NZ-sol} is valid for $\max(1,\rho_{\rm c})\ll\rho\ll n$.  Neither near--far matching nor the homogeneous quadratic equation determines $\chi_h$ or $\chi_{\rm c}$.
	
	It is convenient to express the effective thermal mass at the plateau in terms of $\Xi^2$:
\begin{equation}
	m_{\rm eff}^2
	\equiv \tilde{\beta}^2 K - \tilde{\beta}_{\rm H}^2,
	\qquad
	\Xi^2
	= \frac{r_h^2}{(D-2)^2 K}\,m_{\rm eff}^2.
	\label{eq:m-eff-def}
\end{equation}
	The condition $m_{\rm eff}^2=0$ (\textit{i.e.}, $\tilde{\beta}^2K=\tilde{\beta}_{\rm H}^2$) is the leading marginality condition on the matching plateau; it is not by itself a requirement that the entire black hole have string-scale temperature or curvature.  More generally, with
\begin{equation}
    \beta_{\rm loc}(r)=\beta\sqrt{g_{\tau\tau}(r)},
    \qquad
    m_{\rm eff}^2(r)=
    \frac{\beta_{\rm loc}^2(r)-\beta_{\rm H}^2}{4\pi^2\alpha'^2}
    +\xi_RR(r)+\cdots,
\end{equation}
	the EFT requires the two independent conditions
\begin{equation}
    \boxed{\alpha'|m_{\rm eff}^2(r)|\ll1,
    \qquad \alpha'{\cal R}(r)\ll1.}
    \label{eq:local-Hagedorn-weak-curvature}
\end{equation}
	The first makes the local winding mode nearly marginal and the second keeps the gravitational background weakly curved.  At the formal plateau marginality point, $\Xi^2=0$ and $\lambda_+=0$, $\lambda_-=-1$.

\paragraph{Far-zone solution and matching} \label{subsec:TS-far-zone}
	We now analyse the thermal scalar in the far-zone and show that the	radial equation has exactly the same ODE structure as in the near-zone once we use an appropriate large-$D$ coordinate. In the far-zone it is convenient to use the standard large-$D$ variable
\begin{equation}
	R \equiv \left(\frac{r}{r_h}\right)^{D-2},
	\qquad
	r = r_h R^{1/(D-2)}.
	\label{eq:R-def-TS}
\end{equation}
	Then
\begin{equation}
	\partial_r
	= (D-2)\,\frac{R}{r}\,\partial_R,
	\qquad
	r^{D-2} = r_h^{D-2}R.
\end{equation}
	For any $s$-wave $X=X(R)$ we find
\begin{equation}
	\Box X(R)
	= \frac{1}{r^{D-2}}\partial_r\bigl(r^{D-2}F(r)\partial_r X\bigr)
	\simeq F(r)\,\frac{(D-2)^2}{r_h^2}\,
	\frac{d}{dR}\bigl(R^2 X'(R)\bigr),
	\qquad
	X'(R)\equiv\frac{dX}{dR},
	\label{eq:Box-FZ-TS}
\end{equation}
	where we have used $r^{D-2}=r_h^{D-2}R$ and kept only the leading large-$D$ scaling of the derivatives.
    
    In the matching far-zone (where the potential has already plateaued, but we are not yet at asymptotic infinity) we can approximate the blackening function by its constant value,
\begin{equation}
	F(r) \simeq K,
\end{equation}
	with the same $K$ as in~\eqref{eq:F0-NZ-TS}. With $F(r)\simeq K$ and $\phi_0$ constant, the LO equation~\eqref{eq:chi-eq-LO} becomes
\begin{equation}
	\Box\chi_0 - \bigl(\tilde{\beta}^2 K - \tilde{\beta}_{\rm H}^2\bigr)\chi_0 = 0,
\end{equation}
	or using~\eqref{eq:Box-FZ-TS} and the definition~\eqref{eq:m-eff-def},
\begin{equation}
	K\,\frac{(D-2)^2}{r_h^2}\,
	\frac{d}{dR}\bigl(R^2\chi_0'(R)\bigr)
	- m_{\rm eff}^2\,\chi_0(R) = 0.
\end{equation}
	Dividing by $K(D-2)^2/r_h^2$ and using~\eqref{eq:Xi-def-TS} gives
\begin{equation}
	\frac{d}{dR}\bigl(R^2\chi_0'(R)\bigr) - \Xi^2\,\chi_0(R) = 0,
	\label{eq:chi-FZ-R-form}
\end{equation}
	with \emph{the same} parameter $\Xi^2$ as in the near-zone equation~\eqref{eq:chi-NZ-ODE}, up to $1/D$ corrections.

	Introducing
\begin{equation}
	Y \equiv \ln R,
\end{equation}
	we have
\begin{equation}
	\chi_0'(R)
	= \frac{1}{R}\frac{d\chi_0}{dY},
	\qquad
	\frac{d}{dR}\bigl(R^2\chi_0'(R)\bigr)
	= \frac{d^2\chi_0}{dY^2} + \frac{d\chi_0}{dY}.
\end{equation}
	Equation~\eqref{eq:chi-FZ-R-form} therefore reduces to
\begin{equation}
	\frac{d^2\chi_0}{dY^2}
	+ \frac{d\chi_0}{dY}
	- \Xi^2\,\chi_0(Y) = 0,
	\label{eq:chi-FZ-Y-form}
\end{equation}
	which is identical in form to the near-zone equation~\eqref{eq:chi-NZ-ODE}, upon the replacement $\rho\to Y$. The general far-zone solution is thus
\begin{equation}
	\chi_0^{\rm (FZ)}(R)
	= A_{\rm FZ} R^{\lambda_+} + B_{\rm FZ} R^{\lambda_-},
	\qquad
	\lambda_{\pm}
	= \frac{-1 \pm \sqrt{1+4\Xi^2}}{2},
\end{equation}
	with the same exponents $\lambda_\pm$ as in~\eqref{eq:lambda-pm-def}. Normalisability at large $R$ requires the decaying mode $R^{\lambda_-}$, so we set $A_{\rm FZ}=0$ and write
\begin{equation}
	\chi_0^{\rm (FZ)}(R)
	= \chi_\infty\,R^{\lambda_-},
	\qquad
	\lambda_-<0.
	\label{eq:chi-FZ-sol}
\end{equation}
	
	The near and far descriptions overlap in a region where both sets of coordinates are valid and their approximations hold:
\begin{equation}
	1\ll \rho \ll n,
	\qquad
	1\ll R \ll e^n.
\end{equation}
	Using~\eqref{eq:rho-def-TS} and~\eqref{eq:R-def-TS}, we have
\begin{equation}
	\rho
	= n\ln\frac{r}{r_h},
	\qquad
	\ln R
	= (D-2)\ln\frac{r}{r_h}
	= (n+1)\ln\frac{r}{r_h}
	= \frac{n+1}{n}\,\rho
	= \rho + \mathcal{O}\!\left(\frac{\rho}{n}\right).
\end{equation}
	Therefore, in the overlap region,
\begin{equation}
	Y \equiv \ln R \simeq \rho
	\qquad (n\gg 1).
\end{equation}
    The near-zone solution~\eqref{eq:chi-NZ-sol} behaves as
\begin{equation}
	\chi_0^{\rm (NZ)}(\rho)
	= \chi_h e^{\lambda_-\rho}
	\simeq \chi_h e^{\lambda_- Y},
\end{equation}
	while the far-zone solution~\eqref{eq:chi-FZ-sol} can be written as
\begin{equation}
	\chi_0^{\rm (FZ)}(R)
	= \chi_\infty R^{\lambda_-}
	= \chi_\infty e^{\lambda_- Y}.
\end{equation}
	Smooth matching in the overlap region requires
\begin{equation}
	\chi_\infty = \chi_h,
\end{equation}
	Thus the same decaying mode with exponent $\lambda_-$ controls both zones.  This equality fixes only the relative coefficient of the near- and far-zone representations.  Because the radial equation is linear and homogeneous, regularity and normalisability select the admissible profile (and can impose an eigenvalue condition on the temperature) but do not fix its overall normalization.  Determining that normalization requires nonlinear thermal scalar interactions, fully backreacted thermodynamic boundary data, or microscopic input from the string-scale core.

\paragraph*{Boxed leading-order exterior thermal scalar}
	We summarise the LO thermal scalar result as follows.
\begin{equation}
	\boxed{
		\begin{aligned}
			&\text{\bf Background:} &&
			\text{EGB Boulware--Deser black hole with constant dilaton }
			\phi_0=\phi_\infty,\\[0.4em]
			&\text{\bf Effective mass at the plateau:} &&
			m_{\rm eff}^2 \equiv \tilde{\beta}^2 K - \tilde{\beta}_{\rm H}^2,
			\qquad
			\Xi^2
			= \frac{r_h^2}{(D-2)^2 K}\,m_{\rm eff}^2,
			\\[0.4em]
			&\text{\bf Exponents:} &&
			\lambda_{\pm}
			= \frac{-1 \pm \sqrt{1+4\Xi^2}}{2},
			\qquad
			\lambda_+>0,\ \lambda_-<0\ (\Xi^2>0),\\[0.4em]
			&\text{\bf Near-zone (\(r-r_h=\mathcal{O}(r_h/D)\)):} &&
			\rho = n\ln\frac{r}{r_h},
			\quad
			F_0(\rho) = K\bigl(1-e^{-\rho}\bigr),
			\\[0.2em]
			&&&
			\chi_0^{\rm (NZ)}(\rho)
			= \chi_h\,e^{\lambda_-\rho},
			\qquad 1\ll \rho \ll n,
			\\[0.6em]
			&\text{\bf Far-zone (\(R=(r/r_h)^{D-2}\)):} &&
			R = \Bigl(\frac{r}{r_h}\Bigr)^{D-2},
			\quad
			Y=\ln R,
			\\[0.2em]
			&&&
			\chi_0^{\rm (FZ)}(R)
			= \chi_h\,R^{\lambda_-}
			= \chi_h e^{\lambda_- Y},
			\qquad
			1\ll R \ll e^n,
			\\[0.6em]
			&\text{\bf Hagedorn condition:} &&
			m_{\rm eff}^2 = 0
			\quad\Longleftrightarrow\quad
			\tilde{\beta}^2 K = \tilde{\beta}_{\rm H}^2
			\quad\Longleftrightarrow\quad
			\Xi^2 = 0,\ \lambda_+=0,\ \lambda_-=-1.
		\end{aligned}
}\end{equation}

\subsection{Next-to-leading order fields in the large-\texorpdfstring{$D$}{D} expansion} \label{sec:NLO-fields}
    In this subsection we collect the next-to-leading order corrections to the blackening function, dilaton, and thermal scalar in the large-$D$ expansion. We work around the leading-order Boulware--Deser background
\begin{equation}
	F(r) = F_0(\rho) + \frac{1}{n}F_1(\rho) + \mathcal{O}\!\left(\frac{1}{n^2}\right),
	\qquad
	F_0(\rho) = K\bigl(1 - e^{-\rho}\bigr),
	\qquad
	D = n+3,
\end{equation}
    in the near-zone coordinate
\begin{equation}
	\rho = n\ln\frac{r}{r_h},
\end{equation}
    with $r_h$ the horizon radius of the BD solution. The LO dilaton and thermal scalar profiles are
\begin{equation}
	\phi_0 = \phi_\infty,
	\qquad
	\chi_0^{\rm (NZ)}(\rho) = \chi_h e^{\lambda_-\rho},
	\qquad
	\lambda_\pm = \frac{-1 \pm \sqrt{1+4\Xi^2}}{2},
	\quad
	\lambda_-<0,
\end{equation}
    where
\begin{equation}
	\Xi^2
	= \frac{r_h^2}{n^2 K}\,\bigl(\tilde{\beta}^2 K - \tilde{\beta}_{\rm H}^2\bigr),
	\qquad
	m_{\rm eff}^2 \equiv \tilde{\beta}^2 K - \tilde{\beta}_{\rm H}^2.
\end{equation}
    The corresponding far-zone LO scalar is
\begin{equation}
	\chi_0^{\rm (FZ)}(R)
	= \chi_\infty R^{\lambda_-},
	\qquad
	R \equiv \Bigl(\frac{r}{r_h}\Bigr)^{D-2},
	\qquad
	Y \equiv \ln R,
\end{equation}
    with $\chi_\infty$ fixed by matching to the near-zone.

\subsubsection{NLO metric correction \texorpdfstring{$F_1$}{F1}} \label{subsubsec:NLO-metric}
    At NLO the metric perturbation $F_1(\rho)$ is sourced by the LO thermal scalar stress tensor $T^{(\chi)}_{\mu\nu}[\chi_0]$. The LO dilaton is constant and does not contribute to the Einstein equation at this order. In the \emph{probe} approximation used here, we also neglect the linear dilaton contribution $2\nabla_\mu\nabla_\nu\phi_1 - 2 g_{\mu\nu}\nabla^2\phi_1$ as a source for $F_1$, so that the NLO metric correction is driven purely by the thermal scalar stress tensor; quadratic terms $(\partial\phi_1)^2$ are $\mathcal{O}(\kappa^2)$ and are likewise ignored.

\paragraph{Near-zone equation and solution}
    The combination of Einstein equations $\mathcal{E}^\rho{}_\rho-\mathcal{E}^\tau{}_\tau=0$, expanded to first order in $1/n$ around the LO EGB background, yields a first-order radial equation for $F_1(\rho)$,
\begin{equation}
	F_1'(\rho) + F_1(\rho)
	= \mathcal{S}_{\rm NZ}(\rho),
	\label{eq:F1-NZ-eq-final}
\end{equation}
    where the prime denotes $d/d\rho$ and the source is built from the LO thermal scalar. For a static, purely radial scalar $\chi_0(\rho)$ with canonical kinetic term, the relevant stress tensor combination is
\begin{equation}
	\bigl(T^\rho{}_\rho - T^\tau{}_\tau\bigr)_{\rm LO}
	= g^{\rho\rho}_{(0)}(\partial_\rho\chi_0)^2
	= g^{\rho\rho}_{(0)} \lambda_-^2\,\chi_0(\rho)^2,
\end{equation}
    so that the source can be written as
\begin{equation}
	\mathcal{S}_{\rm NZ}(\rho)
	= \frac{2\kappa_{\rm eff}}{D-2}\,g^{\rho\rho}_{(0)}(\partial_\rho\chi_0)^2
	\equiv
	\mathcal{A}_\chi\,(1 - e^{-\rho})\,e^{2\lambda_- \rho},
\end{equation}
    with $g^{\rho\rho}_{(0)}$ the LO radial inverse metric and $\kappa_{\rm eff}$ an effective gravitational coupling including the Gauss--Bonnet corrections. Using the near-zone scaling of $g^{\rho\rho}_{(0)}$ one finds the explicit coefficient
\begin{equation}
	\mathcal{A}_\chi
	=
	\frac{2\kappa_{\rm eff}}{D-2}
	\frac{n^2}{r_h^2}\,
	K \lambda_-^2 \chi_h^2,
	\label{eq:Achi-def}
\end{equation}
    so that all explicit $\rho$-dependence arises from $F_0(\rho)$ and
$\chi_0(\rho)$.
    With this notation the near-zone equation becomes
\begin{equation}
	F_1'(\rho) + F_1(\rho)
	= \mathcal{A}_\chi\Bigl(e^{2\lambda_- \rho}
	- e^{(2\lambda_- -1)\rho}\Bigr).
\end{equation}
    This is a linear inhomogeneous first-order ODE. The integrating factor is $e^{\rho}$, so
\begin{equation}
	\frac{d}{d\rho}\bigl(e^{\rho}F_1(\rho)\bigr)
	= \mathcal{A}_\chi\Bigl(e^{(2\lambda_- +1)\rho}
	- e^{2\lambda_-\rho}\Bigr).
\end{equation}
    Integrating from $\rho=0$ to $\rho$ gives
\begin{align}
	e^{\rho}F_1(\rho)
	&=
	F_1(0)
	+ \mathcal{A}_\chi
	\int_0^\rho
	\Bigl(e^{(2\lambda_- +1)\sigma}
	- e^{2\lambda_- \sigma}\Bigr)\,d\sigma
	\nonumber\\[0.4em]
	&=
	F_1(0)
	+ \mathcal{A}_\chi
	\left[
	\frac{e^{(2\lambda_- +1)\rho}-1}{2\lambda_- + 1}
	- \frac{e^{2\lambda_- \rho}-1}{2\lambda_-}
	\right].
\end{align}
    We work in the gauge where the horizon position is fixed at $\rho=0$ at NLO, which imposes
\begin{equation}
	F_1(0) = 0.
\end{equation}
    This gives the exact NLO near-zone metric correction
\begin{equation}
	\boxed{
		F_1^{\rm (NZ)}(\rho)
		=
		\mathcal{A}_\chi\,e^{-\rho}
		\left[
		\frac{e^{(2\lambda_- +1)\rho}-1}{2\lambda_- + 1}
		- \frac{e^{2\lambda_- \rho}-1}{2\lambda_-}
		\right].
	}
	\label{eq:F1-NZ-solution-final}
\end{equation}
    Near the horizon, $\rho\to 0$,
\begin{equation}
	F_1^{\rm (NZ)}(\rho)
	=
	\frac{\mathcal{A}_\chi}{2}\,\rho^2
	+ \mathcal{O}(\rho^3),
\end{equation}
    which is regular and compatible with the cigar geometry. For large $\rho$ the dominant behaviour is
\begin{equation}
	F_1^{\rm (NZ)}(\rho)
	\simeq
	\mathcal{A}_\chi
	\left[
	\frac{e^{2\lambda_- \rho}}{2\lambda_- + 1}
	- \frac{e^{(2\lambda_- -1)\rho}}{2\lambda_-}
	\right],
	\qquad
	\rho\to\infty,
	\label{eq:F1-NZ-asympt-final}
\end{equation}
    so that $F_1$ decays faster than the background metric for $\lambda_-<0$, as expected.

\paragraph{Far-zone behaviour and matching}
    In the far-zone we use
\begin{equation}
	R \equiv \left(\frac{r}{r_h}\right)^{D-2},
	\qquad
	Y \equiv \ln R,
\end{equation}
    and approximate the BD background by its constant-curvature form
\begin{equation}
	F(r) \simeq K,
	\qquad
	r - r_h = \mathcal{O}(r_h),
\end{equation}
    as in the LO far-zone analysis. The LO thermal scalar in the far-zone has the same exponent as in the near-zone,
\begin{equation}
	\chi_0^{\rm (FZ)}(R) = \chi_\infty R^{\lambda_-}
	= \chi_\infty e^{\lambda_- Y},
	\qquad
	\lambda_-<0.
\end{equation}

    Linearising the Einstein equation and taking the same combination
$\mathcal{E}^r{}_r - \mathcal{E}^\tau{}_\tau$, one finds
\begin{equation}
	\frac{dF_1^{\rm (FZ)}}{dY} + F_1^{\rm (FZ)}(Y)
	= \mathcal{S}_{\rm FZ}(Y),
    \label{eq:F1-NZ-eq}
\end{equation}
    with
\begin{equation}
	\mathcal{S}_{\rm FZ}(Y)
	=
	\mathcal{A}_\chi^{(\infty)} e^{2\lambda_- Y},
	\qquad
	\mathcal{A}_\chi^{(\infty)}
	\equiv
	\frac{2\kappa_{\rm eff} K \lambda_-^2 \chi_\infty^2}{D-2}
	\times \Bigl[\text{far-zone Jacobian factors}\Bigr].
\end{equation}
    The structure of this equation coincides with the deep near-zone limit of~\eqref{eq:F1-NZ-eq-final}, where $(1-e^{-\rho})\to 1$. Solving Eq.~\eqref{eq:F1-NZ-eq} gives
\begin{equation}
	F_1^{\rm (FZ)}(Y)
	=
	\frac{\mathcal{A}_\chi^{(\infty)}}{2\lambda_- +1}
	e^{2\lambda_- Y}
	+ F_1^{(\infty)} e^{-Y}.
\end{equation}
    Regularity and normalisability in the asymptotic region require the
$e^{-Y}\propto R^{-1}$ mode to be absent, so we set $F_1^{(\infty)}=0$ and obtain
\begin{equation}
	\boxed{
		F_1^{\rm (FZ)}(R)
		=
		\frac{\mathcal{A}_\chi^{(\infty)}}{2\lambda_- +1}\,
		R^{2\lambda_-},
		\qquad
		R\to\infty,
	}
	\label{eq:F1-FZ-asympt-final}
\end{equation}
    with $2\lambda_-<0$.

    The near- and far-zones overlap in
\begin{equation}
	1 \ll \rho \ll n
	\qquad\Longleftrightarrow\qquad
	1 \ll Y=\ln R \ll n,
\end{equation}
    and in this region
\begin{equation}
	Y = \ln R
	= \frac{D-2}{n}\,\ln\frac{r}{r_h}
	= \frac{n+1}{n}\,\rho
	= \rho + \mathcal{O}\!\left(\frac{\rho}{n}\right),
\end{equation}
    so we may identify $Y\simeq\rho$ at leading order in $1/n$. The large-$\rho$ behaviour of~\eqref{eq:F1-NZ-asympt-final},
\begin{equation}
	F_1^{\rm (NZ)}(\rho)
	\simeq
	\frac{\mathcal{A}_\chi}{2\lambda_- +1}\,e^{2\lambda_- \rho}
	+ \mathcal{O}\bigl(e^{(2\lambda_- -1)\rho}\bigr),
\end{equation}
    matches the far-zone form~\eqref{eq:F1-FZ-asympt-final} provided we identify
\begin{equation}
	\mathcal{A}_\chi^{(\infty)} = \mathcal{A}_\chi,
	\qquad
	\chi_\infty = \chi_h.
\end{equation}
    With the matching relations above, Eq.~\eqref{eq:F1-NZ-solution-final} gives the exterior near-zone NLO metric correction, whose overlap behaviour agrees with the far-zone expansion at fixed matching data.

\paragraph{Exterior NLO blackening function}
    Collecting the results, the NLO-corrected blackening function in the large-$D$ expansion is
\begin{equation}
	\boxed{
		F(r)
		=
		F_0(\rho)
		+ \frac{1}{n}\,F_1(\rho)
		+ \mathcal{O}\!\left(\frac{1}{n^2}\right),
		\qquad
		F_0(\rho) = K\bigl(1 - e^{-\rho}\bigr),
	}
\end{equation}
    with
\begin{equation}
	\boxed{
		F_1(\rho)
		=
		\mathcal{A}_\chi\,e^{-\rho}
		\left[
		\frac{e^{(2\lambda_- +1)\rho}-1}{2\lambda_- + 1}
		- \frac{e^{2\lambda_- \rho}-1}{2\lambda_-}
		\right],
		\qquad
		\lambda_- = \frac{-1 - \sqrt{1+4\Xi^2}}{2}<0,
		\quad
		\Xi^2 = \frac{r_h^2}{n^2 K}(\tilde{\beta}^2 K - \tilde{\beta}_{\rm H}^2).
	}
\end{equation}

\subsubsection{NLO dilaton profile in the probe regime} \label{subsubsec:NLO-dilaton}
    We now construct the NLO dilaton profile in the large-$D$ expansion, working in the \emph{probe} regime for the string gas: the thermal scalar and dilaton fluctuate on a fixed EGB BD background, and their backreaction on the metric is included only to linear order in the gravitational coupling $\kappa$. In particular, in the NLO metric analysis we treat $F_1$ as sourced directly by the LO thermal scalar stress tensor and neglect the linear dilaton contribution as a subleading probe field effect. Quadratic dilaton terms $(\partial\phi_1)^2$ are $\mathcal{O}(\kappa^2)$ and are also neglected.

\paragraph{Master equation for \texorpdfstring{$\phi_1$}{phi1}}
    We start from the string frame equations
\begin{align}
	&G_{\mu\nu} + \Lambda g_{\mu\nu} + \alpha H_{\mu\nu}
	+ 2\nabla_\mu\nabla_\nu\phi - 2 \left( \nabla^2\phi \right) g_{\mu\nu}
	+ 2 (\partial\phi)^2 g_{\mu\nu}
	= 2\kappa\, T^{(\chi)}_{\mu\nu},
	\label{eq:EGB-dilaton-Einstein-NLO}\\[4pt]
	&4 \nabla^2\phi - 4 (\partial\phi)^2
	+ R - 2\Lambda + \alpha R_{\rm GB}^2
	= 2\kappa\,\big[(\partial\chi)^2 + m_{\rm th}^2(r)\chi^2\big],
	\label{eq:dilaton-eom-NLO}\\[4pt]
	&\nabla^2\chi - 2\,\partial_\mu\phi\,\nabla^\mu\chi
	- m_{\rm th}^2(r)\,\chi = 0,
\end{align}
    with
\begin{equation}
	m_{\rm th}^2(r) \equiv \tilde{\beta}^2 g_{\tau\tau}(r) - \tilde{\beta}_{\rm H}^2
	= \tilde{\beta}^2 F(r) - \tilde{\beta}_{\rm H}^2.
\end{equation}
    We expand in $1/D$ and in the small parameter $\kappa$:
\begin{equation}
	\phi(x) = \phi_0 + \phi_1(x) + \mathcal{O}(\kappa^2),
	\qquad
	\phi_0 = \phi_\infty,
\end{equation}
    and work to linear order in $\kappa$. On the BD background the geometric combination in~\eqref{eq:dilaton-eom-NLO} vanishes,
\begin{equation}
	\bigl(R - 2\Lambda + \alpha R_{\rm GB}^2\bigr)_{\rm BD} = 0,
\end{equation}
    so to linear order in $\kappa$ and in $\phi_1$ the dilaton equation reduces to
\begin{equation}
	4\nabla^2\phi_1
	= 4\kappa\big[(\partial\chi_0)^2 + m_{\rm th}^2(r)\chi_0^2\big],
\end{equation}
    or
\begin{equation}
	\boxed{
		\nabla^2\phi_1
		= \frac{\kappa}{2} \big[(\partial\chi_0)^2 + m_{\rm th}^2(r)\chi_0^2\big]
	}
	\label{eq:phi1-master-final}
\end{equation}
    on the fixed BD background.

\paragraph{Near-zone equation and solution for \texorpdfstring{$\phi_1'(\rho)$}{phi1'(rho)}}
    For an $s$-wave scalar $X=X(r)$ in the metric $F(r)$, the radial Laplacian is
\begin{equation}
	\Box X
	= \frac{1}{r^{D-2}}\partial_r\bigl(r^{D-2}F(r)\partial_r X\bigr).
\end{equation}
    In the near-zone we use
\begin{equation}
	\rho = n\ln\frac{r}{r_h},
	\qquad
	r = r_h e^{\rho/n},
	\qquad
	\partial_r = \frac{n}{r}\,\partial_\rho,
\end{equation}
    so that to leading order in $1/n$
\begin{equation}
	r^{D-2} \sim r_h^{n+1} e^{\rho},
	\qquad
	\partial_r \sim \frac{n}{r_h}\partial_\rho.
\end{equation}
    With $F(r)\simeq F_0(\rho)$ this gives
\begin{equation}
	\Box X(\rho)
	\simeq \frac{n^2}{r_h^2}\,
	e^{-\rho}\frac{d}{d\rho}
	\Bigl(e^{\rho}F_0(\rho)X'(\rho)\Bigr),
	\qquad
	X'(\rho)\equiv\frac{dX}{d\rho}.
\end{equation}
    The LO near-zone scalar is $\chi_0(\rho)=\chi_h e^{\lambda_-\rho}$. Its radial derivative is
\begin{equation}
	\partial_r\chi_0
	= \lambda_- \chi_h e^{\lambda_-\rho}\frac{n}{r}
	\simeq \lambda_- \chi_h e^{\lambda_-\rho}\frac{n}{r_h},
\end{equation}
    so
\begin{equation}
	(\partial\chi_0)^2
	= g^{rr}(\partial_r\chi_0)^2
	\simeq F_0(\rho)\left(\frac{n}{r_h}\right)^2
	\lambda_-^2 \chi_h^2 e^{2\lambda_-\rho}
	= \frac{n^2}{r_h^2}K(1-e^{-\rho})
	\lambda_-^2 \chi_h^2 e^{2\lambda_-\rho}.
\end{equation}
    To leading order we approximate
$m_{\rm th}^2(r)\simeq m_{\rm eff}^2\equiv \tilde{\beta}^2 K - \tilde{\beta}_{\rm H}^2$, so
\begin{equation}
	m_{\rm th}^2(r)\chi_0^2
	\simeq m_{\rm eff}^2 \chi_h^2 e^{2\lambda_-\rho}
	= \frac{n^2K}{r_h^2}\,\Xi^2\chi_h^2 e^{2\lambda_-\rho},
\end{equation}
    where we used
\begin{equation}
	\Xi^2
	= \frac{r_h^2}{n^2 K}m_{\rm eff}^2
	\quad\Longrightarrow\quad
	m_{\rm eff}^2
	= \frac{n^2 K}{r_h^2}\,\Xi^2.
\end{equation}
    Thus the total source in~\eqref{eq:phi1-master-final} is
\begin{equation}
	(\partial\chi_0)^2 + m_{\rm th}^2\chi_0^2
	\simeq \frac{n^2 K}{r_h^2}\chi_h^2 e^{2\lambda_-\rho}
	\Bigl[(1-e^{-\rho})\lambda_-^2 + \Xi^2\Bigr].
\end{equation}
    Inserting this and the near-zone Laplacian into~\eqref{eq:phi1-master-final} gives
\begin{equation}
	\frac{n^2}{r_h^2}\,
	e^{-\rho}\frac{d}{d\rho}
	\Bigl(e^{\rho}F_0(\rho)\phi_1'(\rho)\Bigr)
	= \kappa\,\frac{n^2 K}{r_h^2}\chi_h^2 e^{2\lambda_-\rho}
	\Bigl[(1-e^{-\rho})\lambda_-^2 + \Xi^2\Bigr].
\end{equation}
    Cancelling $n^2/r_h^2$ and using $F_0(\rho)=K(1-e^{-\rho})$ leads to
\begin{equation}
	e^{-\rho}\frac{d}{d\rho}
	\Bigl(e^{\rho}F_0(\rho)\phi_1'(\rho)\Bigr)
	= \kappa K\chi_h^2 e^{2\lambda_-\rho}
	\Bigl[(1-e^{-\rho})\lambda_-^2 + \Xi^2\Bigr].
	\label{eq:phi1-NZ-eq-final}
\end{equation}
    Defining
\begin{equation}
	Y(\rho)
	\equiv e^{\rho}F_0(\rho)\phi_1'(\rho)
	= K(e^{\rho}-1)\phi_1'(\rho),
\end{equation}
    we obtain
\begin{equation}
	\frac{dY}{d\rho}
	= \kappa K\chi_h^2
	\left[
	(\lambda_-^2 + \Xi^2) e^{(2\lambda_-+1)\rho}
	- \lambda_-^2 e^{2\lambda_-\rho}
	\right].
\end{equation}
    Regularity of $\phi_1$ at the tip requires $\phi_1'$ to remain finite as $\rho\to 0$, hence $Y(\rho)\to 0$ and $Y(0)=0$. Integrating from $0$ to $\rho$ gives
\begin{equation}
	Y(\rho)
	= \kappa K\chi_h^2
	\left[
	\frac{\lambda_-^2 + \Xi^2}{2\lambda_-+1}\bigl(e^{(2\lambda_-+1)\rho}-1\bigr)
	- \frac{\lambda_-}{2}\bigl(e^{2\lambda_-\rho}-1\bigr)
	\right].
\end{equation}
    Dividing by $e^{\rho}F_0(\rho)=K(e^{\rho}-1)$, we obtain the explicit NLO near-zone dilaton derivative
\begin{equation}
	\boxed{
		\phi_1'(\rho)
		=
		\kappa\chi_h^2\,
		\frac{
			\displaystyle
			\frac{\lambda_-^2 + \Xi^2}{2\lambda_-+1}\bigl(e^{(2\lambda_-+1)\rho}-1\bigr)
			- \frac{\lambda_-}{2}\bigl(e^{2\lambda_-\rho}-1\bigr)
		}{
			e^{\rho}-1
		}.
	}
	\label{eq:phi1-prime-NZ-correct}
\end{equation}
    This expression is manifestly regular at $\rho=0$ and decays exponentially as $\rho\to\infty$ because $\lambda_-<0$.

    It is convenient to introduce the quadrature
\begin{equation}
	\mathcal{I}(\rho)
	\equiv
	\int_0^\rho d\sigma\,
	\frac{
		\displaystyle
		\frac{\lambda_-^2 + \Xi^2}{2\lambda_-+1}\bigl(e^{(2\lambda_-+1)\sigma}-1\bigr)
		- \frac{\lambda_-}{2}\bigl(e^{2\lambda_-\sigma}-1\bigr)
	}{
		e^{\sigma}-1
	},
\end{equation}
    so that
\begin{equation}
	\phi_1^{\rm (NZ)}(\rho)
	= \phi_1(0) + \kappa\chi_h^2\,\mathcal{I}(\rho),
	\qquad
	\mathcal{I}'(\rho) = \text{integrand of~\eqref{eq:phi1-prime-NZ-correct}}.
\end{equation}
    The integral $\mathcal{I}(\rho)$ can be written in terms of digamma and Gauss hypergeometric functions by the change of variables $t=e^{-\sigma}$ and the standard integral representation
\begin{equation}
	J(a;\rho)
	\equiv
	\int_0^\rho d\sigma\,\frac{e^{a\sigma}-1}{e^{\sigma}-1},
\end{equation}
    so that
\begin{equation}
	\mathcal{I}(\rho)
	=
	\frac{\lambda_-^2 + \Xi^2}{2\lambda_-+1}\,J(2\lambda_-+1;\rho)
	- \frac{\lambda_-}{2}\,J(2\lambda_-;\rho).
\end{equation}
    We relegate the explicit special-function expression to an appendix.

\paragraph{Far-zone dilaton and matching}
    In the far-zone we use the same variables $R$ and $Y=\ln R$. For an $s$-wave scalar $X=X(R)$ the radial Laplacian reduces to
\begin{equation}
	\Box X(R)
	\simeq K\,\frac{(D-2)^2}{r_h^2}
	\left(\frac{d^2 X}{dY^2} + \frac{dX}{dY}\right),
\end{equation}
    in the matching far-zone where $F(r)\simeq K$ and $R=e^Y$. The LO far-zone thermal scalar is
\begin{equation}
	\chi_0^{\rm (FZ)}(R)
	= \chi_\infty e^{\lambda_- Y}.
\end{equation}
    Repeating the analysis of the source, one finds
\begin{equation}
	(\partial\chi_0)^2 + m_{\rm th}^2\chi_0^2
	\simeq \frac{(D-2)^2K}{r_h^2}\,\chi_\infty^2 e^{2\lambda_- Y}
	(\lambda_-^2 + \Xi^2),
\end{equation}
    so that~\eqref{eq:phi1-master-final} becomes, after cancelling
$K(D-2)^2/r_h^2$,
\begin{equation}
	\frac{d^2\phi_1}{dY^2} + \frac{d\phi_1}{dY}
	= \kappa\chi_\infty^2 e^{2\lambda_- Y}(\lambda_-^2 + \Xi^2).
	\label{eq:phi1-FZ-eq-final}
\end{equation}
    Defining
\begin{equation}
	A_\phi \equiv \kappa\chi_\infty^2(\lambda_-^2 + \Xi^2),
	\qquad
	Z(Y) \equiv e^{Y}\frac{d\phi_1}{dY},
\end{equation}
    we obtain
\begin{equation}
	\frac{dZ}{dY}
	= A_\phi e^{(2\lambda_-+1)Y},
\end{equation}
    so that
\begin{equation}
	Z(Y)
	= Z_0
	+ \frac{A_\phi}{2\lambda_-+1}
	\bigl(e^{(2\lambda_-+1)Y}-1\bigr).
\end{equation}
    Dividing by $e^Y$ and integrating once more, we arrive at
\begin{equation}
	\phi_1^{\rm (FZ)}(Y)
	= \phi_{1,\infty}
	+ B_1 e^{-Y}
	+ B_2 e^{2\lambda_- Y},
\end{equation}
    where
\begin{equation}
	\phi_{1,\infty} \equiv C_0,
	\qquad
	B_1 \equiv -Z_0 + \frac{A_\phi}{2\lambda_- +1},
	\qquad
	B_2 \equiv \frac{A_\phi}{2\lambda_-(2\lambda_- +1)}
	= \frac{\kappa\chi_\infty^2(\lambda_-^2 + \Xi^2)}{2\lambda_-(2\lambda_- +1)}.
\end{equation}
    Since $\lambda_-<0$, both $e^{-Y}$ and $e^{2\lambda_- Y}$ decay as $Y\to\infty$, and the dilaton approaches a constant $\phi_1^{\rm (FZ)}(Y)\to\phi_{1,\infty}$. The constant $\phi_{1,\infty}$ represents an asymptotic shift of the dilaton, fixed by boundary conditions at infinity.

    In the overlap region
\begin{equation}
	1\ll \rho \ll n,
	\qquad
	1\ll Y \ll n,
\end{equation}
    we may identify $Y\simeq\rho$ at leading order in $1/n$. The near-zone solution then tends to
\begin{equation}
	\phi_1^{\rm (NZ)}(\rho)
	\xrightarrow[\rho\to\infty]{}
	\phi_1(0) + \kappa\chi_h^2\,\mathcal{I}(\infty)
	\equiv \phi_{1,\infty},
\end{equation}
    which must coincide with the far-zone constant. This fixes
\begin{equation}
	\phi_{1,\infty}
	= \phi_1(0) + \kappa\chi_h^2\,\mathcal{I}(\infty).
\end{equation}
    Matching the coefficients of $e^{2\lambda_-Y}$ selects $B_2$ in terms of the near-zone data and the ratio $\chi_\infty/\chi_h$, while the coefficient of $e^{-Y}$ selects $B_1$ and thus $Z_0$. The detailed expressions can be written in terms of the special-function representation of $\mathcal{I}(\rho)$ and are relegated to an appendix.

\paragraph{Summary of the NLO dilaton profile}
    At fixed matching data, the exterior NLO dilaton in the large-$D$ probe regime can be summarised as
\begin{equation}
	\boxed{
		\begin{aligned}
			&\text{\bf Near-zone:}&
			\phi_1^{\rm (NZ)}(\rho)
			&= \phi_1(0)
			+ \kappa\chi_h^2\,\mathcal{I}(\rho),
			\quad
			\mathcal{I}'(\rho)=
			\frac{
				\displaystyle
				\frac{\lambda_-^2 + \Xi^2}{2\lambda_-+1}\bigl(e^{(2\lambda_-+1)\rho}-1\bigr)
				- \frac{\lambda_-}{2}\bigl(e^{2\lambda_-\rho}-1\bigr)
			}{
				e^{\rho}-1
			},\\[0.8em]
			&&\phi_1'(\rho)
			&= \kappa\chi_h^2\,
			\frac{
				\displaystyle
				\frac{\lambda_-^2 + \Xi^2}{2\lambda_-+1}\bigl(e^{(2\lambda_-+1)\rho}-1\bigr)
				- \frac{\lambda_-}{2}\bigl(e^{2\lambda_-\rho}-1\bigr)
			}{
				e^{\rho}-1
			},\\[0.9em]
			&\text{\bf Far-zone:}&
			\phi_1^{\rm (FZ)}(Y)
			&= \phi_{1,\infty}
			+ B_1 e^{-Y}
			+ B_2 e^{2\lambda_- Y},
			\qquad
			Y = \ln R,
			\\[0.2em]
			&&B_2
			&= \frac{\kappa\chi_\infty^2(\lambda_-^2 + \Xi^2)}
			{2\lambda_-(2\lambda_- +1)},\\[0.9em]
			&\text{\bf Matching:}&
			\phi_{1,\infty}
			&= \phi_1(0) + \kappa\chi_h^2\,\mathcal{I}(\infty),
			\qquad
			Y\simeq\rho\quad\text{in the overlap region}.
		\end{aligned}
	}
\end{equation}

\subsubsection{NLO thermal scalar profile} \label{subsubsec:NLO-thermal-scalar}
    We finally construct the NLO correction to the thermal scalar in the near- and far-zone. The LO operator and its Green's function structure are universal and control the NLO solution.

\paragraph{LO operator and Green's function}
    At LO the thermal scalar satisfies the constant-coefficient ODE
\begin{equation}
	\mathcal{L}_0[\chi_0] = 0,
	\qquad
	\mathcal{L}_0[y]
	\equiv y''(\rho) + y'(\rho) - \Xi^2 y(\rho),
\end{equation}
    with
\begin{equation}
	\Xi^2
	\equiv \frac{r_h^2}{n^2 K}\bigl(\tilde{\beta}^2 K - \tilde{\beta}_{\rm H}^2\bigr).
\end{equation}
    The characteristic equation has roots
\begin{equation}
	\lambda_{\pm}
	= \frac{-1 \pm \sqrt{1+4\Xi^2}}{2},
	\qquad
	\lambda_+>0,\quad \lambda_-<0,
\end{equation}
    with homogeneous solutions
\begin{equation}
	y_1(\rho) = e^{\lambda_+\rho},
	\qquad
	y_2(\rho) = e^{\lambda_-\rho}.
\end{equation}
    Normalisability in the near-zone selects the decaying mode, so the LO solution is
\begin{equation}
	\chi_0(\rho) = \chi_h e^{\lambda_-\rho},
\end{equation}
    The Wronskian is
\begin{equation}
	W(\rho)
	= y_1 y_2' - y_2 y_1'
	= (\lambda_- - \lambda_+)\,e^{(\lambda_+ + \lambda_-)\rho}
	= (\lambda_- - \lambda_+)\,e^{-\rho},
\end{equation}
    where we used $\lambda_+ + \lambda_- = -1$. For the inhomogeneous equation,
\begin{equation}
	\mathcal{L}_0[\chi_1(\rho)] = S(\rho),
	\label{eq:chi1-inhomog-final}
\end{equation}
    with $\chi_1(\rho)$ regular at the tip and decaying at infinity, the Green's function solution compatible with our boundary conditions can be written as
\begin{equation}
	\boxed{
		\chi_1(\rho)
		= \frac{e^{\lambda_-\rho}}{\lambda_- - \lambda_+}
		\int_\rho^\infty d\sigma\,
		S(\sigma)\,e^{-\lambda_-\sigma}.
	}
	\label{eq:chi1-GF-final}
\end{equation}

\paragraph{Near-zone NLO equation and source}
    We expand the fields to first subleading order in $1/n$:
\begin{equation}
	F(\rho) = F_0(\rho) + \frac{1}{n}F_1(\rho) + \cdots,
	\qquad
	\phi(\rho) = \phi_0 + \phi_1(\rho) + \cdots,
	\qquad
	\chi(\rho) = \chi_0(\rho) + \frac{1}{n}\chi_1(\rho) + \cdots,
\end{equation}
    with
\begin{equation}
	F_0(\rho) = K(1-e^{-\rho}),
	\qquad
	\phi_0 = \phi_\infty,
	\qquad
	\chi_0(\rho) = \chi_h e^{\lambda_-\rho},
\end{equation}
    and insert into the full thermal scalar equation
\begin{equation}
	\Box\chi - 2\,\partial_\mu\phi\,\nabla^\mu\chi
	- m_{\rm th}^2(r)\,\chi = 0,
	\qquad
	m_{\rm th}^2(r)
	\equiv \tilde{\beta}^2 F(r)-\tilde{\beta}_{\rm H}^2.
\end{equation}
    In the near-zone scaling, after factoring out the overall $n^2/r_h^2$, we obtain
\begin{equation}
	e^{-\rho}\frac{d}{d\rho}\Bigl(e^{\rho}F(\rho)\chi'(\rho)\Bigr)
	- \frac{r_h^2}{n^2}\bigl(\tilde{\beta}^2 F(\rho)-\tilde{\beta}_{\rm H}^2\bigr)\chi(\rho)
	- 2 F(\rho)\,\chi'(\rho)\,\phi_1'(\rho) = 0.
\end{equation}
    Using $\mathcal{L}_0[\chi_0]=0$, the NLO equation takes the form
\begin{equation}
	\mathcal{L}_0[\chi_1(\rho)] = S(\rho),
	\qquad
	S(\rho)
	= S_{\rm mass}(\rho)
	+ S_{\rm kin}(\rho)
	+ S_\phi(\rho)
	+ S_{\rm metric}(\rho),
\end{equation}
    where
\begin{itemize}
	\item $S_{\rm mass}$ comes from replacing the constant-curvature
	approximation of the mass term by the full $F_0(\rho)$;
	\item $S_{\rm kin}$ encodes the analogous correction in the kinetic term
	of the radial operator;
	\item $S_\phi$ arises from the NLO dilaton coupling
	$-2\,\partial\phi\cdot\partial\chi$ and involves the corrected
	$\phi_1'(\rho)$ in~\eqref{eq:phi1-prime-NZ-correct};
	\item $S_{\rm metric}$ encodes the explicit dependence on the NLO metric
	correction $F_1(\rho)$.
\end{itemize}
    All these pieces can be written as finite linear combinations of the basis functions
\begin{equation}
	\frac{e^{(p_j-1)\rho}}{1-e^{-\rho}},
\end{equation}
    with a finite set of exponents $p_j$ and coefficients determined algebraically by the LO and NLO profiles.

\paragraph{Integral basis and near-zone solution}
    Inserting the source into~\eqref{eq:chi1-GF-final}, every term in $S(\sigma)$ of the form
\begin{equation}
	S(\sigma)\supset \mathcal{C}_j\,
	\frac{e^{(p_j-1)\sigma}}{1-e^{-\sigma}}
\end{equation}
    contributes a piece proportional to
\begin{equation}
	I(p_j-\lambda_-;\rho)
	\equiv
	\int_\rho^\infty d\sigma\,
	\frac{e^{(p_j-\lambda_- -1)\sigma}}{1-e^{-\sigma}}.
\end{equation}
    It is therefore natural to define
\begin{equation}
	I(p;\rho)
	\equiv
	\int_\rho^\infty d\sigma\,
	\frac{e^{(p-1)\sigma}}{1-e^{-\sigma}},
\end{equation}
    so that the generic contribution is $I(p_j-\lambda_-;\rho)$. Performing the change of variables $t=e^{-\sigma}$, one finds
\begin{equation}
	I(p;\rho)
	= \int_0^{e^{-\rho}}dt\,\frac{t^{-p}}{1-t},
\end{equation}
    with the special-function representation
\begin{equation}
	I(p;\rho)
	=
	\frac{e^{-(1-p)\rho}}{1-p}\,
	{}_2F_1\bigl(1,1-p;2-p;e^{-\rho}\bigr),
\end{equation}
    in terms of the Gauss hypergeometric function ${}_2F_1$. For $\Re p<1$, $I(p;\rho)\to 0$ as $\rho\to\infty$, in agreement with its definition.

    For the set of exponents relevant here (coming from the mass, kinetic, dilaton, and metric contributions), the NLO thermal scalar can be written in the compact form
\begin{equation}
	\boxed{
		\chi_1^{\rm (NZ)}(\rho)
		=
		\frac{e^{\lambda_-\rho}}{\lambda_- - \lambda_+}
		\left[
		C_0\,I(0;\rho)
		+ C_1\,I(2\lambda_-;\rho)
		+ C_2\,I(2\lambda_- -1;\rho)
		+ \sum_j D_j\,I(p_j;\rho)
		\right],
	}
\end{equation}
    where $C_0,C_1,C_2$ encode the combined contributions of the mass, kinetic and dilaton induced sources and the coefficients $D_j$ encode the metric effects. Their explicit dependence on $(\lambda_-,\Xi^2,\kappa,\chi_h,\mathcal{A}_\chi,\ldots)$ can be obtained algebraically; the structure of the solution in terms of the universal integrals $I(p;\rho)$ is model-independent. Including the LO piece, the near-zone thermal scalar up to NLO is
\begin{equation}
	\chi^{\rm (NZ)}(\rho)
	=
	\chi_h e^{\lambda_-\rho}
	+ \frac{1}{n}\,\chi_1^{\rm (NZ)}(\rho)
	+ \mathcal{O}\!\left(\frac{1}{n^2}\right).
\end{equation}

\paragraph{Far-zone NLO thermal scalar and matching}
    In the far-zone we work in the coordinate $X$ defined by
\begin{equation}
	X \equiv \frac{\sqrt{K}}{r_h}\,(r-r_h),
	\qquad
	X=0 \ \Leftrightarrow\ r=r_h,
\end{equation}
	This is a convenient formal origin only; the EFT domain begins at $X=X_{\rm c}>0$ corresponding to $r=r_{\rm c}$. The radial metric is approximately flat, $ds_{\rm rad}^2\simeq (r_h^2/K^2)\,dX^2$. At strict LO one finds
\begin{equation}
	\left(\frac{d^2}{dX^2} - m_{\rm eff}^2\right)\chi_0(X) = 0,
	\qquad
	\chi_0^{\rm (FZ)}(X)
	= \hat{\chi}_h\,e^{-m_{\rm eff} X},
\end{equation}
	Matching relates $\hat{\chi}_h$ to $\chi_h$ but leaves their common overall scale undetermined.

    At NLO, the full thermal scalar equation leads to
\begin{equation}
	\left(\frac{d^2}{dX^2} - m_{\rm eff}^2\right)\chi_1(X)
	= S_\chi^{\rm (FZ)}(X),
\end{equation}
    where the source $S_\chi^{\rm (FZ)}(X)$ is linear in the NLO fields $(\phi_1^{\rm (FZ)},F_1^{\rm (FZ)})$ and quadratic in $\chi_h$. Since $\chi_0(X)\propto e^{-m_{\rm eff}X}$, and both $\phi_1^{\rm (FZ)}$ and $F_1^{\rm (FZ)}$ are linear combinations of decaying exponentials, the source can be decomposed as
\begin{equation}
	S_\chi^{\rm (FZ)}(X)
	=
	\sum_i s_i\,e^{-q_i X},
	\qquad
	q_i>0,
\end{equation}
    with coefficients $s_i$ determined by the near-zone data. For a single term $S(X)=s\,e^{-qX}$ with $q\neq m_{\rm eff}$, a particular
solution is
\begin{equation}
	\chi_1^{\rm (p)}(X)
	= \frac{s}{q^2 - m_{\rm eff}^2}\,e^{-qX},
\end{equation}
    while the homogeneous solution is
\begin{equation}
	\chi_1^{\rm (hom)}(X)
	= A_1\,e^{+m_{\rm eff} X} + B_1\,e^{-m_{\rm eff}X}.
\end{equation}
    Normalisability as $X\to\infty$ requires $A_1=0$, so the full far-zone NLO solution is
\begin{equation}
	\boxed{
		\chi_1^{\rm (FZ)}(X)
		=
		B_1\,e^{-m_{\rm eff}X}
		+ \sum_i \frac{s_i}{q_i^2 - m_{\rm eff}^2}\,e^{-q_i X},
	}
\end{equation}
    with $B_1$ fixed by matching to the near-zone. The only subtlety arises if $q_i=m_{\rm eff}$ for some $i$, in which case a particular solution $X e^{-m_{\rm eff}X}$ appears; this non-generic resonant case can be treated separately but does not occur in the generic parameter regime considered here.

    In the overlap region the near- and far-zone descriptions agree once we relate the radial coordinates through their large-$D$ scaling: we have $Y\simeq \rho$ as above, while $X\propto \rho/n$ near the horizon, so the precise map between $(\rho,Y)$ and $X$ is not needed beyond this scaling information. Matching relates the NLO near- and far-zone coefficients at fixed background and stretched horizon data. The relations retain dependence on the undetermined normalisation $\chi_h$, or equivalently $\chi_{\rm c}$, and do not determine the core contribution to the action.

\paragraph{Summary of the NLO thermal scalar}
    At fixed matching data, the matched exterior thermal scalar profile can be summarised schematically as
\begin{equation}
	\boxed{
		\chi(\rho)
		=
		\chi_h e^{\lambda_-\rho}
		+ \frac{1}{n}\,
		\chi_1(\rho)
		+ \mathcal{O}\!\left(\frac{1}{n^2}\right),
	}
\end{equation}
    with $\chi_1(\rho)$ given in the near-zone by the Green's function representation~\eqref{eq:chi1-GF-final} and the integral basis $I(p;\rho)$, and in the far-zone by a sum of decaying exponentials as in the boxed expression above. The NLO backreaction thus induces a small renormalisation of the leading decay amplitude along the Euclidean cigar, together with a tower of more strongly suppressed tails controlled by the NLO dilaton and metric scales and the large-$D$ structure of the BD background.

\subsubsection{Cutoff-regulated Landau functional and core matching} \label{subsubsec:Feff-LO-NLO}
    The discussion below is a conditional organization of the exterior calculation.  The explicit radial integrals use the same single-scale profile adopted in the preceding NLO subsection, but they do not by themselves determine the complete physical Landau coefficients.  A stretched horizon cutoff and a local core matching functional are required.
    We define the thermal free energy as the generator of connected vacuum diagrams in the Euclidean ensemble. Starting from the Euclidean path integral with period $\beta$,
\begin{equation}
    Z(\beta)\;\equiv\;\int_{\substack{\text{fields periodic in } \tau\\ \tau\sim\tau+\beta}}
    \mathcal{D}g\,\mathcal{D}\phi\,\mathcal{D}\chi\;
    \exp\!\big[-S_E[g,\phi,\chi]\big],
    \label{eq:Z_def}
\end{equation}
    the free energy is
\begin{equation}
    F(\beta)\;\equiv\;-\frac{1}{\beta}\,\log Z(\beta).
    \label{eq:F_def}
\end{equation}
    Equivalently, introducing the Euclidean effective action $\Gamma_E$ via the Legendre transform of the generating functional, the free energy is the on-shell value of $\Gamma_E$ per unit Euclidean time,
\begin{equation}
    F(\beta)\;=\;\frac{1}{\beta}\,\Gamma_E\big[\bar g,\bar\phi,\bar\chi\big],
    \qquad
    \frac{\delta \Gamma_E}{\delta g_{\mu\nu}}=\frac{\delta \Gamma_E}{\delta \phi}
    =\frac{\delta \Gamma_E}{\delta \chi}=0,
    \label{eq:F_as_Gamma}
\end{equation}
    where $(\bar g,\bar\phi,\bar\chi)$ denotes the saddle (or mean field) appropriate to the ensemble.

\paragraph{Dimensionally reduced functional for the thermal scalar}
    Upon integrating out all non-zero Matsubara modes and restricting to the thermal scalar zero mode $\chi_0(\vec x)$, one obtains a static $(D-1)$-dimensional functional.  Spherical reduction then gives the radial one-dimensional functional used below:
\begin{equation}
    Z(\beta)\;=\;\exp\!\big[-\beta\,\mathcal{F}_0\big]\,
    \int \mathcal{D}\chi_0\;
    \exp\!\Big[-\beta\,\mathcal{F}[\chi_0]\Big],
    \label{eq:Z_to_Feff}
\end{equation}
    so that $\mathcal{F}[\chi_0]$ is the effective free-energy functional governing the static mode. In a saddle point (mean-field) treatment one may equivalently write
\begin{equation}
    F(\beta)\;=\;\min_{\chi_0}\,\mathcal{F}[\chi_0]
    \qquad
    (\text{at fixed } \beta \text{ and boundary data}),
    \label{eq:F_min}
\end{equation}
    with the minimum attained at the solution of $\delta \mathcal{F}/\delta \chi_0=0$ subject to the relevant boundary conditions. Collecting the results of the previous subsections, the dimensionally reduced free energy functional~\eqref{eq:Z_to_Feff} evaluated on the thermal scalar zero mode can be written as
\begin{equation}
    \mathcal{F}[\chi_0]
    =
    \mathcal{F}_0
    + \mathcal{F}^{(2)}
    + \mathcal{F}^{(4)}
    + \mathcal{O}(|\chi_h|^6),
\end{equation}
    where $\mathcal{F}_0$ is the $\chi$-independent background contribution.  Both $\mathcal{F}^{(2)}$ and $\mathcal{F}^{(4)}$ contain an exterior contribution from $\rho\geq\rho_{\rm c}$ and a local matching contribution supplied by the unresolved string-scale core.  The exterior calculation alone does not determine the complete coefficients.

\paragraph{Quadratic piece at LO}
    At leading order in $1/D$ the one-dimensional Lagrangian is,
\begin{equation}
    \mathcal{L}_{\rm 1D}^{(0)}(\rho)
    =
    e^{\rho}
    \left[
    F_0(\rho)\,|\chi_0'(\rho)|^2
    + \frac{r_h^2}{n^2}
    \bigl(\tilde{\beta}^2 F_0(\rho) - \tilde{\beta}_{\rm H}^2\bigr)
    |\chi_0(\rho)|^2
    \right],
    \qquad
    F_0(\rho) = K\bigl(1-e^{-\rho}\bigr),
\end{equation}
    and in the plateau approximation $F_0\simeq K$ we have
\begin{equation}
    \mathcal{L}_{\rm 1D}^{(0)}(\rho)
    \simeq
    e^{\rho}
    K\left[
    |\chi_0'(\rho)|^2 + \Xi^2 |\chi_0(\rho)|^2
    \right],
    \qquad
    \Xi^2
    =
    \frac{r_h^2}{(D-2)^2K}\,\bigl(\tilde{\beta}^2 K - \tilde{\beta}_{\rm H}^2\bigr).
\end{equation}
    Projecting onto the normalisable zero mode $\chi_0(\rho)=\chi_h e^{\lambda_-\rho}$ with
\begin{equation}
    \lambda_-
    =
    \frac{-1-\sqrt{1+4\Xi^2}}{2},
    \qquad
    \lambda_-^2 + \lambda_- - \Xi^2 = 0,
\end{equation}
    gives the cutoff-regulated exterior contribution
\begin{align}
    \mathcal{F}^{(2)}_{\rm ext}
    &=
    \mathcal{N}_D K\bigl(\lambda_-^2+\Xi^2\bigr)
    \int_{\rho_{\rm c}}^{\infty}
    d\rho\,e^{(1+2\lambda_-)\rho}|\chi_h|^2
    \nonumber\\
    &=
    \mathcal{N}_D K
    \frac{\lambda_-^2+\Xi^2}{-(1+2\lambda_-)}
    e^{(1+2\lambda_-)\rho_{\rm c}}|\chi_h|^2,
    \label{eq:F2-exact-again}
\end{align}
    where $1+2\lambda_-=-\sqrt{1+4\Xi^2}<0$ on the decaying branch.  This formula applies only when the cutoff lies in the plateau region in which $F_0\simeq K$; otherwise the exact near-zone profile must be retained in the radial integral.

    It is useful to express the result in terms of the field at the matching surface,
\begin{equation}
    \chi_{\rm c}\equiv\chi_0(\rho_{\rm c})
    =\chi_h e^{\lambda_-\rho_{\rm c}}.
\end{equation}
    Using $\Xi^2=\lambda_-^2+\lambda_-$, the same exterior term becomes
\begin{equation}
    \mathcal{F}^{(2)}_{\rm ext}
    =\mathcal{N}_D K(-\lambda_-)
      e^{\rho_{\rm c}}|\chi_{\rm c}|^2.
\end{equation}
    This equality is also obtained directly by integrating the quadratic action by parts and using the scalar equation: the on-shell bulk contribution reduces to the boundary term at $\rho_{\rm c}$.  It is therefore a matching-surface quantity, not a complete free energy of the unresolved cigar.

    Choose any fixed positive normalization $Z_{\rm c}$ of the cutoff amplitude,
\begin{equation}
    u\equiv\sqrt{Z_{\rm c}}\,\chi_{\rm c},
    \qquad Z_{\rm c}>0,
\end{equation}
    and write
\begin{equation}
    \mathcal{F}^{(2)}_{\rm ext}
    =a_2^{\rm ext}(\Xi^2;\rho_{\rm c})|u|^2,
    \qquad
    a_2^{\rm ext}
    =\frac{\mathcal{N}_D K}{Z_{\rm c}}
      (-\lambda_-)e^{\rho_{\rm c}}.
\end{equation}
    The physical coefficient contains the local core contribution,
\begin{equation}
    a_2(\Xi^2;\rho_{\rm c})
    =a_2^{\rm core}(\Xi^2;\rho_{\rm c})
     +a_2^{\rm ext}(\Xi^2;\rho_{\rm c}).
\end{equation}
    Cutoff independence requires the two pieces to run oppositely,
\begin{equation}
    \frac{d}{d\rho_{\rm c}}
    \bigl(a_2^{\rm core}+a_2^{\rm ext}\bigr)=0,
\end{equation}
    after the same ensemble and amplitude normalization have been chosen on both sides.  The marginality condition $a_2(0;\rho_{\rm c})=0$ is consequently a matching condition on $a_2^{\rm core}$; it is not an additive shift of the $u$-independent background term $\mathcal{F}_0$.  Since the core functional is unknown, the exterior calculation alone fixes neither the complete normalization nor the temperature derivative of $a_2$.

\paragraph{Formal quartic coefficient from exterior response}
    At tree level, an exterior quartic term is obtained by solving the linear metric--dilaton response sourced at order $|u|^2$ and substituting that response back into the action.  The NLO profiles displayed above organize representative source and falloff structures.  They do not, however, constitute the complete off-shell two-function gravitational response with ensemble boundary terms.  Consequently they are insufficient to determine a physical quartic coefficient by themselves.

    For a fixed cutoff, fixed boundary data, and fixed normalization of $u$, the exterior contribution is defined formally by
\begin{equation}
    \mathcal{F}^{(4)}_{\rm ext}
    =a_4^{\rm ext}(\Xi^2;\rho_{\rm c})|u|^4,
    \qquad
    a_4^{\rm ext}
    \equiv
    \left.
    \frac{1}{\beta|u|^4}
    S_E^{\rho\geq\rho_{\rm c}}
    \right|_{\text{on shell at order }|u|^4},
    \label{eq:F4-chi-form}
\end{equation}
    where $S_E^{\rho\geq\rho_{\rm c}}$ must include the gravitational boundary terms appropriate to the selected thermodynamic ensemble.  The response kernels generated by the exponential NLO profiles reduce the radial calculation to finite linear combinations of cutoff integrals.

    Writing $z_{\rm c}=e^{-\rho_{\rm c}}<1$, a useful basis is
\begin{align}
    I(p;\rho_{\rm c})
    &\equiv
    \int_0^{z_{\rm c}}dt\,\frac{t^{-p}}{1-t},
    \qquad \Re p<1,
    \\
    J_{\rm ext}(a;\rho_{\rm c})
    &\equiv
    \int_{\rho_{\rm c}}^\infty d\sigma\,
    \frac{e^{a\sigma}-1}{e^\sigma-1},
    \qquad \Re a<1.
    \label{eq:J-master-again}
\end{align}
    The first integral has the regular finite-cutoff representation
\begin{equation}
    I(p;\rho_{\rm c})
    =\frac{z_{\rm c}^{,1-p}}{1-p}
      {}_2F_1\!\left(1,1-p;2-p;z_{\rm c}\right),
    \qquad \Re p<1,
\end{equation}
    while changing variables in the second gives
\begin{align}
    J_{\rm ext}(a;\rho_{\rm c})
    &=\int_0^{z_{\rm c}}dt\,
      \frac{t^{-a}-1}{1-t}
      \nonumber\\
    &=I(a;\rho_{\rm c})+\log(1-z_{\rm c}).
\end{align}
    The subtraction in $J_{\rm ext}$ cancels the apparent $t=1$ endpoint singularity when the cutoff is taken to the formal tip; for $\Re a<1$ one recovers
\begin{equation}
    J_{\rm ext}(a;0)=\psi(1)-\psi(1-a).
\end{equation}
    By contrast, the unsubtracted expression
\begin{equation}
    I(p;0)=\int_0^1dt\,\frac{t^{-p}}{1-t}
\end{equation}
    diverges logarithmically at $t=1$.  It cannot be assigned a finite value merely by writing ${}_2F_1$ at unit argument.  Any cancellation of such endpoint pieces must occur in the complete response, including core and boundary contributions, rather than term by term in an exterior basis.

    The full quartic coefficient therefore has the matching decomposition
\begin{equation}
    a_4(\Xi^2;\rho_{\rm c})
    =a_4^{\rm core}(\Xi^2;\rho_{\rm c})
     +a_4^{\rm ext}(\Xi^2;\rho_{\rm c}),
    \label{eq:a4-exact-final}
\end{equation}
    with cutoff independence requiring
\begin{equation}
    \frac{d}{d\rho_{\rm c}}
    \bigl(a_4^{\rm core}+a_4^{\rm ext}\bigr)=0.
\end{equation}
    A rescaling $u\mapsto c\,u$ also rescales $a_2\mapsto a_2/c^2$ and $a_4\mapsto a_4/c^4$, so a comparison of magnitudes requires one common normalization convention.  More importantly, the sign of the complete $a_4$ is not fixed by positivity of the scalar radial integrals: integrating out the gravitational response and adding the core and boundary functionals can contribute with either sign.

    The present calculation does not supply $a_4^{\rm core}$, the complete two-function response kernel, or all ensemble boundary terms.  We therefore do not claim an exact value, a definite sign, or a Gauss--Bonnet parameter scan for $a_4$.

\paragraph{Cutoff-dependent Landau form}
    Without the core matching functional, the result is the structural expansion
\begin{equation}
    \boxed{
    \mathcal{F}_{\rm eff}[u]
    =
    \mathcal{F}_0
    +\bigl(a_2^{\rm core}+a_2^{\rm ext}\bigr)|u|^2
    +\bigl(a_4^{\rm core}+a_4^{\rm ext}\bigr)|u|^4
    +\mathcal{O}(|u|^6),
    }
\end{equation}
    where every term is evaluated at the same cutoff, with the same normalization of $u$, and in the same ensemble.  The full coefficients are physical only after the cutoff dependence cancels between the core and exterior pieces.
\begin{figure}[H]
    \centering
    \includegraphics[width=\textwidth]{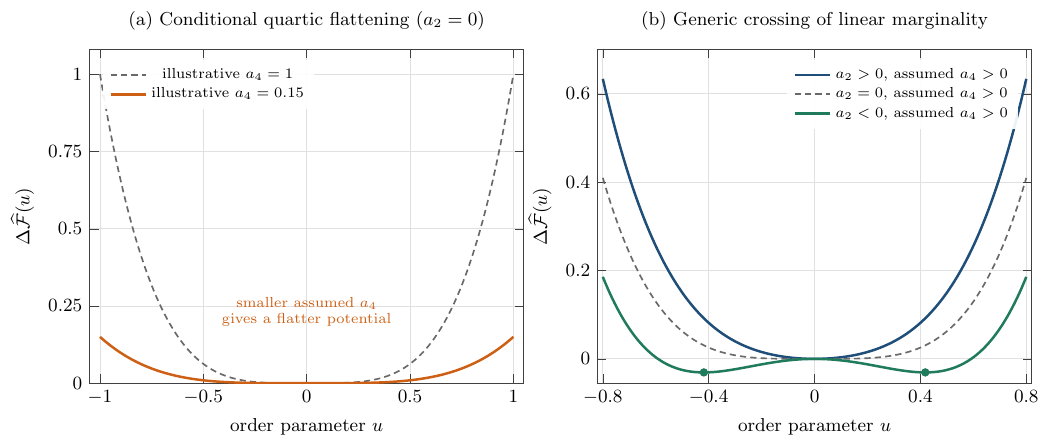}
    \caption{Parameter-independent Landau schematic, not model data.  Panel~(a) shows that, at $a_2=0$ and assuming $a_4>0$, a smaller illustrative quartic coefficient flattens the potential near the origin.  Panel~(b) shows the usual displacement of the minimum as an illustrative quadratic coefficient crosses zero at fixed positive $a_4$.}
    \label{fig:a4_lambdaGB}
\end{figure}
    For orientation only, a generic even Landau expansion can be written as
\begin{equation}
    \mathcal{F}(\Phi)
    =
    a_2\,\Phi^2
    +
    a_4\,\Phi^4
    +a_6\,\Phi^6
    +\cdots,
    \label{eq:LandauExpansion_a4}
\end{equation}
    Here $a_2$ controls linear marginality and $a_4$ is the first nonlinear coefficient if the symmetry excludes odd powers.  Figure~\ref{fig:a4_lambdaGB} illustrates generic shapes under an assumed positive $a_4$.  Standard Landau theory distinguishes an ordinary continuous transition from tricritical behavior by the simultaneous conditions $a_2=a_4=0$ with a positive stabilizing $a_6$~\cite{ChaikinLubensky1995,Goldenfeld1992}.  The phrase ``quasi-tricritical'' is used in other settings for crossover behavior near, but not at, such conditions~\cite{DrozdRzoska2020QuasiTricritical,AnisimovAgayanGorodetskii2000JETPL}.

\subsection{Relation to FZZ duality and limits of the Landau analogy} \label{subsec:FZZ-quasi-tricritical}
    The Euclidean two-dimensional black hole described by the $\mathrm{SL}(2,\mathbb{R})_k/\mathrm{U}(1)$ cigar is related to sine--Liouville theory by FZZ duality.  In that language the winding operator appears in an interaction of the schematic form
\begin{equation}
    V_{\rm SL}(\mathcal{T})
    \;\sim\;
    \mu\,e^{-Q\phi}\cos(Rx).
\end{equation}
    Its expansion motivates an order-parameter description of winding condensation.  This structural analogy does not identify the coefficients of the spacetime cutoff functional in Sec.~\ref{subsubsec:Feff-LO-NLO}.  In particular, expanding the sine--Liouville interaction does not derive either $a_4^{\rm core}$ or $a_4^{\rm ext}$ in the present EGB truncation.

    In our setup the analogous variable is the exterior thermal scalar amplitude $u\propto\chi_{\rm c}$.  Once a matching prescription is chosen, the off-shell functional may be organized as
\begin{equation}
    \mathcal{F}_{\rm eff}[u]
    =
    \mathcal{F}_0
    +a_2|u|^2
    +a_4|u|^4
    +\mathcal{O}(|u|^6),
\end{equation}
    with $a_i=a_i^{\rm core}+a_i^{\rm ext}$.  The quadratic coefficient diagnoses linear marginality only after the core matching condition has been imposed.  The quartic coefficient additionally depends on nonlinear core data, the complete metric--dilaton response, the ensemble boundary terms, and the normalization of $u$.

\paragraph{What the analogy does and does not establish}
    The FZZ description supports the expectation that winding condensation and geometric cap data are coupled.  It also explains why a local polynomial expansion can be useful near a marginal configuration.  It does not imply that the bulk EGB coupling alone fixes the quartic term, nor that a coefficient computed in the worldsheet normalization equals the coefficient of the cutoff amplitude used here.  A quantitative identification would require a common choice of background, ensemble, operator normalization, and matching surface.

    The same caution applies to the shape of an illustrative Landau potential.  Assuming $a_4>0$ produces the familiar continuous displacement of the minimum when $a_2$ crosses zero, but this conditional statement is not evidence that the complete coefficient in the present model is positive.  Likewise, a flatter schematic curve does not establish that the theory realizes a small-$a_4$ regime.

\paragraph{No tricritical claim}
    We have not computed a parameter scan of the complete $a_4$.  We therefore cannot infer its sign, a minimum as a function of the Gauss--Bonnet coupling, or a ``quasi-tricritical window.''  An actual tricritical point would require a common parameter value at which the matched coefficients satisfy $a_2=a_4=0$, together with a computed positive sextic coefficient.  None of these conditions is established by the exterior calculation.

\paragraph{Requirements for a quantitative FZZ comparison}
    A future comparison should first derive the core matching functional in a specified string background and combine it with the exterior response at one cutoff.  The result must then be shown to be cutoff independent and translated to a common normalization of the winding operator.  Only after these steps could one test whether a small or vanishing quartic coefficient is physically realized or whether higher-order interactions change the transition.

    Effects beyond the present approximation---including higher-derivative operators beyond Gauss--Bonnet, string-loop corrections, and higher orders in $1/D$---may alter both core and exterior terms.  They should therefore be incorporated at the level of the full matched functional rather than interpreted as a correction to an already known $a_4$.  Sharpening the map between the HP$+$GB variables and the couplings of sine--Liouville theory remains an interesting open problem.

\section{Numerical evaluation in finite dimensions} \label{sec:numsol}
    We numerically integrate the radial equations of the action~\eqref{eq:action} for $h(\rho)$, $g(\rho)$, $\phi(\rho)$, and $\chi(\rho)$ at finite $D$ in the unit radial gauge~\eqref{eq:unit_radial_gauge}, retaining the dilaton--Gauss--Bonnet contribution~\eqref{eq:phiGB-extra}. The independent equations and the radial constraint are specified below. Appendix~\ref{app:fullnum} records the evolution equations in the large-$D$ metric truncation, and Appendix~\ref{app:subnum} presents a reduced subsystem.

\subsection{Numerical Implementation and results}
    In order to obtain the second-order equations of motion, we adopt the $D$-dimensional Euclidean metric in the unit radial gauge which makes the cylindrical cap geometry more pronounced,
\begin{equation}
    \mathrm{d}s^2 = h(\rho)^2 \mathrm{d}\tau^2 + \mathrm{d}\rho^2 + g(\rho) \mathrm{d}\Omega_{D-2}^2 .
    \label{eq:unit_radial_gauge}
\end{equation}
     With the metric~\eqref{eq:unit_radial_gauge}, the four unknown functions are $h$, $g$, $\phi$, and $\chi$. A convenient independent system uses the $\tau\tau$ and angular metric equations and the thermal scalar equation, while the $\rho\rho$ constraint selects the branch of $\phi'$. The dilaton equation is monitored as the dependent Bianchi consistency equation. We define the trace reduced metric equation as
\begin{align}
    \mathcal{E}_{\mu}{}^\nu &\equiv R_{\mu}{}^\nu + \alpha \left( H_{\mu}{}^\nu + \Delta^{(\phi {\rm GB})}{}_{\mu}{}^{\nu} + \frac{1}{2} R_{\rm GB}^2 \delta_{\mu}{}^{\nu} \right) + 2 \nabla_\mu \nabla^\nu \phi \nonumber
    \\
    & \qquad - \kappa \left( 2 \nabla_\mu \chi \nabla^\nu \chi - 2 \tilde{\beta} \chi^2 g_{\tau\mu} g_{\tau}{}^\nu \right) = 0,
    \label{eq:reduced_eom_metric}
    \\
    \Psi &\equiv R - 2 \Lambda + \alpha R_\mathrm{GB}^2 + 4 \nabla^2 \phi - 4 \left( \partial \phi \right)^2 - 2 \kappa \left[ \left( \partial \chi \right)^2 + m_\mathrm{th} \chi^2 \right] = 0,
    \\
    \Xi &\equiv \nabla^2 \chi - 2 \partial^\rho \chi \partial_\rho \phi - m_\mathrm{th}^2 \chi = 0,
\end{align}
    where $\tilde{\beta}\equiv\beta/(2\pi\alpha')$ and the plots use string units $\alpha'=1$. A linear combination of the equation of motion for $\phi$ and the $\rho\rho$ component, $\mathcal{E}_{\rho}{}^{\rho}$, yields a constraint. The resulting constraint, $\mathcal{E}_{\rho}{}^{\rho} - \Psi/2 = 0$, is
\begin{align}
    0 &= 2 \left( \phi' \right)^2 - \left( \frac{2 h'}{h} + \frac{\left( D - 2 \right) g'}{g} \right) \phi' + \left( D - 2 \right) \left[ \frac{h' g'}{2 h g} + \frac{\left( D - 3 \right) \left( \left( g' \right)^2 - 4 g \right)}{8 g^2} \right] + \Lambda \nonumber
    \\
    & \quad - \kappa \left[ \left( \chi' \right)^2 - \left( \tilde{\beta}^2 h^2 - \tilde{\beta}_\mathrm{H}^2 \right) \chi^2 \right] + \alpha \left( D - 4 \right) \left( D - 3 \right) \left( D - 2 \right) \left\{ \frac{4 g - (g')^2}{32 h g^4} \right. \nonumber
    \\
    & \qquad \left. \times \left[ 8 g h' g' - \left( D - 5 \right) h \left( 4 g - (g')^2 \right) \right] - \left[ \frac{h' \left( 4 g - 3 (g')^2 \right)}{\left( D - 4 \right) h g^2} + \frac{g' \left( 4 g - (g')^2 \right)}{2 g^3} \right] \phi' \right\},
    \label{eq:constraint}
\end{align}
    where $\tilde{\beta}_\mathrm{H}\equiv\beta_\mathrm{H}/(2\pi\alpha')$ and $\tilde\beta_{\rm H}=\sqrt2$ in the string units used numerically. We now impose the formal regularity conditions at the coordinate tip,
\begin{equation}
    h(0) \equiv 0, \qquad g(0) > 0.
    \label{eq:hg_initial}
\end{equation}
    To extract the appropriate initial values, we introduce a small radial displacement~$\epsilon$ from the tip, and expand all equations to leading order in $\epsilon$. For the equation of motion for $\chi(\rho)$, the expansion leads to
\begin{equation}
    \chi''(\rho) = - \frac{\chi'(0)}{\epsilon} + \mathcal{O}(1).
    \label{eq:chi_initial}
\end{equation}
    Since the expression~\eqref{eq:chi_initial} diverges when $\epsilon = 0$, consistency of the expansion requires
\begin{equation}
    \chi'(0) = 0.
\end{equation}
    Expanding the equation of motion for $g(\rho)$ near $\rho=0$ yields an expression for $g''(\rho)$ containing a term proportional to $\alpha^{-1}$. This expression diverges as $\alpha\to0$ for fixed $g(0) > 0$. Setting $g(0) = 0$ to remove this term is incompatible with the condition~\eqref{eq:hg_initial}. This divergence does not, however, imply an inconsistency in the initial assumptions at $\alpha=0$. Rather, it shows that taking the $\alpha \to 0$ limit after performing the initial expansion at nonzero $\alpha$ is not the appropriate procedure. Hence, to determine the initial behavior in the $\alpha \to 0$ limit, we first set $\alpha = 0$ in the equation of motion for $g(\rho)$ and only then expand the resulting equation in the small parameter $\epsilon$. Expanding the equation of motion for $g(\rho)$ with $\alpha = 0$, we obtain
\begin{equation}
    \left. g''(\rho) \right|_{\alpha = 0} = - \frac{g'(0)}{\epsilon} + \mathcal{O}(1).
\end{equation}
    To remove the divergence of $g''(\rho)$ in the $\alpha = 0$ case, it is necessary to impose
\begin{equation}
    g'(0) = 0.
    \label{eq:g'_initial}
\end{equation}
    Although the condition~\eqref{eq:g'_initial} is obtained from the equation of motion with $\alpha = 0$, a family of initial values that remains regular as $\alpha \to 0$ should satisfy the same condition. We thus impose $g'(0) = 0$ also for nonzero $\alpha$. Applying the same expansion procedure to the constraint equation~\eqref{eq:constraint} yields
\begin{equation}
    0 = \frac{4 \phi'(0)}{\epsilon} + \mathcal{O}(1).
\end{equation}
    To eliminate the divergence, we impose
\begin{equation}
    \phi'(0) = 0.
\end{equation}
    The initial values can then be summarized as follows:
\begin{equation}
    h(0) = g'(0) = \phi'(0) = \chi'(0) = 0, \qquad g(0) > 0.
\end{equation}
    These conditions remove the singular terms in the formal near-tip expansion. The finite terms impose additional relations on the Taylor coefficients.

    For numerical initialisation, we evaluate the formal regular Taylor expansions at a small positive coordinate value $\rho=\epsilon$:
\begin{equation}
    \begin{aligned}
        h(\rho) &= h_1 \rho + \mathcal{O}(\rho^3),
        &
        \qquad g(\rho) &= g_0 + g_2 \rho^2 + \mathcal{O}(\rho^4),
        \\
        \phi(\rho) &= \phi_0 + \phi_2 \rho^2 + \mathcal{O}(\rho^4),
        &
        \qquad \chi(\rho) &= \chi_0 + \chi_2 \rho^2 + \mathcal{O}(\rho^4).
    \end{aligned}
\end{equation}
    Here $\epsilon$ is a numerical initialisation parameter. It does not by itself define the physical stretched horizon cutoff $\rho_{\rm c}$ or supply the microscopic core matching data. An EFT interpretation is restricted to $\rho\geq\rho_{\rm c}$ and requires the conditions~\eqref{eq:EFT-validity}. Although the geometry is regular at the tip, the vanishing of $h$ makes a direct numerical initialization at $\rho = 0$ ill-conditioned. We thus initialize the numerical integration at a small but finite radial cutoff, $\rho = \epsilon$, using the regular near-tip expansions
\begin{align}
    & h(\rho) \sim h_1 \rho, \quad g(\rho) \sim g_0 + g_2 \rho^2, \quad \phi(\rho) \sim \phi_0 + \phi_2 \rho^2, \quad \chi(\rho) \sim \chi_0 + \chi_2 \rho^2,
    \\
    & h'(\rho) \sim h_1, \quad g'(\rho) \sim 2 g_2 \rho, \qquad\;\;\; \phi'(\rho) \sim 2 \phi_2 \rho, \qquad\quad \chi'(\rho) \sim 2 \chi_2 \rho.
\end{align}
    Substituting the near-tip expansions into the constraint equation and equation of motion for $\chi$, we obtain at leading order, respectively,
\begin{align}
    0 &= \Lambda - 4 \phi_2 - \frac{\left( D - 2 \right) \left( D - 3 - 2 g_2 \right)}{2 g_0} - \kappa \tilde{\beta}_\mathrm{H}^2 \chi_0^2 \nonumber
    \\
    & \qquad - \frac{\alpha \left( D - 3 \right) \left( D - 2 \right) \left[ \left( D - 4 \right) \left( D - 5 - 4 g_2 \right) + 16 g_0 \phi_2 \right]}{2 g_0^2},
    \\
    0 &= \tilde{\beta}_\mathrm{H}^2 \chi_0 + 4 \chi_2.
\end{align}
    These relations constrain $g_0$ and $\chi_2$ for specified values of the remaining local data. To state the independent conditions at the outer boundary, recall that $\phi$ is the full $D$-dimensional dilaton. The asymptotic string coupling $g_{s,\infty}=e^{\phi_\infty}$ fixes the additive dilaton normalisation. A finite nonzero $\phi_\infty$ is admissible when it is compatible with the coupled background and the assumed weak coupling regime. For a constant dilaton endpoint, the required limits as $\rho\to\infty$ include
\begin{equation*}
    \phi - \phi_\infty \to 0, \qquad \left( \nabla \phi \right)^2 = \left( \phi' \right)^2 \to 0, \qquad \nabla_{\hat a} \nabla_{\hat b} \phi \to 0.
\end{equation*}
    Here primes denote proper radial derivatives and hatted indices denote orthonormal components. A slope that tends to zero can still yield an unbounded $\phi$ if its radial integral diverges.

\paragraph{Coupled AdS asymptotics}
    Consider a string frame AdS endpoint of the coupled equations, with curvature radius $L\equiv\ell_{\rm eff}$ and falloffs
\begin{equation}
    h(\rho) \sim h_\infty e^{\rho / L}, \qquad g(\rho) \sim g_\infty e^{2\rho/L}, \qquad \phi(\rho) \to \phi_\infty, \qquad \chi(\rho) \to 0.
    \label{eq:ads-falloffs}
\end{equation}
    When the dilaton derivatives and all winding sources vanish, Eqs.~\eqref{eq:EinsteinEq}~and~\eqref{eq:DilatonEOM} reduce, respectively, to
\begin{align*}
    0 &= \Lambda + \frac{\left( D - 2 \right) \left( D - 1 \right)}{2 L^2} - \frac{\alpha \left( D - 4 \right) \left( D - 3 \right) \left( D - 2 \right) \left( D - 1 \right)}{2 L^4},
    \\
    0 &= - 2 \Lambda - \frac{\left(D - 1 \right) D}{L^2} + \frac{\alpha \left( D - 3 \right) \left( D - 2 \right) \left( D - 1 \right) D}{L^4}.
\end{align*}
    For $D\geq4$ and $\alpha>0$, these equations require
\begin{equation}
    \alpha = \frac{L^2}{2 \left( D - 3 \right) \left( D - 2 \right)}, \qquad \Lambda = - \frac{\left( D - 1 \right) D}{4 L^2}.
    \label{eq:asymptotic_conditions}
\end{equation}
    The balance fixes $\alpha ( D - 3 ) ( D - 2 ) / L^2 = 1/2$. At $\alpha = 0$, the same pair of equations instead requires
\begin{equation}
    \frac{D-1}{L^2} = - \frac{2 \Lambda}{D} = 0.
    \label{eq:asymptotic_einstein_limit}
\end{equation}
    Thus a finite radius AdS endpoint with constant dilaton and vanishing winding sources requires a positive Gauss--Bonnet coupling. A global winding solution would also require interior matching data and a solution of the coupled boundary value problem with the prescribed far boundary conditions.

\paragraph{Reduced dilaton equation on a prescribed AdS metric}
    To isolate a source of dilaton running, we consider the form of the radial relation~\eqref{eq:reduced_eom_phi} obtained by setting $\Delta^{(\phi{\rm GB})}_{\mu\nu} = 0$ in the metric equation. Define
\begin{gather*}
    m = D - 2, \qquad C_D = \left( m - 2 \right) \left( m - 1 \right) m, \qquad a = \frac{\tilde\beta_{\rm H}^{2}}{\tilde\beta},
    \\
    r = \sqrt{g}, \qquad u = \frac{r'}{r}, \qquad w = \frac{1 - ( r' )^2}{r^2}.
\end{gather*}
    Using $\chi' = -\tilde\beta h \chi$ and $h' / h = \phi' - mu / 2 + a / ( 2 h )$ to eliminate $\chi'$ and $h'$ from the $\rho\rho$ component of Eq.~\eqref{eq:EinsteinEq} gives
\begin{align}
    0 &= \left( \frac{a}{h} - 2 \alpha C_D u w \right) \phi' + \frac{m}{2} u^2 + \frac{\left( m - 1 \right) m}{2 r^2} - \frac{m a u}{2 h} \nonumber
    \\
    & \qquad + \kappa \tilde\beta_{\rm H}^{2} \chi^2 - \Lambda + \alpha C_D w \left[ \frac{m + 3}{2} u^2 + \frac{m - 3}{2 r^2} - \frac{a u}{h} \right].
    \label{eq:reduced-dilaton}
\end{align}
    We now evaluate Eq.~\eqref{eq:reduced-dilaton} in isolation on a prescribed metric with the $h$ and $g$ falloffs of Eq.~\eqref{eq:ads-falloffs}, leaving $\phi(\rho)$ free. Assume relative metric corrections of order $r^{-2}$ with the corresponding derivative falloffs and quadratic winding sources of order $h^{-2}$ or smaller. Then $u\to1/L$ and $w \to - 1 / L^2$. Keeping the constant and $1/h$ terms gives
\begin{align}
    \left( \frac{a}{h} + \frac{2 \alpha C_D}{L^3} \right) \phi' &\simeq K_\alpha + \frac{a}{h} \left( \frac{m}{2 L} - \frac{\alpha C_D}{L^3} \right), \nonumber
    \\
    K_\alpha &= \Lambda - \frac{m}{2 L^2} + \frac{\alpha C_D \left( m + 3 \right)}{2 L^4}.
    \label{eq:outer-balance}
\end{align}
    For fixed $\alpha>0$ and $D\geq5$, the large-radius slope is
\begin{equation}
    \phi' \to q_\alpha = \frac{L^3}{2 \alpha C_D} \left( \Lambda - \frac{m}{2 L^2} \right) + \frac{m + 3}{4 L}.
\end{equation}
    When $q_\alpha\ne 0$, $\phi = q_\alpha \rho + \mathcal{O}(1)$, corresponding to $\phi = q_\alpha L \log( r / \sqrt{g_\infty} ) + \mathcal{O}(1)$ in the areal radius. A finite limit of the reduced relation requires the cancellation
\begin{equation}
    K_\alpha = 0 \quad \Longleftrightarrow \quad \Lambda = \frac{m}{2 L^2} - \frac{\alpha C_D \left( m + 3 \right)}{2 L^4}.
\end{equation}
    At this cancellation, $\phi = \phi_\infty + \mathcal{O}(e^{- \rho / L})$. For $\alpha = 0$, Eq.~\eqref{eq:outer-balance} gives the leading term
\begin{equation}
    \phi \sim \frac{L h_\infty}{a} \left( \Lambda - \frac{m}{2 L^2} \right) e^{\rho/L},
\end{equation}
    which runs downward for $\Lambda < 0$. An expansion about $\alpha = 0$ requires $\epsilon_\rho = 2 \alpha C_D h / ( a L^3 ) \ll 1$. For fixed positive $\alpha$ and $D \geq 5$, the exponential growth of $h$ eventually takes $\epsilon_\rho$ outside this regime. An EFT interpretation of the formal tails requires the bounds~\eqref{eq:EFT-validity} and control of the omitted metric term.

    The cancellation $K_\alpha=0$ characterises the isolated reduced relation on the prescribed metric. Evaluating $K_\alpha$ at the coupled AdS vacuum~\eqref{eq:asymptotic_conditions} gives $K_\alpha=-D/L^2$. For any field $q\in\{h,g,\phi,\chi\}$, the coefficients $q_i$ appearing in the tables, such as $g_0$ and $g_2$, denote the corresponding coefficients in the near-tip Taylor expansion about the formal coordinate origin $\rho = 0$, which are used as shooting seeds for the numerical integration. These tabulated values were chosen to display representative integrations; they are not all fixed by the local equations.  In a genuine boundary value problem the unconstrained metric and dilaton data are selected by the asymptotic conditions, while the thermal scalar normalization remains arbitrary at quadratic order and must be supplied at the stretched horizon or fixed by nonlinear thermodynamic data.  The numerical curves are interpreted only for $\rho\ge\rho_{\rm c}$, outside the string-scale core. In the numerical analysis, we set $\alpha' = 1$ and absorb $\tilde{\beta}$ and $\sqrt{\kappa}$ into $h$ and $\chi$, respectively, through the rescalings $h \to \widetilde{\beta}h$ and $\chi \to \sqrt{\kappa}\chi$.
\begin{figure}[!htbp]
    \centering
    \includegraphics[width=7.4cm]{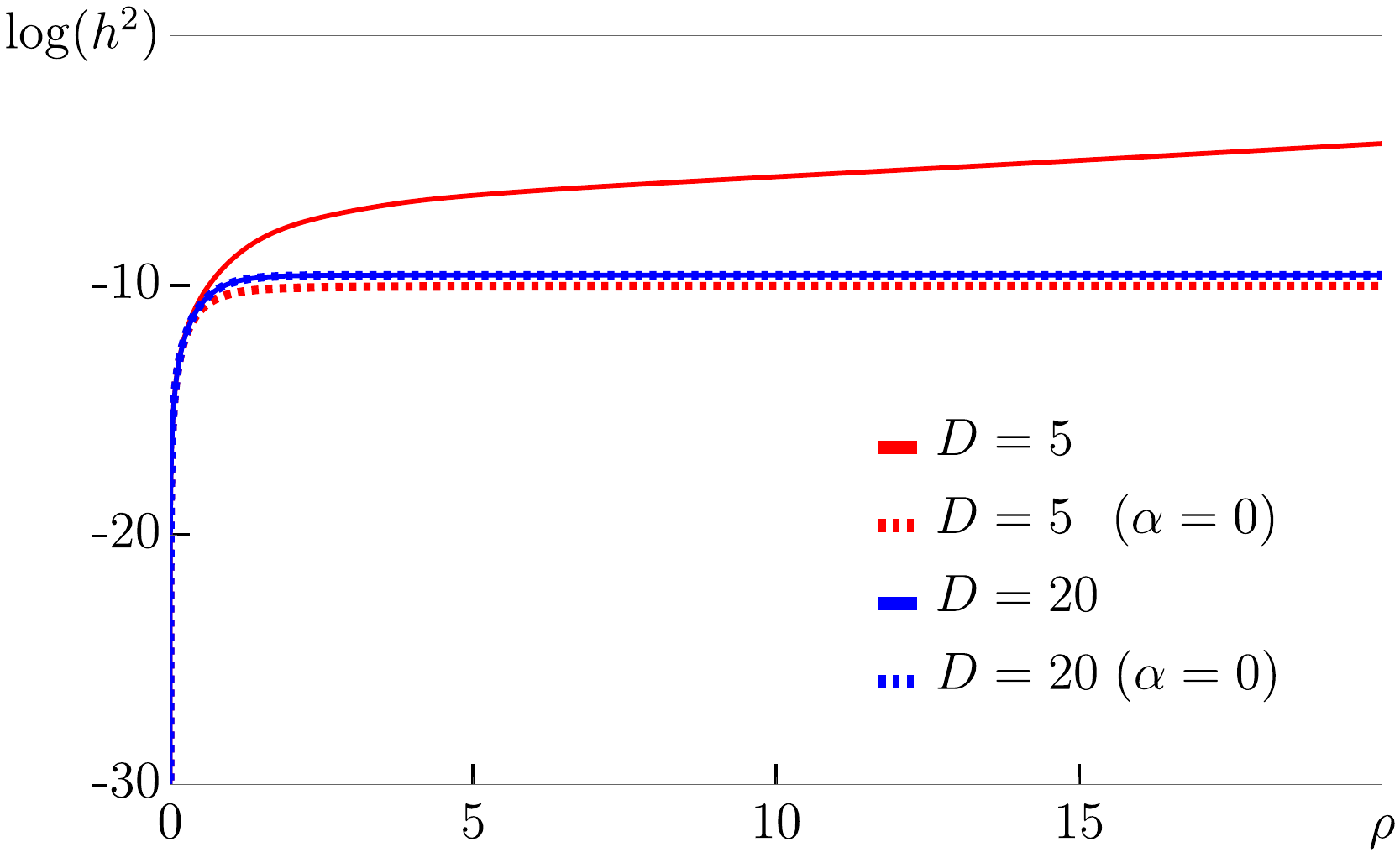}
    \includegraphics[width=7.4cm]{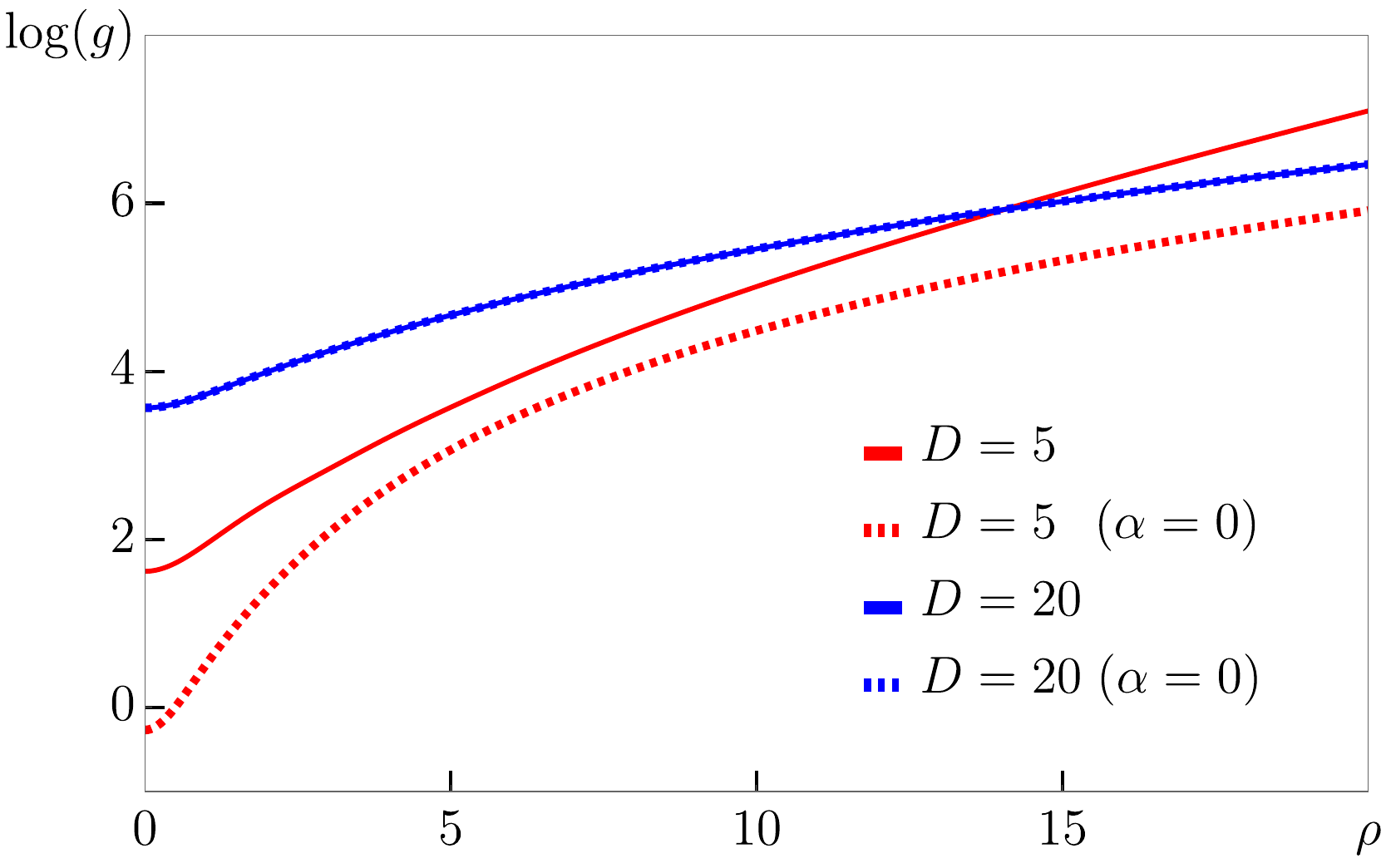}
    \includegraphics[width=7.4cm]{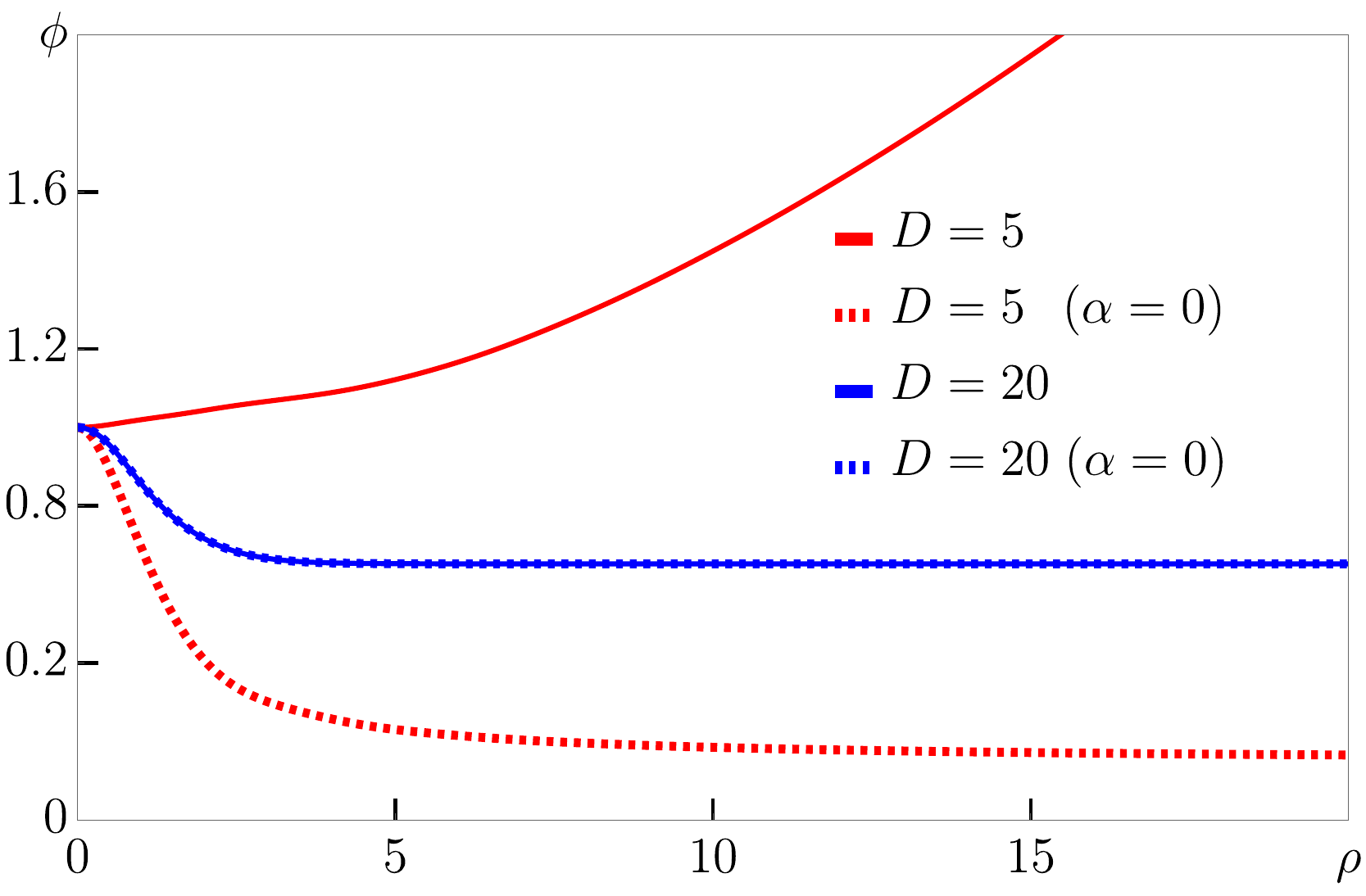}
    \includegraphics[width=7.4cm]{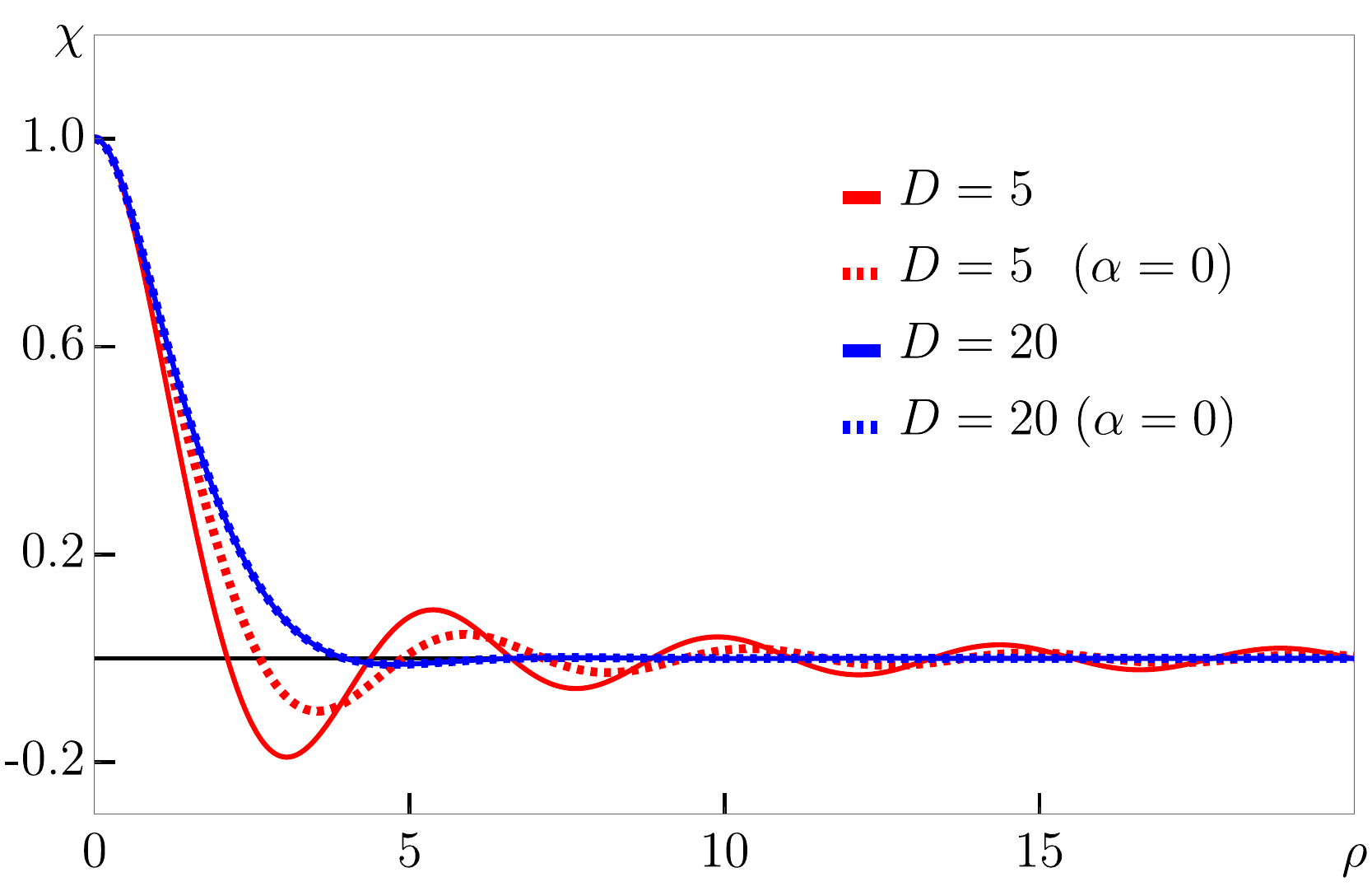}
    \caption{Finite-interval integrations of the radial equations for $\Lambda = 0$, with local data listed in Table~\ref{table:second_order_flat}. No prescribed asymptotic dilaton value has been imposed. Some dilaton profiles flatten over the displayed interval, whereas the $D=5$ profile with nonzero $\alpha$ continues to increase. The plots do not establish asymptotic convergence. Only $\rho\geq0$ is shown; an EFT interpretation requires $\rho\geq\rho_{\rm c}$ and the validity conditions~\eqref{eq:EFT-validity}.}
    \label{fig:second_order_flat}
\end{figure}
\begin{table}[!htbp]
    \centering \renewcommand{\arraystretch}{1.5}
    \resizebox{\textwidth}{!}{%
    \begin{tabular}{|c|c|c|c|c|c|c|c|c|c|c|} \hline
    Parameters & $\ell_\textrm{eff}$ & $\alpha$ & $\Lambda$ & $h_1$ & $g_2$ & $\phi_0$ & $\phi_2$ & $\chi_0$
    \\ \hline
    $D = 5$ & \multirow{2}{*}{$5$} & \multirow{2}{*}{$\frac{\ell_\textrm{eff}^2}{2 ( D - 3 ) ( D - 2 )}$} & \multirow{2}{*}{$0$} & \multirow{2}{*}{$0.01$} & $1.5$ & \multirow{2}{*}{$1$} & \multirow{2}{*}{$-0.01$} & \multirow{2}{*}{$1$}
    \\ \cline{1-1} \cline{6-6}
    $D = 20$ & {} & {} & {} & {} & $9$ & {} & {} & {}
    \\ \hline
    \end{tabular}}
    \caption{Shooting seeds used in Fig.~\ref{fig:second_order_flat}; the dashed lines correspond to $\alpha = 0$.}
    \label{table:second_order_flat}
\end{table}
\begin{figure}[!htbp]
    \centering
    \includegraphics[width=7.4cm]{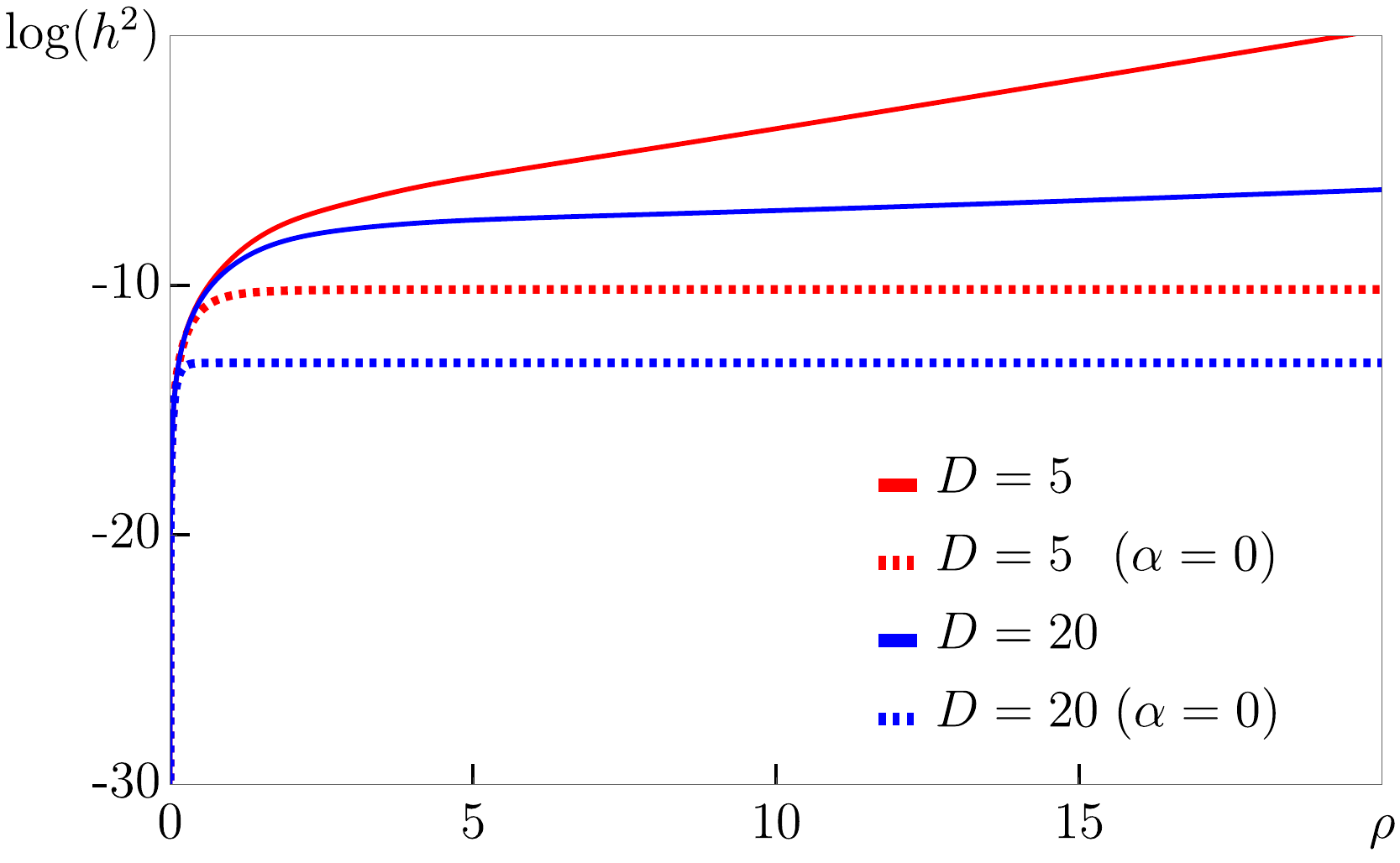}
    \includegraphics[width=7.4cm]{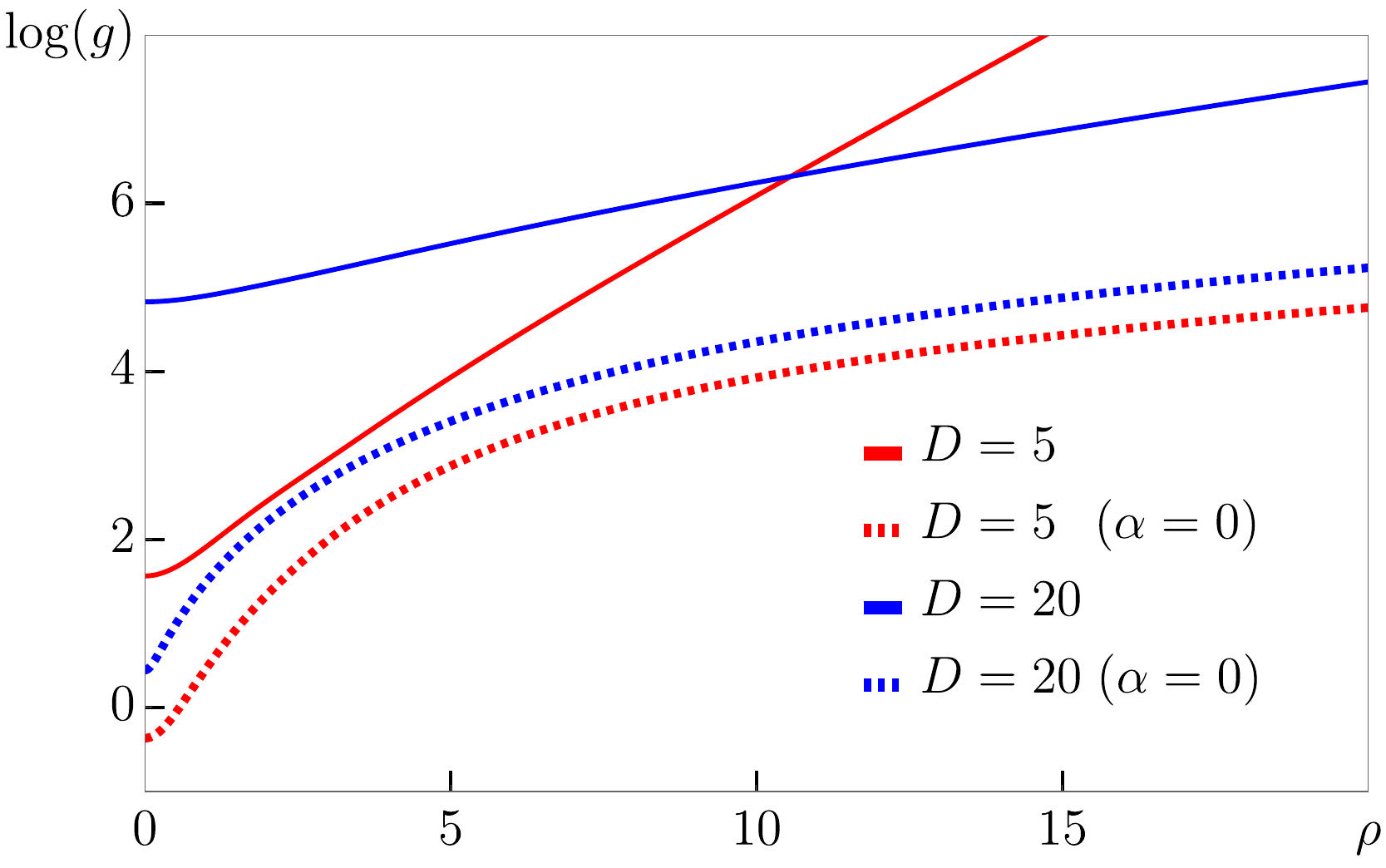}
    \includegraphics[width=7.4cm]{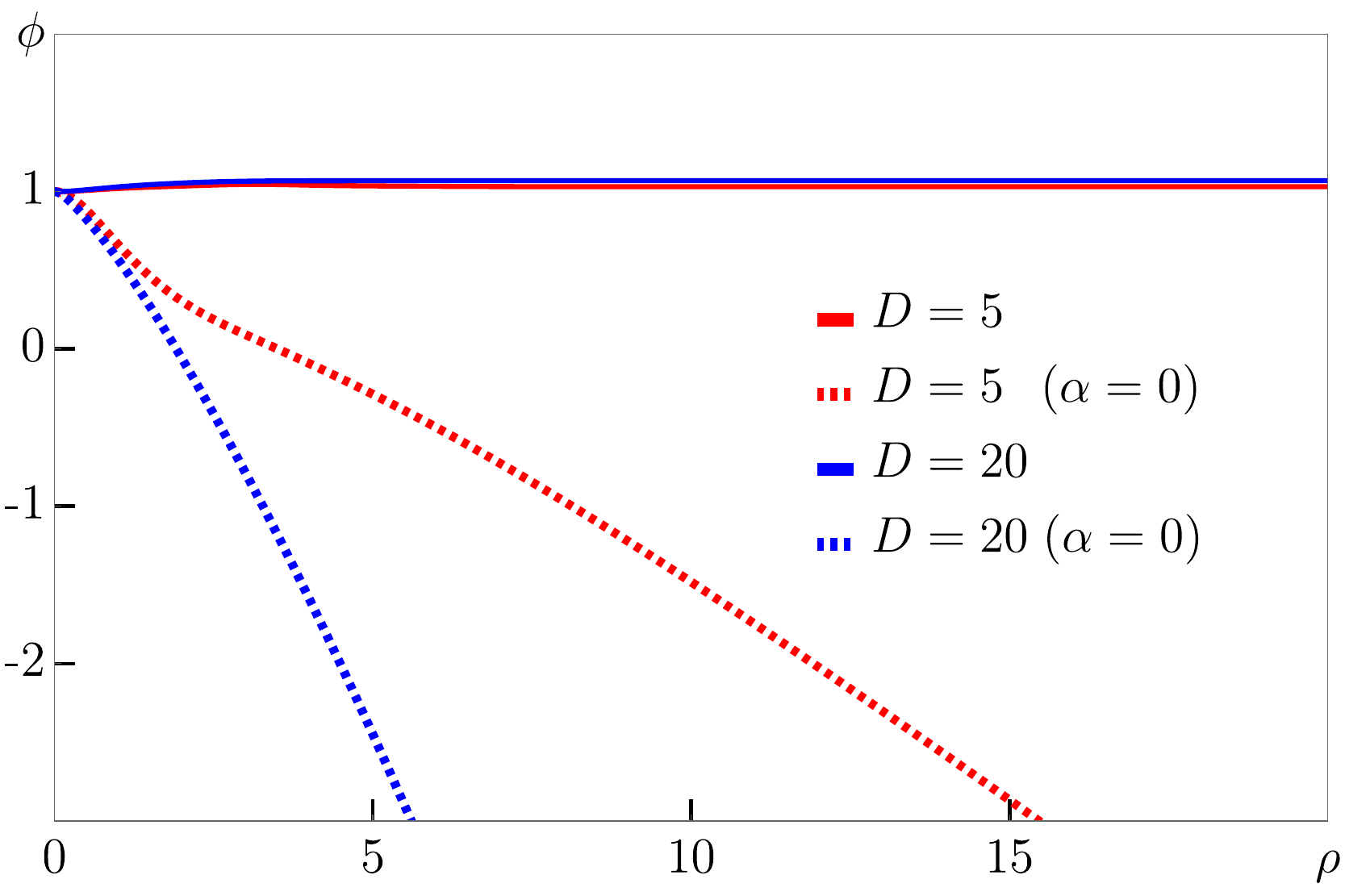}
    \includegraphics[width=7.4cm]{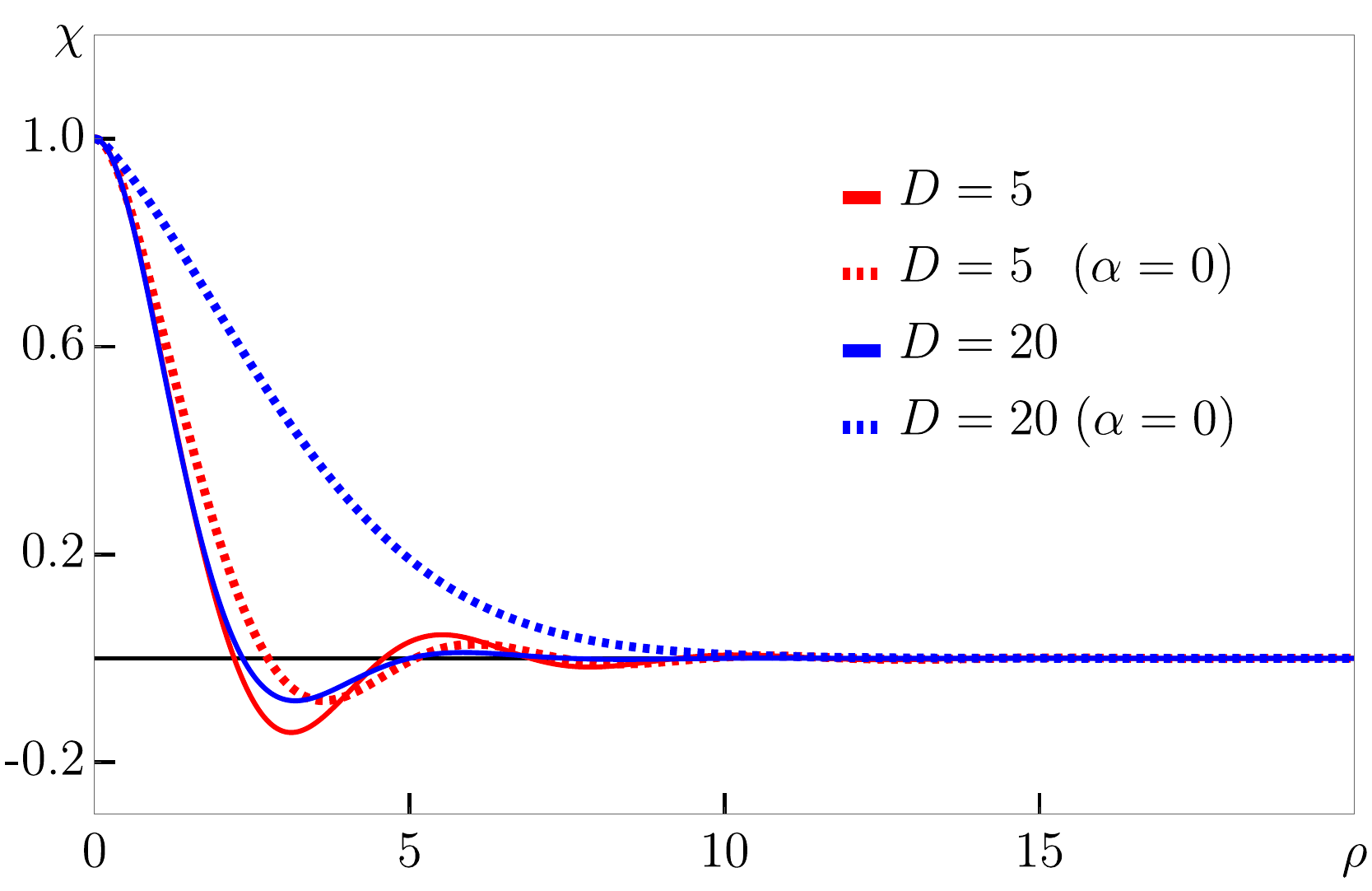}
    \caption{Finite-interval integrations of the full radial equations with $\Lambda<0$, with parameters and local data listed in Table~\ref{table:second_order_ads}. The downward running dilaton curves correspond to the $\alpha=0$ comparison cases, for which the finite-dilaton AdS asymptotic condition is excluded within the adopted ansatz by Eq.~\eqref{eq:asymptotic_einstein_limit}. The nonzero-$\alpha$ trajectories have not been tuned to prescribed asymptotic data. The plots therefore do not establish global asymptotically AdS solutions. Only $\rho\geq0$ is shown; an EFT interpretation requires $\rho\geq\rho_{\rm c}$ and the validity conditions~\eqref{eq:EFT-validity}.}
    \label{fig:second_order_ads}
\end{figure}
\begin{table}[!htbp]
    \centering \renewcommand{\arraystretch}{1.5}
    \resizebox{\textwidth}{!}{%
    \begin{tabular}{|c|c|c|c|c|c|c|c|c|c|c|} \hline
    Parameters & $\ell_\textrm{eff}$ & $\alpha$ & $\Lambda$ & $h_1$ & $g_2$ & $\phi_0$ & $\phi_2$ & $\chi_0$
    \\ \hline
    $D = 5$ & \multirow{2}{*}{$5$} & \multirow{2}{*}{$\frac{\ell_\textrm{eff}^2}{2 ( D - 3 ) ( D - 2 )}$} & \multirow{2}{*}{$- \frac{( D - 1 ) D}{4 \ell_\textrm{eff}^2}$} & \multirow{2}{*}{$0.01$} & $1.5$ & \multirow{2}{*}{$1$} & \multirow{2}{*}{$-0.01$} & \multirow{2}{*}{$1$}
    \\ \cline{1-1} \cline{6-6}
    $D = 20$ & {} & {} & {} & {} & $9$ & {} & {} & {}
    \\ \hline
    \end{tabular}}
    \caption{Shooting seeds used in Fig.~\ref{fig:second_order_ads}; the dashed lines correspond to $\alpha = 0$.}
    \label{table:second_order_ads}
\end{table}
    Figs.~\ref{fig:second_order_flat} and~\ref{fig:second_order_ads} show representative finite-$D$ integrations of the full radial equations. The tables list the shooting seeds, not independently determined horizon observables.  The robust comparison with the analytic calculation is qualitative: the exterior thermal scalar is localized toward the cap and decays outward.  We do not claim a coefficient-by-coefficient match to the $1/D$ expansion, an EFT prediction for $\chi(r_h)$, or a determination of the string-scale cap.

\paragraph{Geometric interpretation in unit radial gauge}
    To interpret the numerical profiles it is useful to recall the geometric meaning of the single ansatz~\eqref{eq:unit_radial_gauge}. We use $f(\rho)\equiv h(\rho)^2$ only as an abbreviation, not as a second metric ansatz. The Euclidean ``cigar'' consists of a smooth $(\rho,\tau)$ cap together with an $S^{D-2}$ whose effective cylinder radius is encoded in $g(\rho)=r(\rho)^2$. Regularity at the formal tip fixes the leading near-tip expansions
\begin{equation}
	f(\rho) = \kappa_\mathrm{E}^2 \rho^2 + f_4 \rho^4 + \cdots,
	\qquad
	g(\rho) = r_h^2 + g_2 \rho^2 + g_4 \rho^4 + \cdots,
\end{equation}
	so the departure from a constant-radius cylinder is governed by the coefficient $g_2$ (and higher Taylor data). In pure Einstein gravity, once $\kappa$ (equivalently $\beta$) is fixed by smoothness, the remaining radial equations relate the coefficients $\{f_4,g_2,\ldots\}$ in a comparatively simple way, and one typically finds that $g(\rho)$ remains nearly constant throughout the cap.
    
    With Gauss--Bonnet corrections, by contrast, the metric equations take the schematic form
\begin{equation}
	G_{\mu\nu} + \Lambda g_{\mu\nu} + \alpha\, H_{\mu\nu} = \cdots ,
\end{equation}
    where the Lanczos tensor $H_{\mu\nu}$ contains curvature-squared combinations. On a 	cigar$\times S^{D-2}$ background these contributions are not controlled solely by the small curvature of the $(\rho,\tau)$ plane near the tip: they receive sizable input from the intrinsic curvature of the transverse sphere, schematically $R_{S^{D-2}}\sim 1/r^2$ and $R_{S^{D-2}}^2\sim 1/r^4$. Consequently, even though the cap is locally flat at $\rho=0$, the Gauss--Bonnet term feeds into the radial constraints through the $S^{D-2}$ curvature and shifts the relations among $\{\kappa,r_h,f_4,g_2,\ldots\}$ by terms controlled by the dimensionless ratio $\alpha/r_h^2$ (and higher powers). In practice this renormalizes $g_2$, thereby changing how rapidly $g(\rho)$ departs from $r_h^2$: the transverse sphere radius can remain effectively ``cylindrical'' deeper into the cap or run more quickly away from its horizon value, depending on the sign and magnitude of
	$\alpha$. This is the precise sense in which the Gauss--Bonnet term modifies the cap-to-cylinder matching and hence the near-horizon background relevant for winding/thermal scalar physics.

\paragraph{Backreaction of the dilaton and winding condensate}
    The functions $g(\rho)$, the dilaton $\phi(\rho)$, and the winding condensate $\chi(\rho)$ are not independent: both $\phi$ and $\chi$ backreact on the geometry and therefore modify the effective cylinder radius encoded in $g(\rho)$. The exact metric equation is~\eqref{eq:EinsteinEq}. In addition to the two-derivative dilaton terms and the winding stress tensor~\eqref{eq:thermal-stress-exact}, it contains the dilaton--Gauss--Bonnet contribution~\eqref{eq:phiGB-extra}. In unit radial gauge the radial constraint is sourced by the local dilaton derivatives and by the localized matter associated with $\chi(\rho)$. A nontrivial dilaton profile therefore shifts the near-tip relations, while a nonzero winding condensate produces additional backreaction concentrated in the cap region where the thermal circle shrinks and the thermal scalar is lightest. As a result, the near-tip expansion
\begin{equation}
    f(\rho) = \kappa^2 \rho^2 + \cdots, \qquad     g(\rho) = r_h^2 + g_2 \rho^2 + \cdots
\end{equation}
    is renormalized by both $\phi$ and $\chi$, shifting the coefficient $g_2$ and hence the rate at which the geometry departs from a constant-radius cylinder. While the winding backreaction primarily reshapes the cap, the dilaton controls how this cap data is propagated and matched to the asymptotic region.

\paragraph{Overview and conventions}
    Figs.~\ref{fig:second_order_flat} and~\ref{fig:second_order_ads} show representative second-order integrations for $D=5$ (red) and $D=20$ (blue), with finite Gauss--Bonnet coupling (solid) and the Einstein limit (dashed).  The shooting seeds are listed in Tables~\ref{table:second_order_flat} and~\ref{table:second_order_ads}.  The proper radial coordinate increases outward, and only the physical half-line $\rho\ge0$ is shown.

\paragraph{Metric function $\log h(\rho)^2$ (top-left)}
    The upper-left panel plots $h(\rho)^2 = f(\rho)$.  The formal regular Taylor expansion is quadratic at the coordinate origin.  Outside the cutoff, the displayed profiles show the expected opening of the Euclidean circle and a modest deformation when $\widetilde\alpha$ is varied.  Their similarity at the two displayed values of $D$ is a qualitative observation, not a fitted large-$D$ coefficient.

\paragraph{Radial function $\log g(\rho)$ (top-right)}
    The upper-right panel displays $\log g(\rho)$, with $r(\rho) = \sqrt{g(\rho)}$ the transverse sphere radius. The formal expansion $g(\rho) = g_0 + g_2 \rho^2 + \cdots$ admits an approximately constant radius over an inner interval satisfying $|g_2| \rho^2 \ll g_0$, provided higher terms remain small. The plots do not establish suppression of the radial derivatives over the entire integration interval.

\paragraph{Dilaton profile $\phi(\rho)$}
    The dilaton profiles in Figs.~\ref{fig:second_order_flat}~and~\ref{fig:second_order_ads} were obtained by integrating the radial equations from the representative local data listed in Tables~\ref{table:second_order_flat} and~\ref{table:second_order_ads}. For $\Lambda=0$, the $D=5$ profile with nonzero $\alpha$ continues to rise toward the outer end of the displayed interval. For $\Lambda < 0$, the $\alpha = 0$ profiles decrease, whereas the nonzero-$\alpha$ profiles remain approximately constant over the displayed interval. The downward running in Fig.~\ref{fig:second_order_ads} is consistent with Eq.~\eqref{eq:asymptotic_einstein_limit}, which excludes an AdS endpoint with a finite constant dilaton and vanishing winding field at $\alpha = 0$, independently of the choice of shooting data. The nonzero-$\alpha$ parameter choices satisfy the necessary compatibility conditions in Eq.~\eqref{eq:asymptotic_conditions}. Establishing convergence to a prescribed $\phi_\infty$ requires solving the coupled boundary value problem with the appropriate interior matching data and the asymptotic conditions for the metric, dilaton, and winding field. On the portion of the exterior interval where the EFT validity conditions~\eqref{eq:EFT-validity} hold, the curves illustrate the local dilaton response.

\paragraph{Winding condensate $\chi(\rho)$}
    The winding profiles are concentrated toward the inner end of the integration interval, and the oscillation amplitudes decrease outward. The plotted normalisation is part of the chosen shooting data. The profiles do not determine $\chi(r_h)$ or provide a quantitative verification of the analytic $1/D$ coefficients. The small amplitude tails are not used to extract asymptotic observables.

\paragraph{}
    The numerical results therefore provide qualitative information about local solutions of the full radial equations. A global physical interpretation requires the boundary value calculation described above.

\section{Discussion} \label{sec:disc}
\begingroup
    In this work, we analysed the exterior Horowitz--Polchinski configuration in a heterotic-inspired EGB truncation. At fixed background and stretched horizon matching data, the large-$D$ calculation determines the exterior thermal scalar decay exponent and relates the near- and far-zone coefficients within the stated approximations. The finite-$D$ integrations provide local diagnostics of the truncated equations and do not establish a global physical solution. The condensate normalisation and the core contribution to the action remain undetermined. Consequently, the calculation does not determine the complete Landau coefficients, a free-energy comparison of the two phases, or the physical correspondence point.

\paragraph{Why large-\texorpdfstring{$D$}{D} simplifies the problem}
    The large-$D$ expansion separates the near and far radial regions and organises the exterior equations in powers of $1/D$. In the plateau approximation, the leading thermal scalar equation reduces to a constant-coefficient radial equation. This determines the exterior decay exponent and the Gauss--Bonnet dependence of the local marginality condition within the adopted truncation. A determination of the physical Hagedorn-temperature shift requires the omitted corrections to the thermal scalar operator and appropriate core matching data.

\paragraph{Higher-curvature corrections beyond Gauss--Bonnet}
    We focused on the Gauss--Bonnet density because it is the distinguished curvature-squared Lovelock combination with second-order gravitational equations in $D>4$.  It is not, by itself, the complete order-$\alpha'$ heterotic action: the antisymmetric tensor, gauge sector, torsionful connections, supersymmetric completion, and curvature corrections to the thermal scalar operator enter at the same derivative order~\cite{Gross:1986iv,Metsaev:1987zx,Bergshoeff:1989de,Chemissany:2007he}.  Type-II strings instead have leading purely gravitational terms of the schematic form $\alpha'^3R^4$~\cite{Gross:1986mw}.  No universal large-$D$ suppression of these omitted operators is assumed; their scaling must be checked term by term.  Extending the calculation therefore requires a consistent enlarged action, a new power counting analysis, and a recomputation of both the background and thermal scalar operator, rather than only adding an effective stress tensor~\cite{Chen:2021qrz}.

\paragraph{Rotating solutions and higher-form fields}
    The framework developed here is flexible and can be extended to include rotation and higher-form gauge fields. Rotating Euclidean black holes admit a natural generalisation of the cigar geometry in which the near-zone metric acquires an off-diagonal $g_{\tau\varphi}$ component and the thermal scalar couples to an effective chemical potential for winding. At large-$D$, the rotational contributions localise sharply near the horizon, so the universal near-zone reduction persists, though with a modified effective potential \cite{Urbach:2022xzw,Seitz:2025wpc}. Similarly, backgrounds supported by NS--NS or R--R fluxes modify the thermal scalar mass term and can induce new channels of instability. Incorporating these ingredients provides a clear path to studying HP-like correspondence transitions in non--supersymmetric flux compactifications, in rotating black branes, and in settings with nontrivial higher-form potentials such as AdS$_5\times$S$^5$~\cite{Brustein:2022uft,Santos:2024ycg}.

\paragraph{Mean-field description and the fuzzball proposal}
    Our results suggest a conditional mean-field organization of the HP correspondence based on the thermal scalar condensate; the complete Landau functional remains dependent on unresolved core matching data. A smooth thermodynamic crossover remains a possible interpretation motivated by the HP correspondence; the present calculation does not determine the nature or location of the transition. In contrast, the fuzzball programme proposes that black hole microstates are described not by a single Euclidean saddle but by a large ensemble of horizonless geometries. The present exterior analysis does not resolve individual microstates or determine the complete thermodynamics of the ensemble. In this light, the thermal scalar may be viewed as an effective order parameter for the coarse-grained distribution of microstate geometries, with the Euclidean black hole representing the universal saddle of the ensemble at large $D$. Understanding the precise relation between these mean-field and microstate perspectives remains an important open question, but the analytic control afforded by large~$D$ provides a promising arena in which to probe it.

    There are several promising directions for future work. Extending our analysis to ${\cal O}(\alpha'^3)$ $R^4$ corrections~\cite{Chen:2021qrz}, to rotating or charged backgrounds, and to black branes with nontrivial fluxes would significantly broaden the applications of the framework. It would also be interesting to determine whether higher-order terms in the large-$D$ and $\alpha'$ expansions can produce genuine multicritical behaviour, potentially realising higher-order generalisations of the HP transition. Finally, the connection between the thermal scalar mean-field description and microstate based approaches such as the fuzzball proposal deserves detailed investigation.

\endgroup

\begin{acknowledgments}
    The authors would like to thank Ryotaku Suzuki and Imtak Jeon for useful discussions and Wonwoo Lee for collaboration  at an early stage of this project. BHL is supported by the National Research Foundation of Korea (NRF) grant RS-2020-NR049598, and Overseas Visiting Fellow Program of Shanghai University. BHL thanks Asia Pacific Center for Theoretical Physics (APCTP) and Korea Institute for Advanced Study (KIAS) for the hospitality during his visit, where a part of this project was done. HL is supported by Basic Science Research Program through the National Research Foundation of Korea (NRF) funded by the Ministry of Education (NRF-2022R1I1A2063176) and the Dongguk University Research Fund of 2025. ST thanks Center for Quantum Spacetime (CQUeST), Sogang University for support for this work. S.T  would also like to thank IITBHU for hospitality where a preliminary version of this work was presented.
\end{acknowledgments}

\appendix
\section{Large-\texorpdfstring{$D$}{D} bookkeeping for the static Euclidean problem}
\label{app:oldresults}

This appendix collects the coordinate identities, overlap limits, and power
counting used in Sec.~\ref{sec:solld}.  Its purpose is to make explicit
which statements are kinematical consequences of the large-$D$ expansion and
which require the equations of motion.  We consider only the static Euclidean
problem studied in this paper.  In particular, no Lorentzian quasinormal mode
problem is introduced here.  The general structure of the near, far, and
overlap regions follows the standard large-$D$ construction
\cite{Emparan:2013moa,Emparan:2014aba,Emparan:2020inr}.

\subsection{Radial regions and useful variables}

We write
\begin{equation}
 D=n+3,\qquad
 \rho=n\log\!\left(\frac{r}{r_h}\right),\qquad
 r=r_h e^{\rho/n},\qquad
 \partial_r=\frac{n}{r}\,\partial_\rho .
 \label{eq:appA-rho-def}
\end{equation}
Close to the horizon,
\begin{equation}
 \frac{r-r_h}{r_h}
 =\frac{\rho}{n}+\frac{\rho^2}{2n^2}
 +O\!\left(\frac{\rho^3}{n^3}\right).
 \label{eq:appA-rho-expand}
\end{equation}
The three radial domains relevant for matched asymptotic expansion are
therefore
\begin{align}
 \text{near region:}\quad&
 r-r_h\ll r_h,\qquad \rho\ll n ,
 \nonumber\\
 \text{large-$D$ scaling layer:}\quad&
 r-r_h=O(r_h/n),\qquad \rho=O(1),
 \nonumber\\
 \text{far region:}\quad&
 r-r_h\gg r_h/n,\qquad \rho\gg1 .
 \label{eq:appA-regions}
\end{align}
The common overlap is
\begin{equation}
 \frac{r_h}{n}\ll r-r_h\ll r_h,
 \qquad 1\ll\rho\ll n .
 \label{eq:appA-overlap}
\end{equation}

Two exponential radial variables occur in the calculation and should not be
identified beyond leading order.  The variable naturally associated with the
geometrical near-zone expansion is
\begin{equation}
 \mathcal R=\left(\frac{r}{r_h}\right)^n=e^\rho ,
 \label{eq:appA-calR}
\end{equation}
whereas the far-zone scalar equation is most conveniently written using
\begin{equation}
 R=\left(\frac{r}{r_h}\right)^{D-2}
  =\left(\frac{r}{r_h}\right)^{n+1},
 \qquad
 Y\equiv\log R=\left(1+\frac1n\right)\rho .
 \label{eq:appA-R}
\end{equation}
Thus $R=e^\rho[1+O(\rho/n)]$ in the overlap, but $\mathcal R$ and $R$
produce different subleading terms.  Keeping this distinction prevents a
pure coordinate correction from being misidentified as an NLO dynamical
effect.

\subsection{Euclidean horizon and the static radial operator}

For a one-function Euclidean background,
\begin{equation}
 ds_E^2=F(r)d\tau^2+\frac{dr^2}{F(r)}
       +r^2d\Omega_{n+1,k}^2,
 \qquad F(r_h)=0,\qquad F'_h>0 ,
 \label{eq:appA-euclidean-metric}
\end{equation}
the near-horizon expansion is
\begin{equation}
 F(r)=F'_h(r-r_h)+O\!\left((r-r_h)^2\right).
 \label{eq:appA-F-expand}
\end{equation}
Introducing the proper radial distance
\begin{equation}
 s=2\sqrt{\frac{r-r_h}{F'_h}},
\end{equation}
the $(\tau,r)$ plane becomes
\begin{equation}
 ds^2_{(\tau,r)}
 =ds^2+\left(\frac{F'_h}{2}\right)^2s^2d\tau^2+O(s^4).
 \label{eq:appA-cigar}
\end{equation}
Regularity fixes the geometrical period to
\begin{equation}
 \beta=\frac{4\pi}{F'_h}.
 \label{eq:appA-period}
\end{equation}
The thermal scalar effective description used in the main text begins at a
stretched horizon cutoff, but the smooth Euclidean background still fixes its
period through Eq.~\eqref{eq:appA-period}.

For any static $s$-wave $X(r)$, Eq.~\eqref{eq:appA-euclidean-metric} gives
the exact identity
\begin{align}
 \nabla^2X
 &=\frac{1}{r^{n+1}}\frac{d}{dr}
   \left(r^{n+1}F\frac{dX}{dr}\right)
 \nonumber\\
 &=\frac{n^2}{r^2}
   \left[F X_{\rho\rho}+(F+F_\rho)X_\rho\right].
 \label{eq:appA-laplacian-rho}
\end{align}
Consequently, an equation of the form
\begin{equation}
 \nabla^2X-2\nabla_\mu\phi\,\nabla^\mu X-M^2(r)X=0
 \label{eq:appA-generic-scalar}
\end{equation}
is equivalent to
\begin{equation}
 F X_{\rho\rho}
 +\left(F+F_\rho-2F\phi_\rho\right)X_\rho
 -\frac{r^2}{n^2}M^2(r)X=0 .
 \label{eq:appA-scalar-rho}
\end{equation}
At leading order in the overlap, $\phi_\rho=0$, $F\to K$, and the
thermal scalar mass approaches
\begin{equation}
 m_{\rm eff}^2=\tilde{\beta}^2K-\tilde{\beta}_H^2.
\end{equation}
Using the definition employed in the main text,
\begin{equation}
 \Xi^2\equiv
 \frac{r_h^2\bigl(\tilde{\beta}^2K-\tilde{\beta}_H^2\bigr)}
      {(D-2)^2K}
 =\frac{r_h^2m_{\rm eff}^2}{(n+1)^2K},
 \label{eq:appA-Xi}
\end{equation}
Eq.~\eqref{eq:appA-scalar-rho} reduces to
\begin{equation}
 X_{\rho\rho}+X_\rho-\Xi^2X
 =O(e^{-\rho})+O\!\left(\frac{1+\rho}{n}\right).
 \label{eq:appA-near-overlap-eq}
\end{equation}
The replacement of $(n+1)^2$ by $n^2$ in an explicitly leading-order
equation changes only relative $O(1/n)$ terms.  The two leading exponents are
\begin{equation}
 \lambda_\pm=\frac{-1\pm\sqrt{1+4\Xi^2}}{2},
 \label{eq:appA-lambda}
\end{equation}
On the stable side, $\Xi^2\ge0$, decay away from the scaling layer selects
$\lambda_-$; the $\lambda_+$ solution is constant at $\Xi^2=0$ and grows
for $\Xi^2>0$.  This selection does not fix the overall scalar amplitude.

\subsection{Far-zone continuation and overlap matching}

In the far-zone plateau, where $F=K+O(1/n)$ and
$\phi=\phi_0+O(1/n)$, the variable $R$ in
Eq.~\eqref{eq:appA-R} gives
\begin{equation}
 \nabla^2X
 =\frac{(n+1)^2K}{r^2}
  \left[
   R^2X_{RR}+\frac{2n+1}{n+1}R X_R
  \right].
 \label{eq:appA-laplacian-R}
\end{equation}
Since $r=r_h[1+O(\rho/n)]$ in the overlap, this becomes
\begin{equation}
 \nabla^2X
 =\frac{(n+1)^2K}{r_h^2}
  \left[
   \frac{d}{dR}\left(R^2\frac{dX}{dR}\right)
   +O\!\left(\frac{1+\rho}{n}\right)
  \right].
 \label{eq:appA-laplacian-R-overlap}
\end{equation}
Here and below, the remainder denotes a relative correction to the
dimensionless differential operator acting on $X$.
The leading overlap equation is therefore
\begin{equation}
 \frac{d}{dR}\left(R^2\frac{dX}{dR}\right)
 -\Xi^2X
 =O(e^{-\rho})+O\!\left(\frac{1+\rho}{n}\right),
 \label{eq:appA-far-overlap-eq}
\end{equation}
with independent leading solutions $R^{\lambda_+}$ and
$R^{\lambda_-}$.  Equations
\eqref{eq:appA-near-overlap-eq} and
\eqref{eq:appA-far-overlap-eq} describe the same overlap behavior because
\begin{equation}
 R^\lambda=e^{\lambda\rho}
 \left[
  1+\frac{\lambda\rho}{n}
  +O\!\left(\frac{\rho^2}{n^2}\right)
 \right].
 \label{eq:appA-kinematic-NLO}
\end{equation}
The term proportional to $\lambda\rho/n$ is the kinematical conversion
between $Y$ and $\rho$; it must be included before comparing genuine NLO
corrections computed in the two coordinates.

Writing the two overlap expansions as
\begin{align}
 X_{\rm near}
 &=A_-e^{\lambda_-\rho}+A_+e^{\lambda_+\rho}+\cdots,
 \nonumber\\
 X_{\rm far}
 &=B_-R^{\lambda_-}+B_+R^{\lambda_+}+\cdots,
 \label{eq:appA-match}
\end{align}
decay sets $A_+=B_+=0$, and matching relates $A_-$ to $B_-$,
including the coordinate correction in
Eq.~\eqref{eq:appA-kinematic-NLO}.  The common normalization remains a
free parameter until a nonlinear solution or an additional normalization
condition is imposed.  Moreover, the overlap equation alone does not replace
the exact far-region boundary value problem: the asymptotic boundary
condition must be imposed on the full far equation once the plateau
approximation ceases to apply.

\subsection{NLO matching without double counting}

The distinction between $\rho$ and $Y=\log R$ becomes essential when the
subleading fields are compared.  Consider one characteristic branch with a
near-zone overlap expansion
\begin{equation}
 X_{\rm near}
 =e^{\lambda\rho}
  \left[
   A_0+\frac{1}{n}\bigl(A_1+a_1\rho\bigr)
   +O(n^{-2})
  \right],
 \label{eq:appA-NLO-near}
\end{equation}
and a far-zone expansion written in its natural variable,
\begin{equation}
 X_{\rm far}
 =R^\lambda
  \left[
   B_0+\frac{1}{n}\bigl(B_1+b_1\log R\bigr)
   +O(n^{-2})
  \right].
 \label{eq:appA-NLO-far}
\end{equation}
Using $Y=(1+1/n)\rho$, the latter becomes
\begin{equation}
 X_{\rm far}
 =e^{\lambda\rho}
  \left[
   B_0+\frac{1}{n}
   \left\{B_1+\bigl(b_1+\lambda B_0\bigr)\rho\right\}
   +O\!\left(\frac{1+\rho^2}{n^2}\right)
  \right].
 \label{eq:appA-NLO-converted}
\end{equation}
Coefficient matching therefore requires
\begin{equation}
 A_0=B_0,\qquad
 A_1=B_1,\qquad
 a_1=b_1+\lambda B_0.
 \label{eq:appA-NLO-match-rule}
\end{equation}
The last term in Eq.~\eqref{eq:appA-NLO-match-rule} is fixed entirely by
the coordinate conversion.  Omitting it would produce a spurious
disagreement between two representations of the same solution.  Conversely,
adding it both to the converted far solution and to the dynamical NLO source
would double count it.  This rule applies branch by branch; if the growing
branch is excluded at leading order, its NLO homogeneous coefficient must
also be excluded unless an inhomogeneous source genuinely regenerates that
behavior.  The remaining constant $A_1=B_1$ can be shifted by an NLO
redefinition of the overall amplitude, so only normalization-independent
combinations provide unambiguous matching tests.

\subsection{Map to unit radial gauge and comparison with numerics}

A change of radial coordinate can put the analytic and numerical metrics in
the same gauge, but it cannot remove a physical difference between their
independent metric functions.  Begin with the general static areal radius
form
\begin{equation}
 ds_E^2=A(r)d\tau^2+\frac{dr^2}{B(r)}
       +r^2d\Omega_{n+1,k}^2 .
 \label{eq:appA-two-function-metric}
\end{equation}
Define proper radial distance $u$ by
\begin{equation}
 du=\frac{dr}{\sqrt{B(r)}}.
 \label{eq:appA-proper-map}
\end{equation}
The metric then takes the unit radial gauge form used in
Sec.~\ref{sec:numsol},
\begin{equation}
 ds_E^2=h(u)^2d\tau^2+du^2+g(u)d\Omega_{n+1,k}^2,
 \qquad
 h(u)^2=A(r(u)),\qquad g(u)=r(u)^2 .
 \label{eq:appA-unit-map}
\end{equation}
The inverse reconstruction is
\begin{equation}
 r(u)=\sqrt{g(u)},\qquad
 A(r(u))=h(u)^2,\qquad
 B(r(u))=\left(\frac{dr}{du}\right)^2
         =\frac{g'(u)^2}{4g(u)} .
 \label{eq:appA-inverse-unit-map}
\end{equation}
Thus a one-function metric, $A=B=F$, obeys
\begin{equation}
 g'(u)=2h(u)\sqrt{g(u)}
 \label{eq:appA-one-function-test}
\end{equation}
when the Euclidean time normalization is the same in both gauges.  A radial
coordinate choice can impose either $g_{uu}=1$ or $g_{\Omega\Omega}=r^2$,
but it cannot in general impose the additional dynamical relation $A=B$.
With matter backreaction the reconstructed functions in
Eq.~\eqref{eq:appA-inverse-unit-map} should therefore be compared
independently.  Equation~\eqref{eq:appA-one-function-test} is a diagnostic
of the one-function truncation, not a gauge identity.

For the exact leading Boulware--Deser reference profile, the map is the
quadrature
\begin{equation}
 u_0(\rho)
 =\frac{r_h}{n}\int_0^\rho
  \frac{e^{\sigma/n}\,d\sigma}
       {\sqrt{F_{\rm BD}^{(0)}(\sigma)}} .
 \label{eq:appA-u-exact-EGB}
\end{equation}
This exact coordinate change is sufficient to compare the analytic reference
metric with unit gauge data.  For generic $\widetilde\alpha\ne0$ it does not
reduce to the elementary hyperbolic map below because $K_h\ne K$ and
$F_{\rm BD}^{(0)}(\rho)\ne K(1-e^{-\rho})$.

For the single-scale surrogate
\begin{equation}
 F_{\rm sur}(\rho)=K(1-e^{-\rho}),
\end{equation}
which is the exact leading profile in the Einstein limit, Eq.~\eqref{eq:appA-proper-map} gives, at fixed $\rho$,
\begin{equation}
 u_0(\rho)
 =\frac{2r_h}{n\sqrt K}\,
  \operatorname{arctanh}\!\sqrt{1-e^{-\rho}} .
 \label{eq:appA-u-of-rho}
\end{equation}
Introducing
\begin{equation}
 x\equiv\frac{n\sqrt K}{2r_h}\,u_0,
 \end{equation}
one obtains
\begin{equation}
 e^{-\rho}=\operatorname{sech}^2x,\qquad
 h_0(u_0)=\sqrt K\,\tanh x,\qquad
 g_0(u_0)=r_h^2
 \left[1+\frac{4}{n}\log\!\cosh x+O(n^{-2})\right].
 \label{eq:appA-unit-cigar}
\end{equation}
In the part of the scaling layer where the constant-coefficient scalar
equation is valid, a branch $X=X_h e^{\lambda\rho}$ becomes
\begin{equation}
 X(u_0)=X_h\bigl(\cosh x\bigr)^{2\lambda}.
 \label{eq:appA-unit-scalar}
\end{equation}
Equations~\eqref{eq:appA-unit-cigar} and
\eqref{eq:appA-unit-scalar} therefore provide direct profiles only for this
single-scale surrogate.  For generic EGB parameters one must use
Eq.~\eqref{eq:appA-u-exact-EGB} with the exact profile.  Beyond leading order, the safest comparison
uses the numerical areal radius $r_{\rm num}=\sqrt g$ and reconstructs
$A_{\rm num}=h^2$ and $B_{\rm num}=g'^2/(4g)$.  Scalars should then be
compared as functions of $r_{\rm num}$, rather than at equal values of the
coordinate $u$.  A coefficient-by-coefficient finite-$D$ test additionally
requires the same field equations, asymptotic time normalization, horizon
radius, and stretched horizon data on both sides; a gauge transformation
alone cannot compensate for a different truncation.

\subsection{Regularity and stretched horizon data}

For a formally smooth Euclidean tip at $u=0$, regular fields have expansions
\begin{align}
 h(u)&=\kappa_\mathrm{E}u+h_3u^3+\cdots,
 &
 g(u)&=r_h^2+g_2u^2+\cdots,
 \nonumber\\
 \phi(u)&=\phi_h+\phi_2u^2+\cdots,
 &
 X(u)&=X_h+X_2u^2+\cdots .
 \label{eq:appA-tip-series}
\end{align}
Absence of a conical singularity requires
\begin{equation}
 \beta\kappa_\mathrm{E}=2\pi,\qquad
 \kappa_\mathrm{E}=\frac{F'_h}{2}
 \label{eq:appA-tip-period}
\end{equation}
for the one-function background.  In unit radial gauge, the static scalar
equation reads
\begin{equation}
 X''+
 \left[
  \frac{h'}{h}
  +\frac{n+1}{2}\frac{g'}{g}
  -2\phi'
 \right]X'
 -\bigl(\tilde{\beta}^2h^2-\tilde{\beta}_H^2\bigr)X=0 .
 \label{eq:appA-unit-scalar-eq}
\end{equation}
If this effective equation were continued all the way to a smooth tip,
Eqs.~\eqref{eq:appA-tip-series} and
\eqref{eq:appA-unit-scalar-eq} would give
\begin{equation}
 X_2=-\frac{\tilde{\beta}_H^2}{4}\,X_h .
 \label{eq:appA-formal-X2}
\end{equation}
In the exterior EFT used here, however, the string-scale tip is not part of
the controlled domain.  The data are imposed instead at
$u=u_{\rm c}>0$ (equivalently $\rho=\rho_{\rm c}>0$), and
Eq.~\eqref{eq:appA-formal-X2} is not used as a physical tip condition.
When the cutoff lies in the region where the decaying plateau solution is
valid, its logarithmic derivative supplies the Robin datum
\begin{equation}
 \left.\frac{1}{X}\frac{dX}{du}\right|_{u_{\rm c}}
 =\left.
   \lambda_-\,\frac{n\sqrt{F(r)}}{r}
  \right|_{r=r_{\rm c}}
 +O(e^{-\rho_{\rm c}})
 +O\!\left(\frac{1+\rho_{\rm c}}{n}\right).
 \label{eq:appA-Robin}
\end{equation}
This fixes the decaying branch but leaves $X(u_{\rm c})$ free at quadratic
order.  If the cutoff is outside the plateau regime, the full near-zone
equation must first be integrated from the overlap to $u_{\rm c}$; replacing
that transport by the constant-coefficient approximation would impose an
uncontrolled boundary condition.

\subsection{Power counting and scope}

The large-$D$ limit used in the manuscript keeps
\begin{equation}
 \widetilde\alpha=n(n-1)\alpha,\qquad
 \ell,\qquad
 \frac{\widetilde\alpha}{\ell^2}
 \label{eq:appA-fixed-data}
\end{equation}
fixed.  Radial derivatives in the scaling layer obey
$\partial_r=O(n/r_h)$.  If the dilaton expansion begins as
\begin{equation}
 \phi=\phi_0+\frac{\phi_1(\rho)}{n^2}+\cdots,
 \label{eq:appA-phi-count}
\end{equation}
then the relevant orders are
\begin{equation}
 P_{\mu\nu\rho\sigma}=O\!\left(\frac{n^2}{r_h^2}\right),
 \qquad
 H_{\mu\nu}=O\!\left(\frac{n^4}{r_h^4}\right),
 \qquad
 \nabla_\mu\nabla_\nu\phi=O\!\left(\frac{1}{r_h^2}\right),
 \qquad
 (\nabla\phi)^2=O\!\left(\frac{1}{n^2r_h^2}\right).
 \label{eq:appA-power-counting}
\end{equation}
Here $P_{\mu\nu\rho\sigma}$ denotes the curvature tensor that contracts the
dilaton Hessian in the string frame Gauss--Bonnet metric equation, while
$H_{\mu\nu}$ denotes the pure Gauss--Bonnet metric tensor.  It follows that
\begin{equation}
 \frac{\alpha\,\Delta_{\mu\nu}^{(\phi{\rm GB})}}
      {\alpha\,H_{\mu\nu}}
 =O(n^{-2})
 \label{eq:appA-GB-ratio}
\end{equation}
under the scaling \eqref{eq:appA-phi-count}.  The absolute orders make the
same hierarchy transparent.  Since
\begin{equation}
 \alpha=\frac{\widetilde\alpha}{n(n-1)},
 \end{equation}
one has
\begin{equation}
 G_{\mu\nu}
 =O\!\left(\frac{n^2}{r_h^2}\right),\qquad
 \alpha H_{\mu\nu}
 =O\!\left(\frac{\widetilde\alpha\,n^2}{r_h^4}\right),\qquad
 \alpha\Delta_{\mu\nu}^{(\phi{\rm GB})}
 =O\!\left(\frac{\widetilde\alpha}{r_h^4}\right).
 \label{eq:appA-absolute-GB-orders}
\end{equation}
For fixed $\widetilde\alpha/r_h^2=O(1)$, the pure Gauss--Bonnet tensor can
therefore compete with the Einstein tensor at leading order, whereas the
dilaton-dependent Gauss--Bonnet tensor is two relative powers of $n$ lower.
The conclusion is tied to the field scaling
\eqref{eq:appA-phi-count}: if a different branch had
$\phi-\phi_0=O(n^{-1})$ or $O(1)$, the order of
$\Delta_{\mu\nu}^{(\phi{\rm GB})}$ would have to be recomputed rather than
assumed.

This distinction also separates two logically different statements.  The
exact Euler--Lagrange equation of the finite-$D$ string frame action always
contains $\Delta_{\mu\nu}^{(\phi{\rm GB})}$.  A relative $1/n$ calculation
may omit it only after declaring the scaling
\eqref{eq:appA-phi-count} and its target accuracy.  Consequently, a numerical
integration of the same LO+NLO truncation may consistently use the shortened
metric equation, while an exact finite-$D$ integration of the displayed
action may not.

Thus the
dilaton--Gauss--Bonnet Hessian term is required by the exact finite-$D$
variation of the displayed string frame action, but it first contributes
beyond the orders retained in the leading large-$D$ metric equation.  This
is the precise sense in which it may be omitted from that truncated
equation; it is not absent from the underlying theory.

Finally, the formulas above concern static Euclidean matching only.
Lorentzian ingoing conditions and quasinormal frequencies constitute a
different boundary value problem and cannot be inferred from
Eqs.~\eqref{eq:appA-near-overlap-eq}--\eqref{eq:appA-match}.  Likewise,
there is no action-independent universal mass function or universal horizon
shift for the dilaton--thermal--scalar system.  Backreaction must be derived
from the chosen action and metric gauge, as done for the controlled
truncations in the main text and in the appendices that follow.
\section{Radial equations used in the truncated analysis} \label{app:fullnum}

Substituting the single unit radial gauge ansatz~\eqref{eq:unit_radial_gauge} into the large-$D$-truncated field equations gives a system for $h(\rho)$, $g(\rho)$, $\phi(\rho)$, and $\chi(\rho)$.  The exact finite-$D$ variation additionally contains $\Delta^{(\phi {\rm GB})}_{\mu\nu}$ in~\eqref{eq:phiGB-extra}; the equations below set it to zero consistently with the relative $D^{-2}$ accuracy stated in~\eqref{eq:phiGB-suppression}. It is, however, retained in the numerical analysis. For compactness in this appendix we use the dimensionless variables
\begin{equation}
    f\equiv\widetilde\beta h,
    \qquad \psi\equiv\sqrt\kappa\,\chi,
    \qquad \alpha'=1,
\end{equation}
and then rename $f$ as $h$ and $\psi$ as $\chi$. A convenient independent evolution set is the $\tau\tau$ metric equation, the angular metric equation, and the thermal scalar equation,
\begin{align}
0={\cal E}_h\equiv{}&
    h''+\frac{[(D-2)g'-4g\phi']h'}{2g}-2h^3\chi^2
    \nonumber\\
&+\alpha(D-3)(D-2)\Bigg\{
    \frac{4g-(g')^2}{2g^2}h''
    -\frac{h'g'}{g^2}g''
    \nonumber\\
&\hspace{9.2em}
    +\frac{[4(D-4)g-(D-6)(g')^2]h'g'}{4g^3}
    \Bigg\},
    \label{eq:full-num-f}\\[0.4em]
0={\cal E}_g\equiv{}&
    g''+\frac{(D-4)(g')^2}{2g}
    +\frac{(h'-2h\phi')g'}{h}-2(D-3)
    \nonumber\\
&+\alpha(D-3)\Bigg\{
    \frac{2[4g-(g')^2]}{hg}h''
    -\left[\frac{4h'g'}{hg}
    -\frac{3(D-4)[4g-(g')^2]}{2g^2}\right]g''
    \nonumber\\
&\hspace{5.2em}
    +\frac{[12(D-4)g-(3D-16)(g')^2]h'g'}{2hg^2}
    \nonumber\\
&\hspace{5.2em}
    -\frac{(D-4)[4g-(g')^2]}{4g^3}
     \left[4(D-5)g-(D-8)(g')^2\right]
    \Bigg\},
    \label{eq:full-num-g}\\[0.4em]
0={\cal E}_\chi\equiv{}&
    \chi''+\left[\frac{h'}{h}+\frac{(D-2)g'}{2g}-2\phi'\right]\chi'
    -(h^2-\widetilde\beta_{\rm H}^{2})\chi .
    \label{eq:full-num-chi}
\end{align}
    The remaining first-order relation is the constraint~\eqref{eq:constraint}, which we denote by ${\cal E}_\phi = 0$; it selects the admissible branch of $\phi'$ and is monitored throughout the numerical integration.  The dilaton equation~\eqref{eq:DilatonEOM} is then a dependent consistency equation by the radial Bianchi identity. The radial system displayed in this appendix is thus
\begin{equation}
    {\cal E}_h = {\cal E}_g = {\cal E}_\chi = {\cal E}_\phi = 0,
\end{equation}
    with $\{h, g, \phi, \chi\}$ as the four radial unknowns.  The dimensionful factors of $\alpha'$ are restored by $\widetilde\beta=\beta/(2\pi\alpha')$ and $\widetilde\beta_{\rm H}=\beta_{\rm H}/(2\pi\alpha')$.

\section{Subsystem for the numerical analysis} \label{app:subnum}
    In this section, we derive the reduced equations for the subsystem~\cite{Brustein:2021qkj, Krishnan:2024zax} from the full equations of motion.

\subsection{Derivation of subsystem equations} \label{subsec:subsyst}
    For later convenience, we recall the complete equations of motion~\eqref{eq:mseqn},
\begin{align}
    0 &= G_{\mu\nu} + \Lambda g_{\mu\nu} + \alpha \left( H_{\mu\nu} + \Delta^{(\phi {\rm GB})}_{\mu\nu} \right) + 2 \nabla_\mu \nabla_\nu \phi - 2 g_{\mu\nu} \nabla^2 \phi + 4 \partial_\mu \phi \partial_\nu \phi - 4 g_{\mu\nu} \left( \partial \phi \right)^2 \nonumber
    \\
    & \quad - 2 \kappa e^{2 \phi} T^{(\chi)}_{\mu\nu},
    \label{eq:equation_g}
    \\
    0 &= R - 2 \Lambda + \alpha R_{\rm GB}^2 + 4 \nabla^2 \phi - 4 \left( \partial \phi \right)^2 - 2 \kappa \left[ 2(\partial \chi)^2 + 2 m_{\rm th}^2 \chi^2 \right],
    \label{eq:equation_phi}
    \\
    0 &= \nabla^2 \chi - 2 \partial_\mu \phi \nabla^\mu \chi - m_{\rm th}^2 \chi,
    \label{eq:equation_chi}
\end{align}
    and reduced equation of motion~\eqref{eq:reduced_eom_metric} from the metric variation,
\begin{equation}
    0 =  R_{\mu}{}^\nu + \alpha \left( H_{\mu}{}^{\nu} + \Delta^{(\phi {\rm GB})}{}_{\mu}{}^{\nu} \right) + 2 \nabla_\mu \nabla^\nu \phi - \kappa \left( 2 \nabla_\mu \chi \nabla^\nu \chi - 2 \tilde{\beta}^2 \chi^2 g_{\tau\mu} g_{\tau}{}^\nu \right).
    \label{eq:reduced_eom_sub}
\end{equation}
    We focus on a subsystem of the equations of motion in the unit radial gauge ansatz~\eqref{eq:unit_radial_gauge}, motivated by the semiclassical picture of the winding condensate, in which $\chi \propto \exp(- S_{\rm NG})$, where $S_{\rm NG}$ denotes the Nambu--Goto action of a string worldsheet wrapping the Euclidean time circle. In this subsystem, we consider the following first-order equation:
\begin{equation}
    \chi' + \tilde{\beta} f \chi = 0.
    \label{eq:subsystem}
\end{equation}
    Substituting~\eqref{eq:subsystem} into~\eqref{eq:equation_chi}, we obtain
\begin{equation}
    h' + \left( \frac{D-2}{4 g} g' - \phi' \right) h - \frac{\tilde{\beta}_{\rm H}^2}{2 \tilde{\beta}} = 0.
    \label{eq:h_p}
\end{equation}
    We now introduce an ansatz in which $\mathcal{X}$ is left undetermined at this stage,
\begin{equation}
    \phi' = h \left( - \kappa \tilde{\beta} \chi^2 + \mathcal{X} \right).
    \label{eq:ansatz}
\end{equation}
    By differentiating Eq.~\eqref{eq:ansatz} and substituting Eqs.~\eqref{eq:subsystem} and~\eqref{eq:h_p}, we find
\begin{equation}
    2 \left[ \phi'' - \left( \phi' \right)^2 \right] - 2 h \mathcal{X}' + \frac{\left( D - 2 \right) g' \phi'}{2 g} - \frac{\tilde{\beta}_{\rm H}^2}{\tilde{\beta}} \mathcal{X} - \kappa \left( 4 \tilde{\beta}^2 h^2 - \tilde{\beta}_{\rm H}^2 \right) \chi^2 = 0.
    \label{eq:phi_pp}
\end{equation}
    By combining the $tt$ and $\theta\theta$ components of Eq.~\eqref{eq:reduced_eom_sub}, one can solve for $h''$ and $g''$. Substituting these expressions into the remaining equations, the second derivatives $h''$ and $g''$ can be eliminated. Differentiating Eq.~\eqref{eq:ansatz} and taking a linear combination with the $\rho\rho$ component of Eq.~\eqref{eq:reduced_eom_sub}, we obtain an expression for $\mathcal{X}'$. Subtracting Eq.~\eqref{eq:equation_phi} from Eq.~\eqref{eq:phi_pp} and substituting the resulting expression for $\mathcal{X}'$, using Eqs.~\eqref{eq:h_p} and~\eqref{eq:ansatz}, the resulting reduced equations for the subsystem are given as follows:
\begin{align}
	0 &= h' + \left[ \frac{\left( D - 2 \right) g'}{4 g} - \phi' \right] h - \frac{\tilde{\beta}_\mathrm{H}^2}{2 \tilde{\beta}},
    \label{eq:reduced_eom_h}
	\\
    0 &= \left[ 1 + \frac{\alpha \left( D - 3 \right)}{2 h g^2} X_1 + \frac{\alpha^2 \left( D - 3 \right)^2 \left( D - 2 \right)}{4 h^2 g^4} X_2 \right] g'' + \frac{\left( D - 4 \right) ( g' )^2}{2 g} + \frac{h' - 2 h \phi'}{h} g' \nonumber
    \\
    & \qquad - 2 \left( D - 3 \right) + \frac{\alpha \left( D - 3 \right)}{2 h^2 g^3} X_3 + \frac{\alpha^2 \left( D - 3 \right)^2 \left( D - 2 \right)}{8 h^2 g^5} X_4,
    \label{eq:reduced_eom_g}
    \\
	0 &= \frac{\alpha \left( D - 3 \right) \left( D - 2 \right) \left( 3 ( g' )^2 - 4 g \right)}{g^2} \left( \phi' \right)^2 - \left\{ \frac{\tilde{\beta}_\mathrm{H}^2}{\tilde{\beta} h} + \frac{\alpha \left( D - 3 \right) \left( D - 2 \right)}{2 \tilde{\beta} h g^3} \left[ \left( D - 1 \right) \tilde{\beta} h ( g' )^3 \right. \right. \nonumber
    \\
    & \qquad \left. \left. - \left( 3 \tilde{\beta}_\mathrm{H}^2 g' + 4 \tilde{\beta} h \right) g g' + 4 \tilde{\beta}_\mathrm{H}^2 g^2 \right] \right\} \phi' - \frac{\left( D - 2 \right)}{2 g} \left( \frac{( g' )^2}{4 g} - \frac{\tilde{\beta}_\mathrm{H}^2 g'}{2 \tilde{\beta} h} + D - 3 \right) \nonumber
    \\
    & \qquad - \frac{\alpha \left( D - 4 \right) \left( D - 3 \right) \left( D - 2 \right)}{2 g^2} \left( - \frac{\left( D + 1 \right) ( g' )^4}{16 g^2} + \frac{\tilde{\beta}_\mathrm{H}^2 ( g' )^3}{4 \tilde{\beta} h g} + \frac{3 ( g' )^2}{2 g} - \frac{\tilde{\beta}_\mathrm{H}^2 g'}{\tilde{\beta} h} + D - 5 \right) \nonumber
    \\
    & \qquad + \Lambda  - \kappa \tilde{\beta}_\mathrm{H}^2 \chi^2,
    \label{eq:reduced_eom_phi}
	\\
	0 &= \chi' + \tilde{\beta} h \chi,
    \label{eq:reduced_eom_chi}
\end{align}
    where
\begin{align}
    X_1 &= - 4 g h' \left( D g' + 4 g \phi' \right) + h \left[ \left( D^2 - 10 \right) \left( 4 g - ( g' )^2 \right) - 8 \left( D - 4 \right) g g' \phi' \right],
    \\
    X_2 &= -32 g^2 ( h' )^2 ( g' )^2 + 4 h g h' \left[ - \left( D - 10 \right) \left( ( g' )^2 - 4 g \right) g' - 12 g \left( ( g' )^2 + 4 g \right) \phi' \right] \nonumber
    \\
    & \qquad + \left[ \left( D^2 + D - 24 \right) \left( ( g' )^2 - 4 g \right)^2 - 32 g g' \left( ( g' )^2 - 4 g \right) \phi' - 64 g^2 ( g' )^2 ( \phi' )^2 \right] h^2,
    \\
    X_3 &= - 32 \kappa \tilde{\beta}^2 h^3 g^3 \chi^2 h' g' + 4 g^2 ( h' )^2 g' \left[ \left( D - 2 \right) g' - 4 g \phi' \right] + \left\{ - 4 \left( D^2 - 10 D + 22 \right) g g' \right. \nonumber
    \\
    & \qquad \left. + \left( D^2 - 6 D + 18 \right) ( g' )^3 + 8 g \left[ 4 \left( D - 3 \right) g - \left( D - 6 \right) ( g' )^2 \right] \phi' \right\} h g h' \nonumber
    \\
    & \qquad + 8 \kappa \tilde{\beta}^2 h^4 g^2 \chi^2 \left[ - \left( D - 3 \right) ( g' )^2 + 4 g \left( D - 3 - g' \phi' \right) \right] + \left\{ - \left( D - 5 \right) \left( D - 4 \right) ( g' )^4 \right. \nonumber
    \\
    & \qquad\quad -32 \left( D - 4 \right) g^3 ( \phi' )^2 + 2 \left[ 6 D^2 - 42 D + 78 - \left( D^2 - 14 D + 34 \right) g' \phi' \right] g ( g' )^2 \nonumber
    \\
    & \qquad \left. + 8 g^2 \left[ - 4 D^2 + 24 D - 38 + \left( D^2 - 12D + 26 + \left( D - 4 \right) g' \phi' \right) g' \phi' \right] \right\} h^2,
    \\
    X_4 &= 2 h g h' \left\{ - 16 \left( D^2 - 15 D + 48 \right) g^2 g' + 8 \left( D - 8 \right) \left( D - 7 \right) g ( g' )^3 \right. \nonumber
    \\
    & \qquad \left. - \left( D^2 - 15 D + 64 \right) ( g' )^5 + 16 g \left[ 8 \left( D - 3 \right) g^2 - 2 \left( D - 7 \right) g ( g' )^2 + ( g' )^4 \right] \phi' \right\} \nonumber
    \\
    & \qquad + h^2 \left\{ - \left( - 4 g + ( g' )^2 \right)^2 \left[ 12 \left( D - 5 \right) \left( D - 4 \right) g - \left( D^2 - 20 D + 72 \right) ( g' )^2 \right] \right. \nonumber
    \\
    & \qquad + 4 g g' \left( 4 g - ( g' )^2 \right) \left[ 4 \left( D - 9 \right) \left( D - 4 \right) g - \left( D^2 - 11 D + 40 \right) ( g' )^2 \right] \phi' \nonumber
    \\
    & \qquad - 16 g^2 \left[ 16 \left( D - 6 \right) g^2 - 16 \left( D - 5 \right) g ( g' )^2 + 3 \left( D - 6 \right) ( g' )^4 \right] ( \phi' )^2 \nonumber
    \\
    & \qquad \left. + 128 g^3 g' \left( - 4 g + ( g' )^2 \right) ( \phi' )^3 \right\} \nonumber
    \\
    & \qquad + 8 g^2 ( h' )^2 g' \left[ - \left( D - 10 \right) ( g' )^3 + 16 g^2 \phi' + 4 g g' \left( D - 6 - g' \phi' \right) \right].
\end{align}
    Eq.~\eqref{eq:reduced_eom_g} is obtained from an appropriate linear combination of the $\tau\tau$ and $\theta\theta$ components of Eq.~\eqref{eq:reduced_eom_sub}. All fields other than $g$ are reduced to first-order equations within this subsystem. When $\alpha = \Lambda = 0$, Eqs.~\eqref{eq:reduced_eom_h},~\eqref{eq:reduced_eom_g},~\eqref{eq:reduced_eom_phi}~and~\eqref{eq:reduced_eom_chi} coincide with the results reported in Refs.~\cite{Brustein:2021qkj, Krishnan:2024zax}.

\subsection{Boundary data for the reduced subsystem} \label{subsec:subnum}
    The origin of the unit radial coordinate is fixed once and for all at the formal Euclidean tip, $\rho=0$; no additional radial offset parameter is introduced.  The regular Taylor data are
\begin{align}
    h(\rho)&=h_1\rho+\mathcal O(\rho^3),&
    g(\rho)&=g_0+g_2\rho^2+\mathcal O(\rho^4),\nonumber\\
    \phi(\rho)&=\phi_0+\phi_2\rho^2+\mathcal O(\rho^4),&
    \chi(\rho)&=\chi_0+\chi_2\rho^2+\mathcal O(\rho^4).
    \label{eq:subsystem-tip-data}
\end{align}
    Thus $h(0)=g'(0)=\phi'(0)=\chi'(0)=0$.  The symbols $g_0$, $\phi_0$, and $\chi_0$ denote Taylor/shooting coefficients; their displayed numerical choices are not boundary conditions derived locally.  For the thermal scalar EFT, the actual integration data are imposed at $\rho=\rho_{\rm c}>0$ and $\chi_{\rm c}=\chi(\rho_{\rm c})$ remains an undetermined normalization at quadratic order.  The reflected branch $\rho<0$ is a coordinate duplication and is not plotted or assigned physical meaning.

    When $g$ is parametrically large and slowly varying in the exterior cap, its logarithmic derivatives and inverse powers are suppressed.  Under this approximation the remaining equations close on $\{h,\phi,\chi\}$ and reduce to
\begin{align}
    0 &= h' - \phi' h - \frac{\tilde{\beta}_\mathrm{H}^2}{2 \tilde{\beta}},
    \\
    0 &= \frac{\tilde{\beta}_{\rm H}^2}{\tilde{\beta} h} \phi' + \kappa \tilde{\beta}_{\rm H}^2 \chi^2 - \Lambda,
    \\
    0 &= \chi' + \tilde{\beta} h \chi .
\end{align}
    This reduced system is used only as a cap region diagnostic.  It does not determine the microscopic boundary condition at the excised string-scale core.

\section{Details of the next-to-leading-order solution} \label{app:NLO-details}
    In this appendix we collect the explicit next-to-leading-order expressions for the metric correction, the dilaton, and the thermal scalar. We work in the conventions of Sec.~\ref{sec:LO-geometry}, with $D=n+3$ and $n\gg1$, near-horizon coordinate
\begin{equation}
	r = r_h \Big(1 + \frac{\rho}{n} + \mathcal{O}(n^{-2})\Big),
	\qquad
	0 \le \rho = \mathcal{O}(1),
\end{equation}
    and the leading-order cigar profile
\begin{equation}
    F_0(\rho)
	=
	K \big(1 - e^{-\rho}\big),
	\qquad
	K \equiv \frac{F'(r_h)}{n} = \mathcal{O}(1).
\end{equation}
    The Gauss--Bonnet coupling and dilaton dressing appear only through the effective gravitational coupling $\kappa_{\rm eff}$ and the LO dilaton $\phi_0=\phi_\infty$ defined in Sec.~\ref{sec:LO-dilaton}.

    Throughout this appendix we drop terms of order $1/n^2$ and higher. All fields are expanded to NLO in the large-$D$ expansion, keeping only the leading nontrivial order in the backreaction parameter $\kappa$:
\begin{equation}
	F(\rho)
	=
	F_0(\rho) + \frac{1}{n} F_1(\rho) + \mathcal{O}(n^{-2}),
	\qquad
	\phi(\rho)
	=
	\phi_0 + \phi_1(\rho) + \mathcal{O}(\kappa^2),
	\qquad
	\chi(\rho)
	=
	\chi_0(\rho) + \frac{1}{n}\,\chi_1(\rho) + \mathcal{O}(n^{-2}) .
\end{equation}

\subsection{Metric correction \texorpdfstring{$F_1(\rho)$}{F1(rho)}} \label{app:metric-F1}

\subsubsection{Near-zone equation and source}
\paragraph{Einstein equation and stress tensor}
    At NLO the $\tau\tau$ and $\rho\rho$ components of the Einstein--Gauss--Bonnet equations give a single first-order equation for $F_1(\rho)$,
\begin{equation}
	F_1'(\rho) + F_1(\rho)
	=
	\mathcal{A}_\chi \big(1-e^{-\rho}\big)\,e^{2\lambda_- \rho},
	\label{eq:F1-ODE-app}
\end{equation}
    where the source comes entirely from the LO winding mode stress tensor,
\begin{equation}
	\big(T^\rho{}_\rho - T^\tau{}_\tau\big)_{\chi}^{\rm LO}
	=
	g^{\rho\rho}_{(0)}\,\big(\partial_\rho\chi_0\big)^2
	=
	\frac{n^2}{r_h^2}\,F_0(\rho)\,\lambda_-^2
	\chi_h^2 e^{2\lambda_-\rho}
	=
	\frac{n^2}{r_h^2}\,K\lambda_-^2
	\chi_h^2
	\big(1-e^{-\rho}\big)e^{2\lambda_-\rho}.
\end{equation}
    Matching to the Einstein equations fixes the overall coefficient
\begin{equation}
	\boxed{
		\mathcal{A}_\chi
		=
		\frac{2\kappa_{\rm eff}}{D-2}
		\frac{n^2}{r_h^2}\,
		K \lambda_-^2 \chi_h^2,
	}
	\label{eq:Achi-def-app}
\end{equation}
    in the conventions where $F_1$ is defined by the expansion $F(r)=F_0(\rho)+n^{-1} F_1(\rho)$ and the LO operator is as in Sec.~\ref{sec:LO-geometry}.

\subsubsection{Closed form solution and asymptotics}
\paragraph{Near-zone solution}
    The ODE~\eqref{eq:F1-ODE-app} is elementary. Integrating with an integrating factor $e^{\rho}$ and imposing $F_1(0)=0$ to preserve the smoothness of the cigar tip gives
\begin{align}
	F_1(\rho)
	&=
	e^{-\rho}
	\mathcal{A}_\chi
	\int_0^\rho
	d\sigma\,
	e^{\sigma}
	\big(1-e^{-\sigma}\big)e^{2\lambda_- \sigma}
	\label{eq:F1-integral-app}
	\\
	&=
	\mathcal{A}_\chi\,e^{-\rho}
	\left[
	\frac{e^{(2\lambda_-+1)\rho}-1}{2\lambda_-+1}
	- \frac{e^{2\lambda_-\rho}-1}{2\lambda_-}
	\right].
	\label{eq:F1-closed-app}
\end{align}
    Near the tip one finds the expansion
\begin{equation}
	F_1(\rho)
	=
	\mathcal{A}_\chi
	\left[
	\frac{1}{2}\rho^2
	+ \mathcal{O}(\rho^3)
	\right],
	\qquad
	\rho \to 0,
\end{equation}
    while for $\rho\to\infty$ (still within the overlap region
$1\ll\rho\ll n$) we have
\begin{equation}
	F_1(\rho)
	=
	\mathcal{A}_\chi
	\left[
	\frac{e^{2\lambda_-\rho}}{2\lambda_-+1}
	- \frac{e^{(2\lambda_- -1)\rho}}{2\lambda_-}
	+ \mathcal{O}\big(e^{-\rho}\big)
	\right],
	\qquad
	\rho\to\infty.
\end{equation}
    The leading decaying piece $e^{2\lambda_-\rho}$ is precisely what is required by matching to the far-zone deformation of the Boulware--Deser solution; its coefficient is fixed in terms of the thermal scalar hair amplitude $\chi_h$ by~\eqref{eq:Achi-def-app}.

\paragraph{Far-zone matching}
    In the far-zone $r-r_h=\mathcal{O}(r_h)$ the metric is governed by the linearised Boulware--Deser equation around the asymptotic background. Writing
\begin{equation}
	F(r)
	=
	F_{\rm BD}(r)
	\big[1 + \varepsilon\,\delta f(r)\big],
	\qquad
	\varepsilon\sim n^{-1},
\end{equation}
    the linearised equation is solved by
\begin{equation}
	\delta f(r)
	=
	\mathcal{C}_1\, \delta f_{\rm reg}(r) + \mathcal{C}_2\,\delta f_{\rm dec}(r),
\end{equation}
    where $\delta f_{\rm reg}$ is regular at the horizon and $\delta f_{\rm dec}\sim r^{-(D-3)}$ is the decaying mode. In the overlap region $1\ll\rho\ll n$ we have $r-r_h\sim r_h \rho/n$ and
\begin{equation}
	F(\rho)
	=
	K\big(1-e^{-\rho}\big)
	+ \frac{1}{n}F_1(\rho)
	=
	F_{\rm BD}(r)
	\big[1 + \varepsilon\,\delta f(r)\big],
\end{equation}
    which yields an explicit identification between the integration constant $\mathcal{C}_2$ in $\delta f$ and the thermal scalar amplitude $\chi_h$ via~\eqref{eq:F1-closed-app}. We do not reproduce the straightforward but lengthy algebra here.

\subsection{Dilaton correction \texorpdfstring{$\phi_1(\rho)$}{phi1(rho)}}\label{app:phi1-details}
\subsubsection{Near-zone equation and integral representation}
\paragraph{Master equation and source}
    The master equation for $\phi_1$ in the probe regime is
\begin{equation}
	\nabla^2\phi_1
	= \kappa\big[(\partial\chi_0)^2 + m_{\rm th}^2(r)\chi_0^2\big],
	\label{eq:phi1-master-app}
\end{equation}
    with $\chi_0(\rho)=\chi_h e^{\lambda_-\rho}$ and $m_{\rm th}^2(r)\simeq m_{\rm eff}^2=\tilde{\beta}^2 K-\tilde{\beta}_{\rm H}^2$ in the near-zone. Using the near-zone Laplacian as in Sec.~\ref{subsubsec:NLO-dilaton}, one finds the first-order equation~\eqref{eq:phi1-NZ-eq-final} for $\phi_1'(\rho)$ and the solution~\eqref{eq:phi1-prime-NZ-correct}. Integrating once more gives
\begin{equation}
	\phi_1^{\rm (NZ)}(\rho)
	= \phi_1(0) + \kappa\chi_h^2\,\mathcal{I}(\rho),
\end{equation}
    with the quadrature
\begin{equation}
	\mathcal{I}(\rho)
	=
	\int_0^\rho d\sigma\,
	\frac{
		\displaystyle
		\frac{\lambda_-^2 + \Xi^2}{2\lambda_-+1}\bigl(e^{(2\lambda_-+1)\sigma}-1\bigr)
		- \frac{\lambda_-}{2}\bigl(e^{2\lambda_-\sigma}-1\bigr)
	}{
		e^{\sigma}-1
	}.
\end{equation}

\paragraph{Special function form}
    The integral $\mathcal{I}(\rho)$ can be expressed in terms of the auxiliary integrals
\begin{equation}
	J(a;\rho)
	\equiv
	\int_0^\rho d\sigma\,\frac{e^{a\sigma}-1}{e^{\sigma}-1},
\end{equation}
    which admit the representation
\begin{equation}
	J(a;\rho)
	=
	\psi(1) - \psi(1-a)
	- \frac{e^{-(1-a)\rho}}{1-a}\,
	{}_2F_1\big(1,1-a;2-a;e^{-\rho}\big) - \log(1 - e^{-\rho}),
	\label{eq:J-hypergeom-app}
\end{equation}
    where $\psi$ is the digamma function and ${}_2F_1$ is the Gauss hypergeometric function. In terms of $J$,
\begin{equation}
	\mathcal{I}(\rho)
	=
	\frac{\lambda_-^2 + \Xi^2}{2\lambda_-+1}\,J(2\lambda_-+1;\rho)
	- \frac{\lambda_-}{2}\,J(2\lambda_-;\rho).
\end{equation}

\subsubsection{Far-zone behaviour}
\paragraph{Asymptotics and matching}
    In the far-zone, the equation
\begin{equation}
	\frac{d^2\phi_1}{dY^2} + \frac{d\phi_1}{dY}
	= \kappa\chi_\infty^2 e^{2\lambda_- Y}(\lambda_-^2 + \Xi^2)
\end{equation}
    has the general solution
\begin{equation}
	\phi_1^{\rm (FZ)}(Y)
	= \phi_{1,\infty}
	+ B_1 e^{-Y}
	+ B_2 e^{2\lambda_- Y},
\end{equation}
    with
\begin{equation}
	B_2
	= \frac{\kappa\chi_\infty^2(\lambda_-^2 + \Xi^2)}
	{2\lambda_-(2\lambda_- +1)}.
\end{equation}
    The constant $\phi_{1,\infty}$ is fixed by matching to $\phi_1^{\rm (NZ)}(\rho)$ in the overlap region via
\begin{equation}
	\phi_{1,\infty}
	= \phi_1(0) + \kappa\chi_h^2\,\mathcal{I}(\infty).
\end{equation}
    The coefficient $B_1$ is fixed by the absence of $e^{-Y}$ term in the overlap with the near-zone solution.

\subsection{Thermal scalar correction \texorpdfstring{$\chi_1(\rho)$}{chi1(rho)}} \label{app:chi1-details}
\subsubsection{Linearised equation and source decomposition}
\paragraph{LO operator and NLO source}
    In the near-zone the covariant thermal scalar equation
\begin{equation}
	\Box\chi - 2\,\partial_\mu\phi\,\partial^\mu\chi - m_{\rm th}^2(r)\,\chi = 0,
\end{equation}
    reduces at LO to
\begin{equation}
	\mathcal{L}_0[\chi_0]
	\equiv
	\Big(\partial_\rho^2 + \partial_\rho - \Xi^2\Big)\chi_0(\rho)
	= 0,
	\qquad
	\chi_0(\rho) = \chi_h e^{\lambda_- \rho},
\end{equation}
    with $\lambda_\pm$ as in Sec.~\ref{subsubsec:NLO-thermal-scalar}. To NLO we obtain
\begin{equation}
	\mathcal{L}_0[\chi_1(\rho)]
	=
	S(\rho),
	\label{eq:chi1-ODE-app}
\end{equation}
    with the source $S(\rho)$ encapsulating the deviation of the true geometry and dilaton from the constant-$F$ background. It is convenient to decompose $S(\rho)$ into four pieces,
\begin{equation}
	S(\rho)
	=
	S_{\rm kin}(\rho)
	+ S_{\rm mass}(\rho)
	+ S_{\rm met}(\rho)
	+ S_{\phi}(\rho),
	\label{eq:S-decomp-app}
\end{equation}
    corresponding respectively to:
\begin{itemize}
	\item the deviation of the radial kinetic operator from its LO form,
	\item the $\rho$-dependence of the thermal mass $m_{\rm th}^2(r)$,
	\item the explicit NLO metric correction $F_1(\rho)$, and
	\item the NLO dilaton gradient.
\end{itemize}
    Starting from the exact radial equation
\begin{equation}
	e^{-\rho}\partial_\rho\big(e^{\rho} F(\rho)\,\partial_\rho\chi\big)
	- \frac{r_h^2}{n^2}\Big[
	\tilde{\beta}^2 F(\rho) - \tilde{\beta}_{\rm H}^2
	\Big]\chi(\rho)
	- 2 F(\rho)\,\phi'(\rho)\chi'(\rho)
	= 0,
\end{equation}
    and subtracting the LO equation evaluated with $F_0\to K$ and $\phi_0'(\rho)=0$, one finds, after some algebra, the following expressions:
\begin{align}
	S_{\rm kin}(\rho)
	&=
	- e^{-\rho}\partial_\rho
	\Big\{
	e^{\rho}\big[F_0(\rho)-K\big]\chi_0'(\rho)
	\Big\},
	\label{eq:S-kin-app}
	\\
	S_{\rm mass}(\rho)
	&=
	- \frac{r_h^2}{n^2}\,\tilde{\beta}^2
	\big[F_0(\rho)-K\big]\chi_0(\rho),
	\label{eq:S-mass-app}
	\\
	S_{\rm met}(\rho)
	&=
	- e^{-\rho}\partial_\rho
	\Big\{
	e^{\rho} F_1(\rho)\chi_0'(\rho)
	\Big\}
	+ \frac{r_h^2}{n^2}\,\tilde{\beta}^2 F_1(\rho)\chi_0(\rho),
	\label{eq:S-met-app}
	\\
	S_{\phi}(\rho)
	&=
	2 F_0(\rho)\,\chi_0'(\rho)\,\phi_1'(\rho),
	\label{eq:S-phi-app}
\end{align}
    where $F_0(\rho)$, $F_1(\rho)$, and $\phi_1(\rho)$ are given in Eqs.~\eqref{eq:F1-closed-app} and~\eqref{eq:phi1-prime-NZ-correct}. All four contributions are proportional to $\chi_h$ and, after expressing everything in terms of exponentials, can be written in the generic form
\begin{equation}
	S(\rho)
	=
	\sum_j \mathcal{C}_j
	\frac{e^{(p_j-1)\rho}}{1-e^{-\rho}}
	+ \sum_k \widetilde{\mathcal{C}}_k\,e^{q_k\rho},
	\label{eq:S-generic-app}
\end{equation}
    with coefficients $\mathcal{C}_j$, $\widetilde{\mathcal{C}}_k$ that are rational functions of $\lambda_\pm$ and the parameters $(K,\tilde{\beta},\tilde{\beta}_{\rm H},\kappa_{\rm eff},r_h)$.

\subsubsection{Green's function and integral basis}
\paragraph{Near-zone Green's function}
    The homogeneous operator $\mathcal{L}_0$ has two linearly independent solutions $y_1=e^{\lambda_+\rho}$, $y_2=e^{\lambda_-\rho}$ with Wronskian
\begin{equation}
	W(\rho)
	= (\lambda_- - \lambda_+)\,e^{-\rho}.
\end{equation}
    Imposing regularity at the tip and decay in the far-zone, the particular solution of~\eqref{eq:chi1-ODE-app} compatible with our boundary conditions is precisely the Green's function expression~\eqref{eq:chi1-GF-final},
\begin{equation}
	\chi_1(\rho)
	=
	\frac{e^{\lambda_-\rho}}{\lambda_- - \lambda_+}
	\int_\rho^\infty d\sigma\,
	S(\sigma)\,e^{-\lambda_-\sigma}.
\end{equation}

\paragraph{Integral basis and special functions}
    Using the generic structure~\eqref{eq:S-generic-app}, all contributions to $\chi_1(\rho)$ can be organized in terms of the integrals
\begin{equation}
	I(p;\rho)
	\equiv
	\int_\rho^\infty d\sigma\,
	\frac{e^{(p-1)\sigma}}{1-e^{-\sigma}}
	=
	\int_0^{e^{-\rho}}dt\,\frac{t^{-p}}{1-t},
\end{equation}
    and
\begin{equation}
	J(a;\rho)
	\equiv
	\int_0^\rho d\sigma\,\frac{e^{a\sigma}-1}{e^{\sigma}-1}.
\end{equation}
    We already recorded the special function representation of $J(a;\rho)$ in Eq.~\eqref{eq:J-hypergeom-app}; the corresponding representation of $I(p;\rho)$ is
\begin{equation}
	I(p;\rho)
	=
	\frac{e^{-(1-p)\rho}}{1-p}\,
	{}_2F_1\big(1,1-p;2-p;e^{-\rho}\big),
	\label{eq:I-hypergeom-app}
\end{equation}
    valid for $\Re p<1$ and extended elsewhere by analytic continuation. For large $\rho$,
\begin{equation}
	I(p;\rho)
	=
	\frac{e^{-(1-p)\rho}}{1-p}
	+ \mathcal{O}\big(e^{-(2-p)\rho}\big),
\end{equation}
    so that $I(p;\rho)\to 0$ in the overlap region as expected.

\subsubsection{Structure of the solution and coefficients}
\paragraph{General form and matching}
    Inserting~\eqref{eq:S-decomp-app} into the Green's function expression and performing the integrals using~\eqref{eq:I-hypergeom-app} we obtain the general structure
\begin{equation}
	\chi_1^{\rm (NZ)}(\rho)
	=
	\frac{e^{\lambda_-\rho}}{\lambda_- - \lambda_+}
	\left[
	C_0\,I(0;\rho)
	+ C_1\,I(2\lambda_-;\rho)
	+ C_2\,I(2\lambda_- -1;\rho)
	+ \sum_j D_j\,I(p_j;\rho)
	\right],
\end{equation}
	with coefficients $C_0,C_1,C_2,D_j$ that are rational functions of $(\lambda_\pm,K,\tilde{\beta},\tilde{\beta}_{\rm H},\kappa_{\rm eff},r_h)$. A boundary condition at the stretched horizon fixes one linear combination of the $C_i$, while excluding the growing far-zone mode fixes the relative integration constants.  Matching then relates the NLO near- and far-zone coefficients; it does not determine the absolute normalization inherited from the string-scale core.

    Equations~\eqref{eq:phi1-prime-NZ-correct}, \eqref{eq:chi1-GF-final}, and \eqref{eq:F1-closed-app} give the exterior NLO expressions within the adopted probe and radial approximations, at fixed matching data. The radial dependence is expressed in terms of exponentials and the functions defined in~\eqref{eq:J-hypergeom-app}--\eqref{eq:I-hypergeom-app}. The expressions do not determine the absolute condensate normalisation or the unresolved core contribution.

\newpage
\bibliographystyle{JHEP}
\bibliography{Refs}

\end{document}